\documentclass[lettersize,journal]{IEEEtran}
\usepackage[utf8]{inputenc} 
\usepackage[T1]{fontenc}
\usepackage{url}
\usepackage{ifthen}
\usepackage{cite}
\usepackage[cmex10]{amsmath} 
\usepackage[flushleft]{threeparttable} 
\usepackage{textcase}
\usepackage[tablename=Table]{caption}
\usepackage{blindtext}
\usepackage{floatrow}
\usepackage{soul}
\usepackage{changes}
\usepackage{gensymb}
\usepackage{cite}
\usepackage{amsmath,amssymb,amsfonts}
\usepackage{textcomp}
\usepackage{graphicx}
\usepackage{amssymb}
\usepackage{amsfonts}
\usepackage{amsmath}
\usepackage{epsfig}
\usepackage{color}
\usepackage{fancybox}
\usepackage{textcomp}
\usepackage{multirow}
\usepackage{setspa ce}
\usepackage{psfrag}
\usepackage{booktabs}
\usepackage{float}
\usepackage[caption = false]{subfig}
\usepackage{algorithm}
\usepackage{algpseudocode}
\usepackage{mathtools, nccmath, bigints, amsfonts}
\usepackage{multirow}
\usepackage{array}
\newcolumntype{L}{>{\centering\arraybackslash}m{3cm}}

\usepackage{mathrsfs}

\newcommand{\gb}[1]{{\textcolor[rgb]{0,0.3,0.5}{#1}}}
\newfloatcommand{capbtabbox}{table}[][0.4\textwidth]

\newtheorem{Remark}{Remark}
\usepackage{verbatim}
\usepackage{graphicx}
\usepackage{cite}

\makeatletter 
\newcommand\citecolor[1]{\@namedef{keycolor#1}{\color{blue}}}
\makeatother

    \def\Complex{{\rm\rule[.23ex]{.03em}{1.1ex}\kern-.3em{C}}}

    \newcommand{\be}{\begin{equation}} \newcommand{\ee}{\end{equation}}
    \newcommand{\bea}{\begin{eqnarray}} \newcommand{\eea}{\end{eqnarray}}
    \newcommand{\benum}{\begin{enumerate}} \newcommand{\eenum}{\end{enumerate}}

    \newcommand{\qa}{{\bf a}}
        \newcommand{\qb}{{\bf b}}
        \newcommand{\qc}{{\bf c}}

        \newcommand{\qf}{{\bf f}}
        \newcommand{\qg}{{\bf g}}
        \newcommand{\qh}{{\bf h}}

        \newcommand{\qm}{{\bf m}}
        \newcommand{\qn}{{\bf n}}

        \newcommand{\qu}{{\bf u}}
        \newcommand{\qv}{{\bf v}}
        
        \newcommand{\qx}{{\bf x}}
        \newcommand{\qy}{{\bf y}}

        \newcommand{\qA}{{\bf A}}
        \newcommand{\qB}{{\bf B}}

        \newcommand{\qH}{{\bf H}}
        \newcommand{\qI}{{\bf I}}
        
        \newcommand{\qK}{{\bf K}}

        \newcommand{\qV}{{\bf V}}
        \newcommand{\qW}{{\bf W}}

        \newcommand{\qzero}{{\bf 0}}

        \newcommand{\qSigma}{{\boldsymbol \Sigma}}

        \newcommand{\qPi}{{\boldsymbol \Pi}}

        \newcommand{\qlambda}{{\boldsymbol \lambda}}
        \newcommand{\qgamma}{{\boldsymbol \gamma}}

        \newcommand{\qmu}{{\boldsymbol \mu}}

        \newcommand{\calN}{{\mathcal N}}

        \newcommand{\Ex}{{\sf E}}

        \newcommand*{\argmin}{\operatornamewithlimits{argmin}\limits}

\def\BibTeX{{\rm B\kern-.05em{\sc i\kern-.025em b}\kern-.08em
    T\kern-.1667em\lower.7ex\hbox{E}\kern-.125emX}}

\begin{document}

\title{{Graph Neural Network Aided MU-MIMO Detectors}}

\author{Alva~Kosasih, Vincent~Onasis, Vera~Miloslavskaya, Wibowo~Hardjawana,\IEEEmembership{~Member,~IEEE}, Victor~Andrean, and Branka~Vucetic,\IEEEmembership{~Life~Fellow,~IEEE}  \thanks {The material in this paper was submitted in part to the 2022 IEEE Wireless Communications and Networking Conference (WCNC) \cite{2021AKosasih_WCNC_GEPNet}. } \thanks{ A. Kosasih, V. Onasis, V. Miloslavskaya, W. Hardjawana,  and B. Vucetic are with Centre of Excellence in Telecommunications, The University of Sydney, Sydney, Australia. email:\{alva.kosasih,vera.miloslavskaya,wibowo.hardjawana,branka.vucetic\} \ @sydney.edu.au.  V. Andrean is with HedgeDesk, San Diego, CA, USA. e-mail:  victor@thehedgedesk.com.}}

\maketitle

\begin{abstract}
Multi-user multiple-input multiple-output  (MU-MIMO) systems can be used to meet high throughput requirements of 5G and beyond networks. A base station serves many users in an uplink MU-MIMO system, leading to a substantial multi-user interference (MUI). Designing a high-performance detector for dealing with a strong MUI is challenging. This paper analyses the performance degradation caused by the posterior distribution approximation used in the state-of-the-art message passing (MP) detectors in the presence of high MUI. We develop a graph neural network  based framework to fine-tune the MP detectors' cavity distributions and thus improve the posterior distribution approximation in the MP detectors. We then propose two novel neural network  based detectors which rely on the expectation propagation (EP) and Bayesian parallel interference cancellation (BPIC), referred to as the GEPNet and GPICNet detectors, respectively. The GEPNet detector maximizes detection performance, while GPICNet detector balances the performance and complexity. We provide proof of the permutation equivariance property, allowing the detectors to be trained only once, even in the systems with dynamic changes of the number of users. The simulation results show that the proposed GEPNet detector performance approaches maximum likelihood  performance in various configurations and GPICNet detector doubles the multiplexing gain of BPIC detector.
\end{abstract}


\section{Introduction}

Multiuser  multiple-input multiple-output (MU-MIMO) technique is one of the key
technologies to enable a high throughput in 5G and beyond networks \cite{2020JZhang_JSAC_Beyond5G,Borges2021}. The usage of a high number of multiple transmit and receive antennas ensures a high spectral efficiency \cite{1998Foschini_WCommun_MultiuserMIMO,2017Bjornson_BOOK_MMIMO}, and therefore a high throughput. 
One of the challenging problems in uplink MU-MIMO systems 
is to design a practical base station detector that can achieve a high reliability performance in the presence of a strong multi-user interference (MUI). The MUI is caused by multiple user antennas simultaneously sending information to multiple base station antennas. The state-of-the-art practical high performance MU-MIMO  detectors can be classified into three categories \cite{2020_CZhang_JSEL_AI5G}: the linear, non-linear, and neural network (NN) based detectors.

The linear detectors (e.g., minimum-mean-square-error (MMSE) detector) are attractive due to their low complexity, however, their performance is far from  the near maximum  likelihood (ML) performance demonstrated by the sphere decoding \cite{ML,2019SphereDecoder,Hassibi_2005}, whose usage is restricted to the systems with a small number of users due to the computational complexity issue.
The non-linear detectors can be classified as interference cancellation (IC), tree search and message passing (MP) detectors \cite{2020_CZhang_JSEL_AI5G}. 
The MP detectors \cite{2009Donoho_ProcSci_AMP,2011_Goldberger_TInf_GTA,Ma-17ACCESS,Jespedes-TCOM14} can achieve an excellent performance with a reasonable complexity. They  use Gaussian distributions to approximate the posterior probability distribution of the transmitted symbols conditioned on the received signal. For simplicity, the latter distribution is further referred to as the posterior distribution. The performance of these detectors was shown to degrade away from the  ML performance in the presence of ill-conditioned channel matrices and/or when the number of user transmit antennas (users) is equal to the number of the base station receive antennas. The latter case is referred to as a high MUI scenario. 
For instance, the approximate message passing (AMP)  \cite{2009Donoho_ProcSci_AMP} and the Bayesian parallel interference cancellation (BPIC) \cite{AKosasih} detectors perform very poorly in the presence of ill-conditioned channel matrices and high MUI.
The problem of ill-conditioned channel matrices has been partially resolved by the orthogonal AMP (OAMP) detector  \cite{Ma-17ACCESS} that integrates the AMP with the linear MMSE filtering. 
The best MP detector, the expectation propagation (EP) detector  \cite{Jespedes-TCOM14}, significantly outperforms the OAMP detector by introducing regularization parameters in the MMSE filter that are iteratively adjusted according to the channel matrix and MUI level. 
However, there is still a significant gap to the ML performance exceeding 2 dB at the symbol  error  rate  (SER) of $10^{-4}$ for a $16\times 16$ MU-MIMO  configuration.
 
The NN based detectors have recently been proposed in   \cite{Corlay_2018,2018HHE_Globecom_OAMPNet,2020_HHe_TSP_OAMPNet,2019H_He_physical,2019_Samuel_TSP_Detnet,2021_KPratik_TSP_REMimo,AScotti_GNN_2020} to  address the performance limitation of the MP detectors in the case of ill-conditioned channel matrices and/or high MUI. This is done by unfolding the iteration process of the MP detectors into NN layers and optimizing NN parameters. 
For example, the OAMPNet detector \cite{2018HHE_Globecom_OAMPNet,2020_HHe_TSP_OAMPNet} combines the OAMP and NN to deal with the ill-conditioned channel matrices. This results in a significant performance improvement compared to the conventional OAMP detector. 
Another example of an NN based detector is the high performance recurrent equivariant (RE)-MIMO detector   \cite{2021_KPratik_TSP_REMimo}. The RE-MIMO detector unfolds the  AMP detector and integrates it with a  self-attention NN to significantly improve  the performance of the  conventional AMP detector.
Unfortunately, the RE-MIMO requires a large number
of NN parameters, which are difficult to optimize. Different from the RE-MIMO, the graph neural network (GNN) detector \cite{AScotti_GNN_2020} is based on the belief propagation algorithm and utilizes a pair-wise Markov random field (MRF) model. The use of pair-wise model results in significant reduction of the number of NN parameters. 
Nevertheless, a significant performance gap remains between the state-of-the-art practical detectors and the ML detector in the presence of ill-conditioned channel matrices and/or high MUI. 
As will be shown in this paper, the gap is due to the inaccuracy of the Gaussian approximation of the posterior distribution in practical detectors. 
To the best of the authors' knowledge, the inaccuracy of the Gaussian approximation in the state-of-the-art detectors has not been addressed in prior work.

In this paper, we first analyse  the state-of-the-art MP detectors, specifically, the EP, OAMP and BPIC
detectors, which use the posterior distribution approximation based on the independent Gaussian approximation and the Gaussian cavity distribution  in the detection process. We show that the use of this posterior distribution approximation leads to the loss of some MUI information in these detectors, and therefore to a severe performance degradation in a high MUI scenario. Guided by this analysis, we propose a GNN based framework to improve the detection performance by fine-tuning the cavity distribution. Compared to the conventional Gaussian cavity function parameterized only by mean and variance, the GNN based framework incorporates additional parameters into the cavity function to capture the MUI information. 
The framework is first used to improve the
posterior distribution approximation in the EP detector \cite{Jespedes-TCOM14}. We refer to the newly developed detector as the graph EP network (GEPNet) detector. 
We then demonstrate how the proposed GNN-based framework can also be used to improve the recently proposed  low-complexity BPIC detector \cite{AKosasih}. We refer to this detector as graph parallel interference cancellation network (GPICNet) detector. The choice of the BPIC detector is motivated by its low computational complexity as it does not need to  perform any matrix inversion operation in contrast to the EP detector.
A mathematical proof of the permutation equivariance for both GEPNet and GPICNet detectors is then provided. The permutation equivariance paves a way to the robustness to the changes in the number of users, which allows the proposed detectors to be trained only once even in the systems with dynamic changes of the number of users.
Simulation results show that the GEPNet detector is able to significantly outperform the EP \cite{Jespedes-TCOM14}, RE-MIMO \cite{2021_KPratik_TSP_REMimo}, OAMPNet \cite{2018HHE_Globecom_OAMPNet}, GNN \cite{AScotti_GNN_2020},  OAMP \cite{Ma-17ACCESS}, AMP \cite{2009Donoho_ProcSci_AMP}, BPIC \cite{AKosasih} detectors by more than $4$ dB at the SER of $10^{-4}$ for $64\times 64$ MU-MIMO configuration and achieve a near ML performance for  various  system  configurations. The GPICNet detector substantially improves the performance and doubles the multiplexing gain of its predecessor BPIC detector. 

The main contributions of this paper:
\begin{itemize}
\item We analyse the MP detectors by evaluating the accuracy of the joint posterior distribution approximation and investigating the connection between the accuracy of the posterior distribution approximation  and the cavity distribution, used to calculate the symbol estimates in the MP detectors. We show that the state-of-the-art posterior  distribution approximation is inaccurate in high MUI scenarios, which leads to inaccurate symbol estimates. 
\item We develop a GNN-based framework that can be used to improve the posterior distribution approximation by fine-tuning  the cavity distributions in the MP detectors \cite{2009Donoho_ProcSci_AMP,2011_Goldberger_TInf_GTA,Ma-17ACCESS,Jespedes-TCOM14}. 
In contrast to \cite{AScotti_GNN_2020}, the proposed GNN-based framework   is designed to allow processing of prior knowledge generated by the MP detectors. 
\item We propose two novel NN based detectors that use the developed GNN-based framework to improve the posterior distribution approximations in the EP and BPIC. The first detector, named GEPNet, is proposed  to ensure an excellent detection performance, while the second detector, named GPICNet, is proposed to balance the performance and complexity, since it does not need to perform  matrix inversion as opposed to the GEPNet detector.
\item We show that it suffices to train the proposed detector only once with two different numbers of users by mathematically proving that the proposed   detectors are equivariant to the user permutations and experimentally demonstrating that the proposed detectors are robust to the changes in the number of users. This is in contrast to the GNN  \cite{AScotti_GNN_2020} and OAMPNet \cite{2018HHE_Globecom_OAMPNet}, which need to be trained individually for each number of users according to \cite{AScotti_GNN_2020,2018HHE_Globecom_OAMPNet}, and the RE-MIMO detector \cite{2021_KPratik_TSP_REMimo}, which need to be trained with all possible numbers of users according to \cite{2021_KPratik_TSP_REMimo}.  
\end{itemize}

This paper is organized as follows. A system model for an MU-MIMO system  is introduced in Section II. It is followed by a brief overview of the EP detector in Section III. The analysis of the posterior distribution approximation in the EP and other MP detectors is given in Section \ref{Sect_Analysis_Gaussian}.  In Section \ref{sFramework}, we present the proposed GNN-based framework with the GEPNet and GPICNet detectors, prove that the proposed detectors are permutation equivariant, and analyse the computational complexity of the MU-MIMO detectors.  
In Section \ref{SectTrainingAndRobust}, we discuss the training method and robustness of the proposed detectors. In Section \ref{Simulation}, we show the simulation results.  Finally, Section \ref{sConclusion} concludes the paper.

{\bf Notations}: $\qI_n$ denotes an identity matrix of size $n$.  For any matrix $\mathbf{A}$, the notations $\mathbf{A}^{T}$  and $\mathbf{A}^{\dagger}$ stand for transpose and pseudo-inverse of $\mathbf{A}$, respectively.  $\|\cdot\|$ denotes the Frobenius norm of a vector or a matrix.   $q^*$ denotes the complex conjugate of a complex number $q$. Let $\qx = [x_1, \cdots, x_K]^T$ and $\qc = [c_1, \cdots, c_K]^T$.
 ${\Ex}[\qx]$ is the mean of random vector $\qx$, and ${\mathrm{Var} }[\qx] = {\Ex}\big[\left(\qx-{\Ex}[\qx]\right)^2\big]$ is its variance.   $\calN(x_k: c_k,v_k)$ represents a single variate Gaussian distribution  for a  random variable $x_k$ with mean $c_k$ and variance $v_k$. $[K]=\{1,2,\dots,K\}$ is the set of all natural numbers up to $K$.

\section{System Model}
\label{sectSystemModel}

\begin{figure}
\centering
{\includegraphics[scale=0.35]{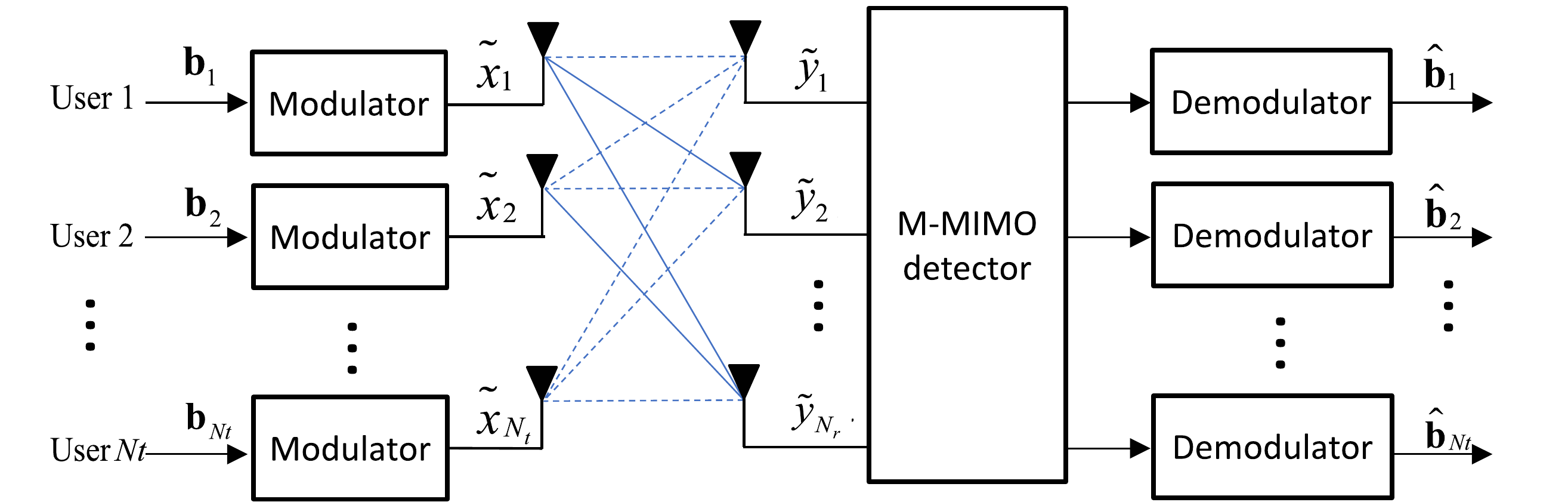}}
\caption{The MU-MIMO system}
\label{up_MU-MIMO}
\end{figure}

We consider an MU-MIMO system used to transmit information streams generated by $N_t$ single-antenna users.  The streams are received by a base station, which is equipped with $N_r\geq N_t$ antennas to simultaneously serve the users. The system is  depicted in  Fig. \ref{up_MU-MIMO}. User $k$ maps $\log_2(\tilde{M})$ bits of its information stream $\qb_k$ to a symbol $\tilde{x}_k \in \tilde{\Omega}$  using a quadrature amplitude modulation (QAM) technique, where $\tilde{\Omega} = \{ s_1, \dots, s_{\tilde{M}}\}$ is a constellation set of $\tilde{M}$-QAM  and $s_m$ is the $m$-th constellation point. The transmitted symbols are uniformly distributed, and the corresponding received signal is given by 
\begin{equation} \label{eII_1a}
\tilde{\qy} = \tilde{\qH} \tilde{\qx} + \tilde{\qn},
\end{equation}
where $\tilde{\qx} = [\tilde{x}_1, \cdots, \tilde{x}_{N_t}]^T$, $\tilde{\qy}=[\tilde{y}_1, \ldots, \tilde{y}_{N_r}]^{T}$,  $\tilde{\qH}=[\tilde{\qh}_1,\dots, \tilde{\qh}_k, \ldots, \tilde{\qh}_{N_t}]  \in \mathbb{C}^{N_r \times {N_t}} $ is the coefficient  matrix of complex memoryless Rayleigh fading channels between ${N_t}$ transmit and $N_r$ receive antennas,  $\tilde{\qh}_k$ is the $k$-th column vector of matrix $\tilde{\qH}$ that denotes wireless channel coefficients between the receive antennas and the $k$-th transmit antenna, where each coefficient follows a Gaussian distribution with zero mean and variance of $1/N_r$, and $\tilde{\qn} \in \mathbb{C}^{N_r}$ denotes the additive white Gaussian noise (AWGN) with a zero mean and covariance matrix $\tilde{\sigma}^2 \qI_{N_r}$. 
For convenience, the complex-valued variables are transformed into real-valued variables. Accordingly, we define $\qx = [\mathcal{R}(\tilde{\qx})^T \quad \mathcal{I}(\tilde{\qx})^T]^T  \in \mathbb{R}^K $, $\qy = [\mathcal{R}(\tilde{\qy})^T \quad \mathcal{I}(\tilde{\qy})^T]^T \in \mathbb{R}^N$, $\qn = [\mathcal{R}(\tilde{\qn})^T \quad \mathcal{I}(\tilde{\qn})^T]^T \in \mathbb{R}^N$, and 
$\qH=\begin{bmatrix}
  \mathcal{R}(\tilde{\qH}) & -\mathcal{I}(\tilde{\qH}) \\ 
  \mathcal{I}(\tilde{\qH})  & \mathcal{R}(\tilde{\qH}) 
\end{bmatrix} \in \mathbb{R}^{N \times K}$,
where $K = 2 N_t$,  $N=2N_r$, $\mathcal{R}(\cdot)$ and $\mathcal{I}(\cdot)$ are the real and imaginary parts, respectively.
Therefore, we can rewrite \eqref{eII_1a} as 
\begin{equation} \label{eII_1}
\qy = \qH \qx + \qn.
\end{equation}
Note that the covariance matrix of $\qn$ is  $\sigma^2 \qI_{N}$ with $\sigma\triangleq    \tilde{\sigma}/\sqrt{2}$,  the energy per transmit antenna in the real-valued system is $\mathit{E}_s \triangleq \tilde{\mathit{E}_s}/2$, where $\tilde{\mathit{E}_s}=1$, and the real-valued constellation is  $\Omega = \{ \mathcal{R}(s_m) | s_m\in \tilde{\Omega} \}$ with $| \Omega  |= M \triangleq  \sqrt{\tilde{M}}$.  
The SNR of the system is defined as SNR $ = 10\log_{10}\frac{\Ex \left[ \|\qH\qx \|^2 \right]}{\Ex \left[ \|\qn \|^2 \right]} $ dB. 
 We consider the system model from \eqref{eII_1} for the rest of the paper. 
We assume that a detector has a perfect knowledge of $\qH, E_s$, and $\sigma^2$. The ML detector finds $\hat{\qx}=\argmin_{\qx\in \Omega^K} p(\qy|\qx)$, where $p(\qy|\qx)\propto \exp\left(-\frac{\|\qy-\qH\cdot\qx\|^2}{2\sigma^2}\right)$ due to \eqref{eII_1}. Given a received vector $\qy$, the posterior distribution $p(\qx|\qy)$ can be expressed as  $p(\qx|\qy)=p(\qy|\qx)\cdot p(\qx)/p(\qy)\propto p(\qy|\qx)$ since the transmitted symbols are uniformly distributed, and therefore \begin{equation}
    p(\qx |\qy)=\frac{1}{Z}\exp\left(-\frac{\|\qy-\qH\cdot\qx\|^2}{2\sigma^2}\right),
    \label{eq_p_true}
\end{equation}
where $Z$ is a normalization constant such that $\sum_{\qx\in \Omega^K}p(\qx |\qy)=1$. We further refer to $p(\qx |\qy)$ in \eqref{eq_p_true} as the true posterior distribution.   
The maximization of $p(\qx|\qy)$ corresponds to the maximum a posteriori (MAP) detection.   Thus, the ML detection solution coincide with the MAP detection solution and is given by 
\begin{equation}
    \hat{\qx}=\argmin_{\qx\in \Omega^K} \|\qy-\qH\cdot\qx\|^2.
    \label{eq:ML}
\end{equation}
Unfortunately, the complexity of solving \eqref{eq:ML} increases rapidly with the number of users $K$.

\section{Conventional MP Detectors}

In this section we briefly review the conventional MP detectors, specifically, the EP \cite{Jespedes-TCOM14} and the BPIC \cite{AKosasih} detectors. Both detectors consist of two modules: the observation and estimation modules.
The goal of the two modules is to iteratively approximate the posterior probability distribution of the transmitted symbols using tunable parameters and calculate the symbol estimates based on the approximated probability distribution.  At each iteration, the observation module is used to calculate the approximated probability distribution using tunable parameters while the estimation module updates the tunable parameters and calculates the symbol estimates.

\subsection{The Expectation Propagation Detector}

In the following, we explain the observation and estimation modules in the EP.

\subsubsection{The Observation Module}

The posterior probability distribution of the transmitted symbols conditioned on the received signal in \eqref{eII_1} can be expressed as
\begin{flalign} \label{eq_EP:Posterior_ori}
    & p(\qx|\qy) = \frac{p(\qy|\qx) }{ p(\qy)} \cdot p(\qx) \propto \underbrace{\mathcal{N} \left( \qy: \qH \qx ,  \sigma^2 \qI_{N} \right) }_{p(\qy|\qx)}    \underbrace{ \prod_{k=1}^{K} p(x_{k}) }_{p(\qx)},
\end{flalign}
where $x_k$ is treated as a continuous random variable, $p(x_k) = \frac{1}{M} \sum_{x \in \Omega } \delta(x_k-x)$ is a priori probability density function  of $x_k$, $\delta$ is the Dirac delta function, and $p(\qy)$ is omitted as it is not related to random variable $x_k$.   
By multiplying system model \eqref{A2} by ${\qH}^\dagger=\left( \qH^T \qH \right)^{-1}\qH^T$, which is a preudoinverse matrix of $\qH$, one obtains that $p(\qy|\qx)\propto \mathcal{N}   \left( \qx: {\qH}^\dagger \qy, \sigma^{2} \left( \qH^T \qH \right)^{-1} \right)$ for a given received vector $\qy$ \cite{Jespedes-TCOM14}. The EP scheme uses this to iteratively approximate $p(\qx|\qy)$ as 
\begin{flalign}\label{eq_EP:Post_approx}
p^{(t)}(\qx|\qy) \propto & p(\qy | \qx) \cdot \chi^{(t)}(\qx)\notag \\
 \propto & \mathcal{N}  \left( \qx: {\qH}^\dagger \qy, \sigma^{2} \left( \qH^T \qH \right)^{-1} \right)  \cdot \notag \\ 
  & \mathcal{N}  \left(\qx: (\qlambda^{(t-1)})^{-1} \qgamma^{(t-1)}, (\qlambda^{(t-1)})^{-1}  \right)\notag \\
\propto &\mathcal{N}  \left( \qx:\qmu^{(t)}, \qSigma^{(t)} \right),
\end{flalign}
where $t$ denotes the iteration index,  $\chi^{(t)}(\qx)$ is an approximation of $p(\qx)$ obtained from the exponential family \cite{Jespedes-TCOM14},  $\qlambda^{(t)} $ is a $K \times K$ diagonal matrix with diagonal elements $\lambda_{k}^{(t)}>0 $  and $ \qgamma^{(t)} = [\gamma_{1}^{(t)}, \dots, \gamma_{K}^{(t)}]^T$. Both  $\lambda_{k}^{(t)}$ and  $\gamma_k^{(t)}$  are real  valued tunable parameters with  $\lambda_{k}^{(0)}=1/\mathit{E}_s $ and $\gamma_k^{(0)}=0$.
Both $\qlambda^{(t)}$ and $\qgamma^{(t)}$ are derived in Section \ref{secEP_EstimModule}.
Note that $p(\qy|\qx)$ in \eqref{eq_EP:Post_approx} is approximated by treating $\qx$ as a random real-valued vector. The product of two Gaussians in \eqref{eq_EP:Post_approx} is computed by using the Gaussian product property\footnote{The product of two Gaussians results in another Gaussian, $\mathcal{N}(\qx:\qa,\qA) \cdot \mathcal{N}(\qx:\qb,\qB)  \propto \mathcal{N} (\qx:(\qA^{-1}+\qB^{-1})^{-1}(\qA^{-1} \qa + \qB^{-1} \qb),(\qA^{-1}+\qB^{-1})^{-1}$.}, given in Appendix A.1 of \cite{Rasmussen-BOOK}. Accordingly, we can obtain the covariance and mean of $p^{(t)}(\qx|\qy) $ as
\begin{subequations} \label{eA1_a0102}
            \begin{align}
&\qSigma^{(t)} =  {  \left( \sigma^{-2} \qH^T \qH+ \qlambda^{(t-1)} \right)}^{-1}, \label{eA1_a01}\\
& \qmu^{(t)} =\qSigma^{(t)} {\left( \sigma^{-2} \qH^T \qy + \qgamma^{(t-1)}\right)}. \label{eA1_a02}
            \end{align}
        \end{subequations}
 
To reduce the computational complexity, the EP   approximates $p^{(t)}(\qx|\qy) $ by a product of independent Gaussian distributions as
 \begin{flalign}\label{Gaussian_approx}
q^{(t)}(\qx) = \prod_{k=1}^K \underbrace{\mathcal{N}  \left( x_k: \mu_k^{(t)}, \Sigma_k^{(t)} \right)}_{q^{(t)}(x_k)},
 \end{flalign}
where $\mu_k^{(t)}$ is the $k$-th element in vector $\qmu^{(t)} $  and $\Sigma_k^{(t)}$ is the $k$-th diagnonal element  in covariance matrix $\qSigma^{(t)}$, obtained from \eqref{eA1_a0102}. As a consequence of using such an approximation, the EP losses some MUI information corresponding to the off-diagonal elements of covariance matrix $\qSigma^{(t)}$. We refer to expression \eqref{Gaussian_approx} as the independent  Gaussian  approximation (IGA), since the off-diagonal elements of the covariance matrix $\qSigma^{(t)}$ are assumed to be zero. 

 The EP then calculates the cavity distribution \cite{Jespedes-TCOM14} based on $q^{(t)}(x_k)$ as
 \begin{flalign}\label{eA1_a0304raw}
q^{(t) \backslash k	}(x_k) \triangleq \frac{q^{(t)}(x_k)}{\chi^{(t)}(x_k)}  &\propto  \frac{\mathcal{N}  \left( x_k:\mu_k^{(t)}, \Sigma_k^{(t)} \right)}   {\mathcal{N} \left(x_k:(\gamma_k^{(t-1)}/\lambda_k^{(t-1)}), ( 1/\lambda_k^{(t-1)})\right)} 
\notag\\ & \propto \mathcal{N}  \left( x_k: x_{{\rm obs},k}^{(t)}, v_{{\rm obs},k}^{(t)} \right), k\in [K],
 \end{flalign}
 where 
 \begin{subequations} \label{eA1_a0304}
            \begin{align}
&v_{{\rm obs},k}^{(t)} =  \frac{\Sigma_{k}^{(t)} }{1- \Sigma_{k}^{(t)}  \lambda_{k}^{(t-1)}},  \label{eA1_a03}\\
&x_{{\rm obs},k}^{(t)}  = v_{{\rm obs},k}^{(t)}  {\left(\frac{\mu_{k}^{(t)}}{\Sigma_{k}^{(t)}}-\gamma_{k}^{(t-1)}\right)}.  \label{eA1_a04} 
            \end{align}
        \end{subequations}
Vector $ \qx^{(t)}_{\rm obs} = [x^{(t)}_{{\rm obs},1}, \dots,x^{(t)}_{{\rm obs},K}]^T $ and $K \times K$ diagonal matrix $\qV^{(t)}_{\rm obs}$ with the main diagonal $[v^{(t)}_{ {\rm obs},1},\dots,v^{(t)}_{ {\rm obs},K}]^T$ are then forwarded to the estimation module.

\subsubsection{The Estimation Module} 
\label{secEP_EstimModule}

In this module, the EP calculates the soft symbol estimate and its variance based on the cavity  distribution given in \eqref{eA1_a0304raw}  as  
 \begin{subequations}\label{eA1_b0102_EP}
            \begin{equation}\label{eA1_b01_EP}
\hat{x}_k^{(t)}=   \sum_{a\in \Omega}  a \times q^{(t) \backslash k	}(x_k =a) ,
            \end{equation}
            \begin{equation}
v_k^{(t)} =  \sum_{a\in \Omega}  \left(x_k  -\hat{x}_k^{(t)}\right)^2 \times q^{(t) \backslash k	}(x_k =a), 
            \end{equation}
\end {subequations}
where $k\in [K]$. 
The EP detection is finished once the maximum number of iterations  $T$ has been reached. Hard estimate of the  transmitted symbols is then made  from $\hat{\qx}^{(T)}$ by converting its back into the complex domain and comparing  its Euclidean distances from the elements of the constellation set $\tilde{\Omega}$. Thus, the detection result for the $k$-th user is given by a hard estimate of $\hat{x}^{(t)}_k$, which is the mean of the cavity distribution $q^{(t) \backslash k}(x_k)$ discretized on $\Omega$ as shown in \eqref{eA1_b01_EP}. Therefore, the accuracy of 
the detection result directly depends on the cavity distribution. In the case of $t < T$, the approximate posterior distribution ${p}^{(t+1)}(\qx|\qy)$  is obtained by replacing $\chi^{(t)}(\qx)$ in \eqref{eq_EP:Post_approx} by \cite{Jespedes-TCOM14}
 \begin{flalign} \label{eq:EP_inference_recon}
\chi^{(t+1)}(\qx)
 &\propto  \frac{\mathcal{N}  \left( \qx:  \hat{\qx}^{(t)}, \qV^{(t)} \right)}{\mathcal{N} \left( \qx: \qx^{(t)}_{{\rm obs}},\qV^{(t)}_{{\rm obs}}\right) } \notag \\
 & =  \mathcal{N}  \left(  \qx: (\qlambda^{(t)})^{-1} \qgamma^{(t)},  (\qlambda^{(t)})^{-1}   \right),
\end{flalign} 
where the vector $\hat{\qx}^{(t)}= [\hat{x}_1^{(t)}, \dots,\hat{x}_K^{(t)}]^T $, the $K \times K$ diagonal matrix  $\qV^{(t)}$ is specified by its diagonal $[v_1^{(t)},\dots,v_K^{(t)}]^T$, and 
\begin{subequations} \label{eA1_b0304}
            \begin{align}
&\qlambda^{(t)} = (\qV^{(t)})^{-1} -  (\qV^{(t)}_{{\rm obs}})^{-1},  \label{eA1_b03}\\
&\qgamma^{(t)} =  (\qV^{(t)})^{-1} \hat{\qx}^{(t)} - (\qV^{(t)}_{{\rm obs}})^{-1}   \qx^{(t)}_{{\rm obs}}. \label{eA1_b04}
            \end{align}
\end{subequations}
Note that $\qlambda^{(t)}$ in \eqref{eA1_b03} may yield a negative value, which should not be the case as it is inverse variance term. Therefore, when $\lambda_k^{(t)} <0$, the EP assigns $\lambda_k^{(t)}=\lambda_k^{(t-1)}$ and $\gamma_k^{(t)}=\gamma_k^{(t-1)}$. Finally, a damping calculation is performed for $(\qlambda^{(t)},\qgamma^{(t)})$ as
\begin{subequations} \label{eq:damping}
\begin{align}
   \qlambda^{(t)} &= (1-\eta)\qlambda^{(t)}+\eta \qlambda^{(t-1)}, \label{damping_lambda} \\
   \qgamma^{(t)} &= (1-\eta)\qgamma^{(t)}+\eta\qgamma^{(t-1)}\label{damping_gamma},
\end{align}
\end{subequations}
where $\eta\in[0, 1]$ is a weighting coefficient.
The estimation module sends the parameters $(\qgamma^{(t)},\qlambda^{(t)})$ to the observation module for the next iteration.

\subsection{The Bayesian Parallel Interference Cancellation Detector} \label{sBPIC}

In this section, we explain the BPIC detector \cite{AKosasih}. The BPIC detector does not need to perform any matrix inversion in contrast to the EP detector, and therefore it has substantially lower computational complexity. Similarly to the EP detector, the BPIC detector consists of the observation and estimation modules.

\subsubsection{The Observation Module} \label{sObs_BPIC}

Note that the observation module of the BPIC detector is referred to as the Bayesian symbol observation module in
\cite{AKosasih}. Similarly to the EP, the joint posterior probability of the transmitted symbols  is approximated by a product of conditionally independent Gaussian functions as $p(\qx|\qy) \approx \prod_{k=1}^K \mathcal{N}  \left( x_k: \mu_{k}^{(t)}, \Sigma_{k}^{(t)} \right)$. Differently from the EP, the mean and variance are defined as in \cite{AKosasih}  
 \begin{subequations}\label{PIC_estim}
				\begin{equation}\label{mean_PIC}
\mu_{k}^{(t)} = \frac{\qh_k^T \left( \qy-\qH \hat{\qx}_{\backslash k}^{(t-1)} \right)}{\qh_k^T\qh_k},
				\end{equation}
            	\begin{equation}\label{var_PIC}
\Sigma_{k}^{(t)} =  \frac{1}{(\qh_k^T\qh_k)^2} \left( \sum_{j\in [K]\setminus\{k\}} s_j^2  v_{j}^{(t-1)} + \qh_k^T\qh_k  \sigma^2 \right),
           		 \end{equation}
\end {subequations}
 where $ \hat{\qx}_{\backslash k}^{(t-1)}   = \left[\hat{x}_{1}^{(t-1)}  , \dots, \hat{x}_{k-1}^{(t-1)} , 0, \hat{x}_{k+1}^{(t-1)}, \dots, \hat{x}_{, K}^{(t-1)} \right]^T $ are the symbol estimates in the $(t-1)$-th iteration,    $s_j =\qh_k^T\qh_j$, and $v_{j}^{(t-1)}$ is the variance of the symbol estimates in  iteration $t-1$.  
At the first iteration, we initialise $\hat{\qx}^{(0)}  = \left[\hat{x}_{1}^{(0)}  , \dots, \hat{x}_{K}^{(0) }\right]^T = \mathbf{0}$ indicating that the PIC is inactive and thus we have $\mu_{k}^{(1)} =\frac{\qh_k^T }{\qh_k^T\qh_k} \qy$, which is the expression of the matched filter  \cite{MRC-LDPC}.  The mean $\qmu^{(t)}$ and variance ${\qSigma}^{(t)}$ of the approximate posterior distribution are then forwarded to the estimation module. 

\subsubsection{The Estimation Module} \label{sEst_BPIC}

Similarly to the EP, the BPIC computes the soft symbol estimate $\hat{x}^{(t)}_k$ of the $k$-th user and its variance $v^{(t)}_k$ in \eqref{eA1_b0102_EP} using $\mathcal{N}  \left( x_k: \mu^{(t)}_k, {\Sigma}^{(t)}_k \right)$ as $q^{(t) \backslash k}(x_k)$.
We then apply the decision  statistics  combining (DSC) concept before sending the estimation results back to the observation module. The DSC exploits the conditionally uncorrelated symbol estimates in the early iterations. According to \cite{Branka_PIC_book}, the symbol estimates are combined using a weighting coefficient 
\begin{equation}\label{DSC_coef}
\rho_{k}^{(t)} =  \frac{e_k^{(t-1)}}{e_k^{(t)}+e_k^{(t-1)}}, 
\end{equation}
 where  $e_k^{(t)}$ is defined as the instantaneous square error of the $k$-th symbol estimate, which can be computed by using a linear filter such as matched filter,
\begin{flalign}\label{DSC_error}
 e_k^{(t)}  =  \left\| \frac{\qh_k^T}{\qh_k^T\qh_k} \left(  \qy - \qH \hat{\qx}^{(t)} \right)\right\|^2.
\end{flalign}
The value of a linear combination of the symbol estimates in two consecutive iterations is assigned to the current symbol estimate \begin{subequations}\label{DSC}
\begin{equation}\label{DSC_mean}
\hat{x}_{k}^{(t)} \leftarrow \left( 1-\rho_{k}^{(t)} \right)  \hat{x}_{k}^{(t-1)}   +   \rho_{k}^{(t)}   \hat{x}_{k}^{(t)},
\end{equation}
\begin{equation}\label{DSC_Var}
v_{k}^{(t)} \leftarrow \left( 1-\rho_{k}^{(t)} \right)  v_{k}^{(t-1)}   +   \rho_{k}^{(t)}   v_{k}^{(t)}. 
\end{equation}
 \end{subequations}
Note that in the first iteration we do not perform the DSC scheme.

\section{Analysis of the Posterior Distribution Approximation}
\label{Sect_Analysis_Gaussian}

In this section, we first analyse the accuracy of the IGA-based posterior distribution approximation  in the state-of-the-art MP detectors, specifically, the EP \cite{Jespedes-TCOM14}, OAMP \cite{Ma-17ACCESS} and BPIC \cite{AKosasih} detectors. We then link the inaccuracy of the IGA-based  posterior distribution approximation in a high MUI scenario with the inability of the symbol estimator, used in these detectors, to deal with the correlated non-Gaussian residual noise.

\subsection{Accuracy of the Posterior Distribution Approximation}

We first explain the importance of having the accurate posterior distribution approximation, in the state-of-the-art MP detectors. Unfortunately, the exact true posterior distribution \eqref{eq_p_true} and its maximum can be found only for small $K$ due to rapidly increasing computational complexity. The MP detectors such as the EP, OAMP and BPIC enable detection for high $K$ by iteratively approximating the true posterior distribution $p(\qx |\qy)$ by a factorizable distribution $q^{(t)}(\qx)$, 
where $t$ is the iteration index. Specifically, these detectors find $q^{(t)}(\qx)$ in form of the IGA \eqref{Gaussian_approx}, which means that $q^{(t)}(\qx)$ is given by the  product  of  Gaussian  distributions. Since any Gaussian distribution is conveniently defined just by the first two moments, i.e., the  mean  and  variance,  $q^{(t)}(\qx)$ is defined by the mean $\qmu^{(t)}\in\mathbb{R}^K$ and diagonal covariance matrix $\qSigma^{(t)}\in\mathbb{R}^{K\times K}$. Therefore, we characterize the similarity between $q^{(t)}(\qx)$ and $p(\qx|\qy)$ by the average difference between the moments of $p(\qx|\qy)$ and $q^{(t)}$ computed as $\Delta^{(t)}_{\mu}={\sf E} [\|\qmu_{{\rm true}}-\qmu^{(t)}\|]$ and $\Delta^{(t)}_{\Sigma}={\sf E} [\|{\rm diag}(\qSigma_{{\rm true}})-{\rm diag}(\qSigma^{(t)})\|]$, 
where $\qmu_{{\rm true}}$ is the mean and $\qSigma_{{\rm true}}$ is the covariance matrix of the true posterior distribution $p(\qx|\qy)$ from \eqref{eq_p_true}, function ${\rm diag}(\cdot )$ returns the main diagonal of a given matrix, and the mean is taken over realizations of transmitted symbols $\qx$, channel matrix $\qH$, noise $\qn$. Note that $\qmu_{{\rm true}}$ and $\qSigma_{{\rm true}}$ are computed by representing the true posterior distribution as a set $\{[\qx,p(\qx|\qy)]\}_{\qx\in\Omega^K}$ 
of possible outcomes $\qx\in\Omega^K$ and their probabilities $p(\qx|\qy)$. 
Table \ref{table_analysis} shows the values of $\Delta^{(T)}_{\mu}$ and $\Delta^{(T)}_{\Sigma}$, where $t = T$ corresponds to the last iteration. Note that the SNR values are set so that the
ML detector can achieve the SER of $10^{-3}$. 
It can be seen that the values of $\Delta^{(T)}_{\mu}$ and $\Delta^{(T)}_{\Sigma}$ increase with the transmit-to-receive antennas ratio $K/N$. Therefore, $q^{(T)}(\qx)$ diverges from $p(\qx|\qy)$ with increasing $K/N$.

\begin{table*}\small
\caption{The accuracy of the posterior distribution approximation with $4$-QAM modulation scheme}
\centering
\small
\renewcommand{\arraystretch}{1.1}
\begin{tabular}{|l|l|c|c|c|c|c|} \hline 
Detector                                   & Configuration                & SER & $\Delta^{(T)}_{\mu}$ & $\Delta^{(T)}_{\Sigma} $ & $r^{(T)}$ & $c^{(T)}$ \\ \hline \hline 
\multirow{3}{*}{EP \cite{Jespedes-TCOM14}} & $N=16$, $K=8$, SNR=$9$dB     & $0.0014$ & $0.0134$ & $0.0051$ & $0.9993$ & $0.0146$ \\ \cline{2-7}
                                           & $N=16$, $K=12$, SNR=$11$dB   & $0.0018$ & $0.0192$ & $0.0071$ & $0.9978$ & $0.0230$ \\ \cline{2-7}
                                           & $N=16$, $K=16$, SNR=$12.5$dB & $0.0025$ & $0.0281$ & $0.0095$ & $0.9947$ & $0.0390$ \\ \hline \hline 
\multirow{3}{*}{OAMP \cite{Ma-17ACCESS}}   & $N=16$, $K=8$, SNR=$9$dB     & $0.0039$ & $0.5152$ & $0.1142$ & $0.9880$ & $0.0180$ \\ \cline{2-7}
                                           & $N=16$, $K=12$, SNR=$11$dB   & $0.0083$ & $0.6587$ & $0.1351$ & $0.9695$ & $0.0420$ \\ \cline{2-7}
                                           & $N=16$, $K=16$, SNR=$12.5$dB & $0.0204$ & $1.5917$ & $0.3838$ & $0.9336$ & $2.5314$ \\ \hline \hline 
\multirow{3}{*}{BPIC \cite{AKosasih}}      & $N=16$, $K=8$, SNR=$9$dB     & $0.0037$ & $0.5116$ & $0.1100$ & $0.9955$ & $0.0150$ \\ \cline{2-7}
                                           & $N=16$, $K=12$, SNR=$11$dB   & $0.0138$ & $0.6474$ & $0.1519$ & $0.9741$ & $0.0328$ \\ \cline{2-7}
                                           & $N=16$, $K=16$, SNR=$12.5$dB & $0.0408$ & $0.8412$ & $0.1787$ & $0.9054$ & $0.0528$ \\ \hline
\end{tabular}
\label{table_analysis}
\end{table*}

We further consider the detection as an optimization problem with the objective function $p(\qx|\qy)$ to be maximized over $\qx\in\Omega^K$, and evaluate how close the approximate maximum $p(\qx_{\rm est}^{(t)}|\qy)$ is to the true maximum $p(\qx_{\rm ML}|\qy)$ by computing $r^{(t)}={\Ex}[p(\qx^{(t)}_{\rm est}|\qy)/p(\qx_{\rm ML}|\qy)]$, where $\qx_{\rm est}^{(t)}$ is an  estimate of transmitted symbols found by a detector in the $t$-th iteration, $\qx_{\rm ML}$ is the ML detection solution, $p(\qx^{(t)}_{\rm est}|\qy)$ and $p(\qx_{\rm ML}|\qy)$ are calculated by substituting $\qx^{(t)}_{\rm est}$ and $\qx_{\rm ML}$ into \eqref{eq_p_true}, and the mean is taken over $\qx$, $\qH$, and $\qn$. Table \ref{table_analysis} shows that $r^{(T)}$ decreases with the transmit-to-receive antennas ratio $K/N$ for the EP, BPIC and OAMP detectors, while the SER increases. Therefore, we conclude that the accuracy of the posterior distribution approximation deteriorates with increasing $K/N$ for the EP, BPIC and OAMP detectors.

\subsection{Role of the Cavity Distribution}



The EP, AMP, OAMP and BPIC detectors employ
the element-wise 
symbol estimator that is defined by \eqref{eA1_a0304raw} and \eqref{eA1_b0102_EP}. 
 The  symbol estimator is optimal only for uncorrelated Gaussian residual noise 
\cite{XYuan2014PA,2020_Khani_TWC_Adaptive}. 
The residual noise of the $k$-th user is given by $\varepsilon^{(t)}_k = (x_{{\rm obs},k}^{(t)}-x_{{\rm ML},k})/{\sqrt{v^{(t)}_{{\rm obs},k}}}$, where $x_{{\rm ML},k}$ is the $k$-th symbol  
of the ML solution $\qx_{\rm ML}$, $x_{{\rm obs},k}^{(t)}$ and $v^{(t)}_{{\rm obs},k}$ are defined by \eqref{eA1_a0304},  and $k\in [K]$. 

To evaluate the correlation between the residual noise of different users, we generated $100000$ random realizations of symbols $\qx$, noise $\qn$ and channel matrix $\qH$. We applied the EP, OAMP and BPIC detectors with $T=10$ iterations and calculated $\varepsilon_k^{(t)}$, $k\in [K]$, for each realization. 
We then computed the Pearson correlation coefficient $C^{(t)}_{k,k'}$ between the normalized residual noise 
for each $(k,k')\in[K]^2$ pair of users. 
As a result, we obtain $K\times K$ matrix $C^{(t)}$ with the unit diagonal. Since we are interested only in the off-diagonal elements of the matrix $C^{(t)}$, we compute coefficient $c^{(t)}=\|C^{(t)}-\qI_{K}\|$. Table \ref{table_analysis} shows the value of $c^{(t)}$ for the EP, OAMP and BPIC detectors, $t=T$. Since $c^{(T)}$ increases with the transmit-to-receive antennas ratio $K/N$, we conclude that the user noise correlation increases with  $K/N$. 

\begin{figure*}
\centering
\subfloat[The BPIC detector]
{\includegraphics[scale=0.31]{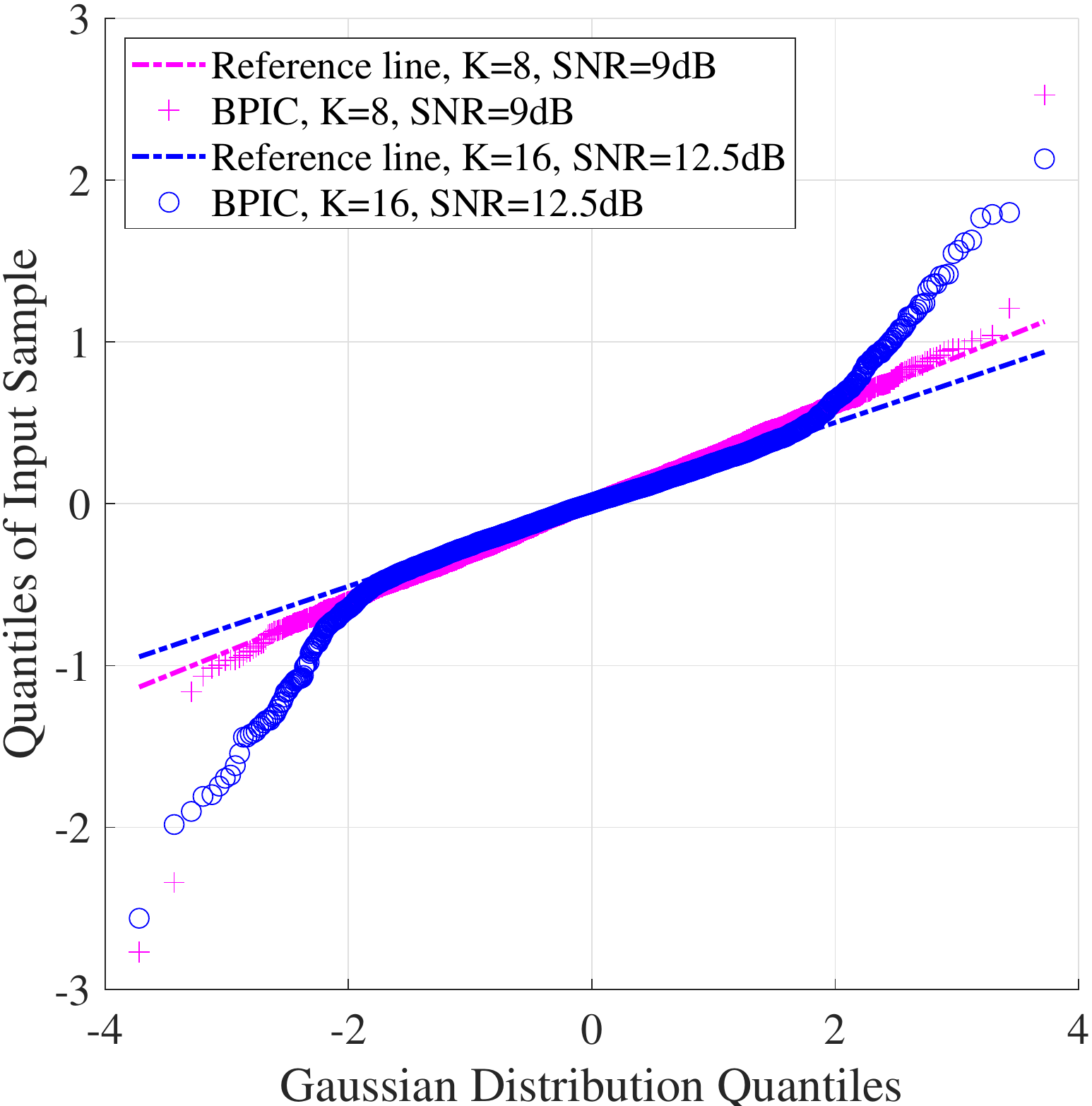}}\hfill
\centering
\subfloat[The OAMP detector]
{\includegraphics[scale=0.31]{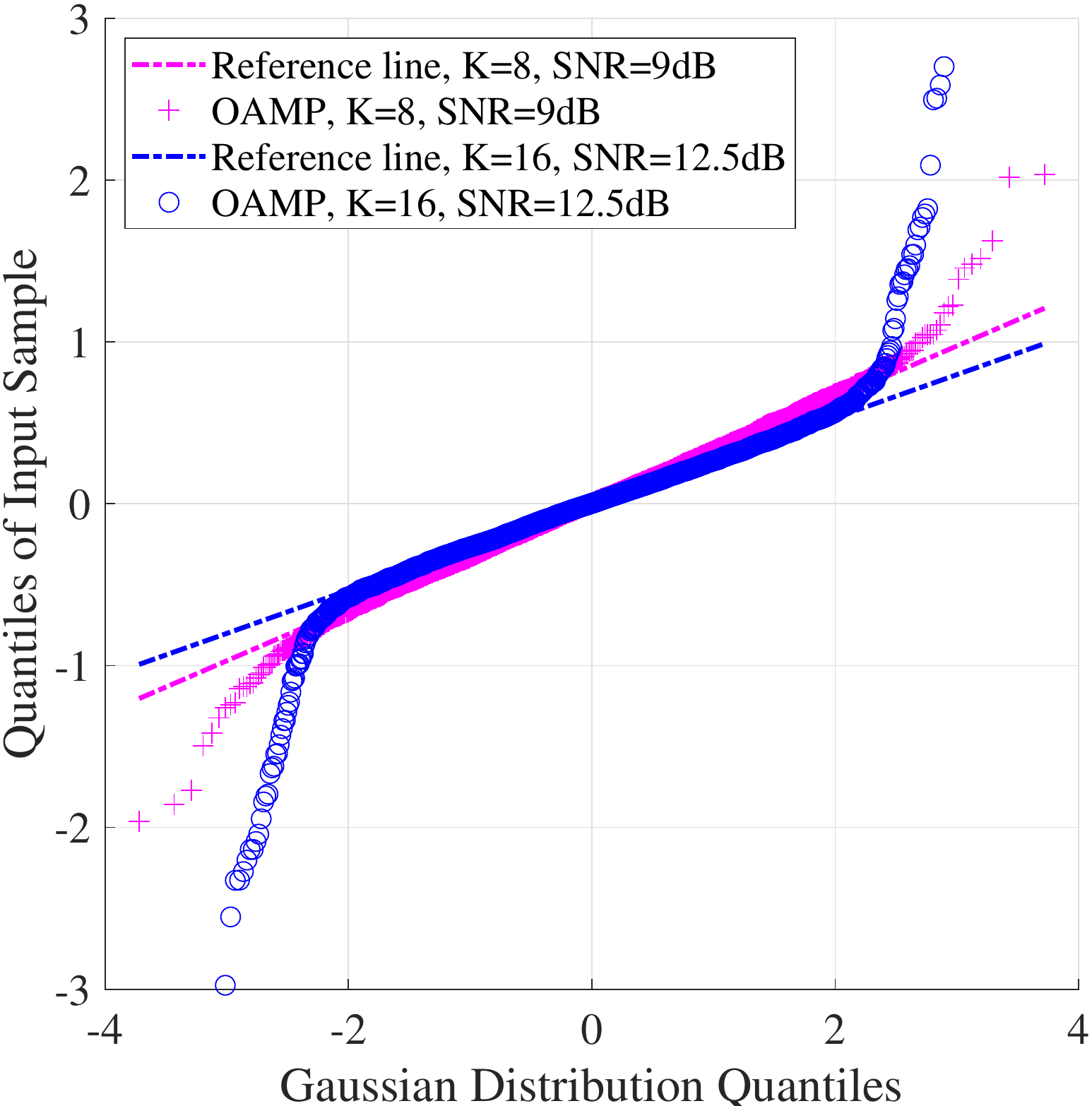}}\hfill
\centering
\subfloat[The EP detector]
{\includegraphics[scale=0.31]{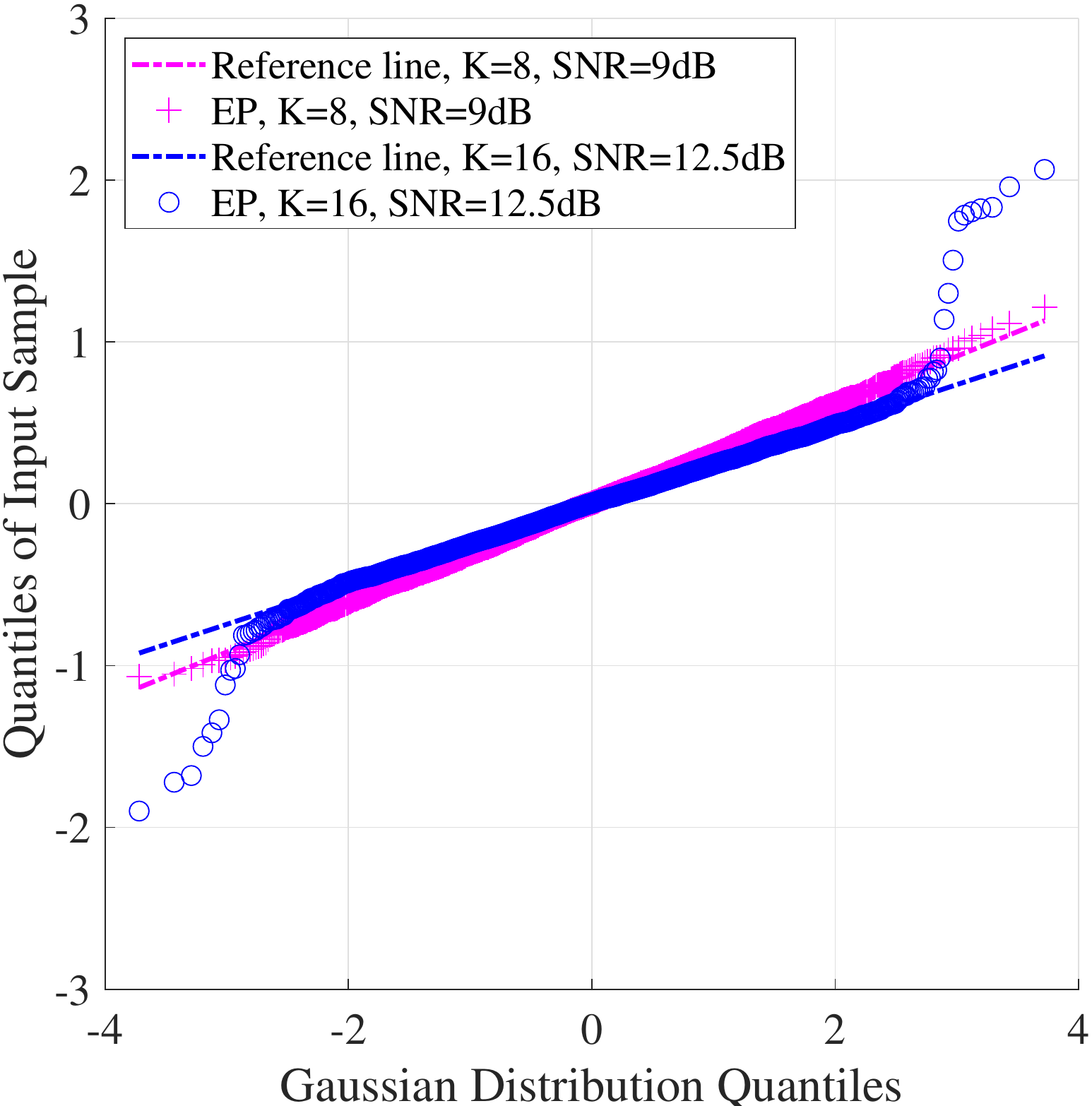}}
\caption{The QQ plots for the residual noise in the conventional MP detectors for $4$-QAM and $N=16$}
\label{QQ_plot}
\end{figure*}

To evaluate the normality of the residual noise for the EP, BPIC and OAMP detectors, we use the quantiles
to quantiles (QQ) plots shown in Fig. \ref{QQ_plot}, where the reference lines are given by Gaussian distributions. 
It can be seen that all six QQ curves fall along the corresponding reference lines in the middle of the plots, but curve off in the extremities. Such a behavior of the QQ curves means that the residual noise distributions of the EP, BPIC and OAMP have 
more extreme values than 
the Gaussian distribution. Since the divergence of the QQ curves from the reference lines increases with $K/N$ for the EP, BPIC and OAMP, we conclude that the residual noise distributions of all three detectors  diverge from the Gaussian distribution with increasing $K/N$. This causes a severe performance degradation in the EP, BPIC and OAMP detectors. 

\subsection{Conclusion}
It is evident that the posterior distribution accuracy metrics in Table \ref{table_analysis} and the residual noise correlation and normality metrics in Table \ref{table_analysis} and Fig. \ref{QQ_plot}, respectively, demonstrate a similar behaviour.
This similarity implies that the accuracy of the posterior distribution approximation can be improved by adjusting the  symbol estimator to deal better with the correlated and non-Gaussian residual noise. Since the main operation in the symbol estimator is the computation of the cavity distribution, we conclude that  an improvement of the cavity distribution paves a way for the improvement of the accuracy of the posterior distribution approximation. 

\section{The Proposed GNN-based Framework}
\label{sFramework}

Motivated by the analysis in Section \ref{Sect_Analysis_Gaussian}, we develop a GNN-based framework to improve the accuracy of the posterior distribution approximation by adjusting the cavity distribution 
in the MP detectors that rely on the IGA, such as the EP \cite{Jespedes-TCOM14}, AMP \cite{2009Donoho_ProcSci_AMP}, BPIC \cite{AKosasih}, Gaussian tree approximation \cite{2011_Goldberger_TInf_GTA}, and OAMP \cite{Ma-17ACCESS} detectors. As later shown in Section \ref{Simulation}, the resulting GNN aided MP detectors can  achieve a significant performance improvement compared to the corresponding classical MP detectors. We provide a description of our framework in terms of the EP and BPIC detectors.

\subsection{The Graph Expectation Propagation Network Detector}
 
To address the issue of the posterior distribution approximation inaccuracy in the EP detector, explained in Section \ref{Sect_Analysis_Gaussian}, we adjust the cavity distribution of the original EP detector using the GNN \cite{GGNN,KYoon2019,AScotti_GNN_2020}, as illustrated in Fig.  \ref{F2}. Specifically, the GNN produces the adjusted cavity distribution of the $k$-th transmitted symbol over the constellation set $\Omega$, $k\in [K]$. Therefore, the proposed detector is referred to as the GEPNet detector. Compared to the EP detector, the GEPNet detector has an additional module referred to as GNN module.

\subsubsection{The GNN Module}

\begin{figure}
\centering
\subfloat[The MP in the $t$-th iteration ]
{\includegraphics[scale=0.45]{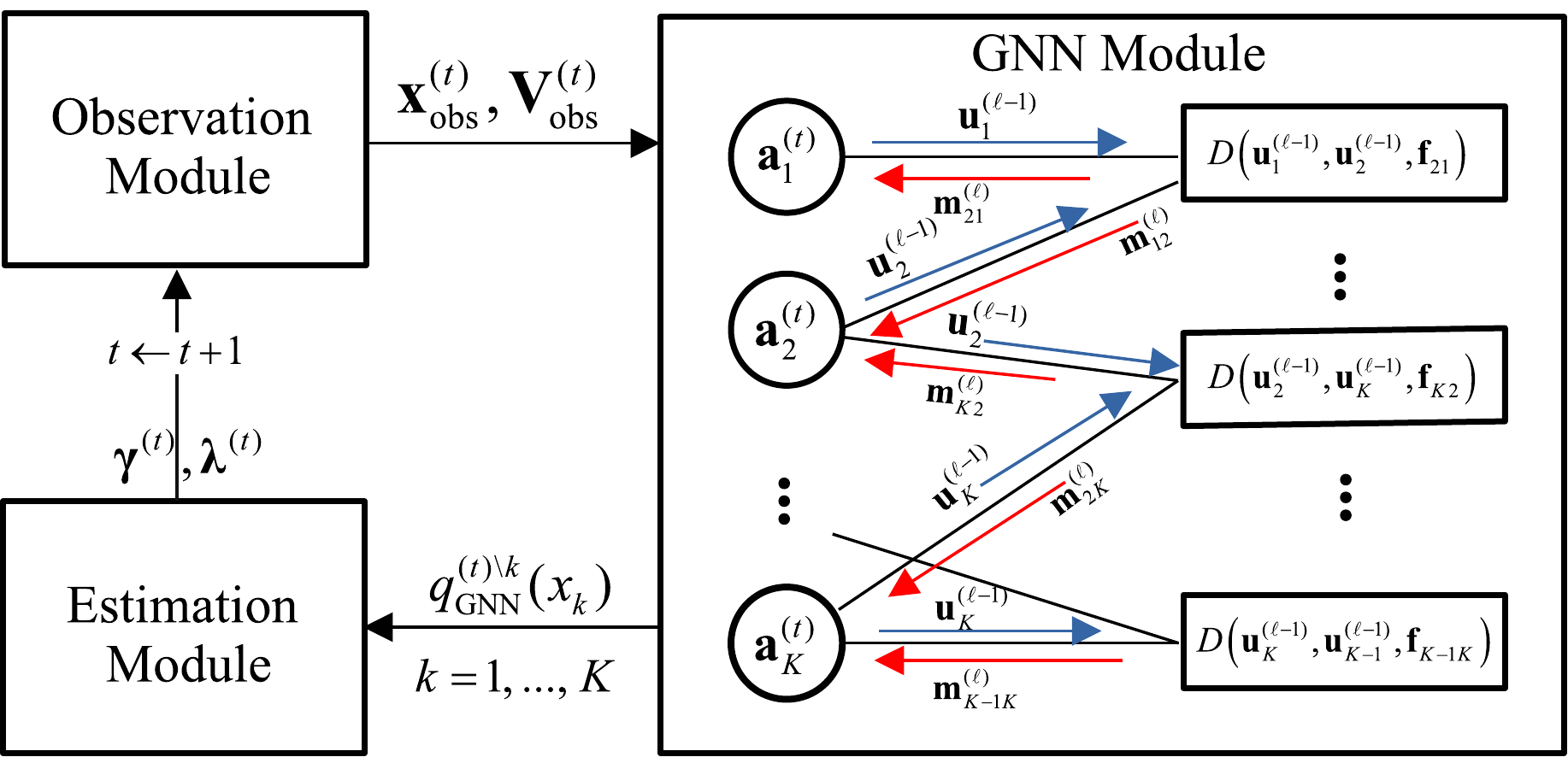}}\hfill
\centering
\subfloat[The iterations in the GEPNet detector]
{\includegraphics[scale=0.35]{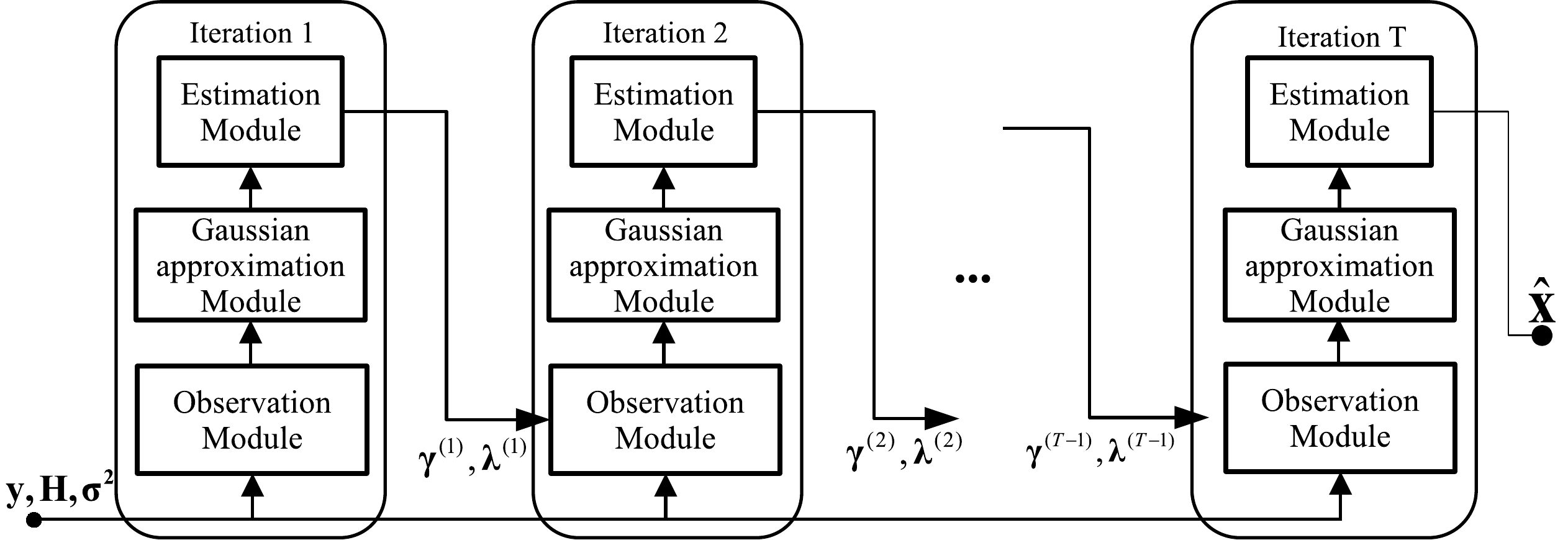}}
\caption{The GEPNet detector model}
\label{F2}
\end{figure}

To obtain an improved cavity distribution, we use a GNN parameterized according to a pair-wise MRF. The variable and factor nodes of the underlying graph are displayed as circles and rectangles in Fig. \ref{F2}a, respectively. We use a pair-wise MRF representation in which a factor node is connected  to two variable nodes. The variable and factor nodes compute self and pair potentials \cite{MRF_1980}, respectively. 

We treat the EP's cavity $q^{(t)\backslash k}(x_k)$ in \eqref{eA1_a0304raw} as the prior knowledge for the variable node $k$. The cavity $q^{(t)\backslash k}(x_k)$ is given by a Gaussian distribution with mean $x_{{\rm obs,k}}^{(t)}$ and variance $v_{{\rm obs,k}}^{(t)}$, defined in \eqref{eA1_a0304}. These mean and variance are concatenated to yield a variable node attribute 
 \begin{equation}\label{concat}
 \qa_k^{(t)} = \left[ x_{\rm obs,k}^{(t)}, v_{\rm obs,k}^{(t)} \right], k\in [K].
 \end{equation}
 The factor node attribute  $\mathbf{f}_{jk} \triangleq \left[ \qh_k^T \qh_j, \sigma^2 \right] $ is obtained by extracting the information from the pair-potential function, given in \cite[Eq. (6)]{AScotti_GNN_2020} as
 \begin{equation}\label{psi}
\psi(x_k,x_j) = {\sf exp} \left(- \frac{1}{{\sigma}^2} \qh_k^T \qh_j x_k x_j\right),
\end{equation}
while the information from the self-potential function, given in \cite[Eq. (5)]{AScotti_GNN_2020} as
            \begin{equation}\label{phi}
\phi(x_k) ={ \sf exp} \left( \frac{1}{{\sigma}^2} \qy^T \qh_k x_k - \frac{1}{2}  \qh_k^T \qh_k x_k^2  \right) p(x_k),
            \end{equation}
is incorporated into the message 
 \begin{equation}\label{GNN_init}
 \qu_k^{(0)} = \qW_1 \cdot [\qy^T \qh_k, \qh_k^T \qh_k, \sigma^2]^T + \qb_1,
 \end{equation}
 where the learnable matrix $\qW_1 \in \mathbb{R}^{N_u \times 3}$ and learnable vector $\qb_1 \in \mathbb{R}^{N_u }$ represent a single layer NN, and $N_u$ is the message size. We use $N_u = 8$. 
 As described in Fig. \ref{F2}a, the 
 messages $\qu^{(0)}_k$, $ k\in [K]$ are sent to the neighboring factor nodes. The factor nodes then commence the following iterative MP between the factor and variable nodes:
 
 \textit{\textbf{Factor to variable}:}
 Each factor node has a multi-layer perceptron (MLP) with two hidden layers of sizes $N_{h_1}$ and $N_{h_2}$ and an output layer of size $N_u$.  In this work, we set  $N_{h_1}=64$ and $N_{h_2}=32$. The rectifier linear unit (ReLU) activation function is used at the output of each hidden layer. 
 
 For any pair of variable nodes $k$ and $j$, there is a factor node connecting them. This factor node first concatenates its attribute  $\mathbf{f}_{jk}$ with the received messages $\qu_k^{(\ell-1)}$ and $\qu_j^{(\ell-1)}$, and then computes
 \begin{equation}\label{factor_to_var}
 \qm_{jk}^{(\ell)} = {\mathcal{D}} \left( \qu_k^{{(}\ell-1{)}}, \qu_j^{{(}\ell-1{)}}, \mathbf{f}_{jk} \right),
 \end{equation}
 where $\mathcal{D}$ is a MLP. 
Finally, the messages $\qm_{jk}^{(\ell)}$, $k,j \in [K]$, $j \neq k$, are sent to the neighboring variable nodes, as illustrated in Fig. \ref{F2}a. 
 
 \textit{\textbf{Variable to factor}:}
 The $k$-th variable node then sums all the incoming messages $\qm_{jk}^{(\ell)}$ from its neighbouring factor nodes and concatenates their sum with the node attribute $\qa_k^{(t)}$ as \begin{equation}
 \label{eq:m_k}
 \qm_k^{(\ell)}  = \left[\sum_{j\in [K]\setminus\{k\}} \qm_{jk}^{(\ell)},  \qa_k^{(t)} \right].
 \end{equation}
 The concatenated vector is  used to compute the message $\qu_k^{(\ell)}$ as
 \begin{subequations}\label{GRU_MLP}
            \begin{equation}\label{GRU}
 \qg_k^{(\ell)} = {\mathcal{U}} \left( \qg_k^{(\ell-1)}, \qm_k^{(\ell)}  \right), 
            \end{equation}
            \begin{equation}\label{MLP}
 \qu_k^{(\ell)}= \qW_2 \cdot \qg_k^{(\ell)} + \qb_2,
            \end{equation}
\end {subequations}
where function $\mathcal U$ is specified by the gated recurrent unit (GRU) network \cite{KYoon2019}, whose current and previous hidden states are $\qg_k^{(\ell)} \in \mathbb{R}^{N_{h_1} }$ and  $\qg_k^{({\ell}-1)} \in \mathbb{R}^{N_{h_1} }$, respectively, $\qW_2 \in \mathbb{R}^{N_u \times N_{h_1}}$ is a learnable matrix, and  $\qb_2\in \mathbb{R}^{N_u }$ is a learnable vector. 
Note that the GNN message computations \eqref{GNN_init}-\eqref{GRU_MLP} can be performed in parallel over $K^2-K$ factor nodes and $K$ variable nodes.
The message  \eqref{MLP} is  then sent to the neighbouring factor nodes to continue the  MP iterations. 

After $L$ rounds of the MP, a readout process yields an adjusted cavity distribution $q_{\rm GNN}^{(t)\backslash k}$ 
 \begin{subequations}\label{Readout}
				\begin{equation}\label{Readout1}
\{d_{k,a}\}_{a\in\Omega}  = {\mathcal{R}} \left(  \qu_k^{(L)} \right),
				\end{equation}
            	\begin{equation}\label{Readout2}
q_{\rm GNN}^{(t)\backslash k}(a)  = \frac{{\sf{exp}} \left(d_{k,a} \right)}  {\sum_{b\in \Omega}  {\sf{exp}} \left(d_{k,b} \right)}, a\in \Omega.
           		 \end{equation}
\end {subequations}
In this work, we set $L=2$.
The readout function ${\mathcal{R}} $ is given by an MLP with two hidden layers of sizes $N_{h_1}$ and $N_{h_2}$, and ReLU activation at the output of each hidden layer. The output size of  ${\mathcal{R}} $ is the cardinality $M$ of real-valued constellation set $\Omega$.
We then assign
\begin{equation}\label{GNN_reset_index}
 \qg_k^{(0)}  \leftarrow   \qg_k^{(L)} \text { and }\qu_k^{(0)} \leftarrow   \qu_k^{(L)}, k\in [K], 
\end{equation}
in order to use the GRU hidden state $\qg_k^{(L)}$  and message $\qu_k^{(L)}$ as the starting point for the next GEPNet iteration.
 \begin{Remark}
In the EP detector, the cavity function $q^{(t) \backslash k}(x_k)\propto \mathcal{N}  \left( x_k: x_{{\rm obs},k}^{(t)}, v_{{\rm obs},k}^{(t)} \right)$ 
is parameterized only by mean $x_{{\rm obs},k}^{(t)}$ and variance $v_{{\rm obs},k}^{(t)}$. 
 In the GEPNet detector, we use the GNN to incorporate  additional parameters    $\qy^T\qh_k,\qh_k^T\qh_k,\qh_k^T\qh_j,\sigma^2$ into  the cavity function, where $j\in [K]\setminus\{k\}$. 
 The GNN 
 enables the GEPNet detector to capture the MUI information  using the factor node attribute.
 \end{Remark}

 \subsubsection{The GEPNet Algorithm}    

The GEPNet detector consists of three modules, i.e., the observation, GNN, and estimation modules. 
The computations in the observation module remain the same as those in the EP detector. 
The difference between the estimation modules of the EP and GEPNet is that $\eqref{eA1_b0102_EP}$ is replaced by
 \begin{subequations}\label{eA1_b0102}
            \begin{equation}
\hat{x}_k^{(t)}=   \sum_{a\in \Omega}  a \times  q_{\rm GNN}^{(t)\backslash k}(x_k=a) ,
            \end{equation}
            \begin{equation}
v_k^{(t)} =  \sum_{a\in \Omega}  \left(x_k  -\hat{x}_k^{(t)}\right)^2 \times q_{\rm GNN}^{(t)\backslash k}(x_k =a).
            \end{equation}
\end{subequations}
The complete GEPNet's pseudo-code is given in Algorithm \ref{A1}.

\begin{algorithm}
\caption{GEPNet detector }
\label{A1}
\begin{small}
\begin{algorithmic}[1]
\State {\textbf{Input: } $\qH,\qy,\sigma^2,\mathit{E}_s ,L,T$} 
\State {Initialization:  $\qgamma^{(0)} = \qzero ,  \qlambda^{(0)} =\frac{1}{\mathit{E}_s }\textbf{I}_K, \eta = 0.7,\qg_k^{(0)}=\mathbf{0}$}
	\For {$t=1,\dots,T$} 
	
	    \Statex \textbf{\quad\, The Observation Module:}
		\State{Compute  $\qSigma^{(t)}$ and $\qmu^{(t)}$  in \eqref{eA1_a0102}}
		\State{Compute  $v_{{\rm obs},k}^{(t)}$  and $x_{{\rm obs},k}^{(t)}$ in \eqref{eA1_a0304}, $k\in [K]$}
		\Statex \textbf{\quad\, The GNN Module:}
		\State{Compute \eqref{concat}}
		\If {$t=1$}
				\State{Compute $\qu_k^{(0)}$ in \eqref{GNN_init}, $k\in [K]$}
		\EndIf
		\For {$l=1,\dots,L$}
				\State{Compute $ \qm_{jk}^{(\ell)}$ in \eqref{factor_to_var}}, $ j,k\in [K],j\neq k$
			\State{Compute $\qm^{(\ell)}_k$ in \eqref{eq:m_k}, $k\in [K]$}	\State{Compute $\qg_k^{(\ell)} $ and  $\qu_k^{(\ell)}$  in \eqref{GRU_MLP}}, $ k\in [K]$
		\EndFor
		
		\State{Compute $q_{\rm GNN}^{(t)\backslash k}(x_k)$  in \eqref{Readout}}, $ k\in [K]$
		\State{Compute \eqref{GNN_reset_index}}	
		\Statex \textbf{\quad\, The Estimation Module:}
		\State{Compute $\hat{x}_k^{(t)}$  and $v_k^{(t)}$  in \eqref{eA1_b0102}}, $ k\in [K]$
		\State{Compute $ \qlambda^{(t)} $ and $\qgamma^{(t)}$   in \eqref{eA1_b0304}}
		\If {$\lambda^{(t)}_k <0 $}
			\State{ $\lambda_k^{(t)}=\lambda_k^{(t-1)}$ and $\gamma_k^{(t)}=\gamma_k^{(t-1)}, k\in [K]$}
		\EndIf
        \State{Smoothen $\qlambda^{(t)}$ and $\qgamma^{(t)}$ using \eqref{eq:damping}}
	\EndFor
\State {\textbf{Return: }  Hard symbol estimates from $\left[\hat{x}_1^{(T)}, \dots,\hat{x}_K^{(T)} \right]$}
\end{algorithmic}
\end{small}
\end{algorithm}

\subsection{The Graph Parallel Interference Cancellation Network Detector}
\label{sectGPICNet}

In this section, we describe the proposed GNN-based framework in the context of the BPIC detector \cite{AKosasih}. The proposed GNN-aided BPIC detector is referred to as the GPICNet detector. Similarly to the BPIC detector, the GPICNet detector  does not need to perform any matrix inversion operation, and therefore has a lower complexity than the GEPNet. The structure of the GPICNet detector is the same as the GEPNet detector. Compared to the BPIC detector, the GPICNet detector has an additional GNN module, explained as follows.

The GNN module takes the outputs of the observation module (Section \ref{sObs_BPIC}) and use\gb{s} them as the node attributes, i.e., 
\begin{equation}\label{concat2}
 \qa_k^{(t)} = \left[\mu^{(t)}_{k}, \Sigma_{k}^{(t)}\right], k\in [K].
 \end{equation}
The rest of the computations in the GNN module of the GPICNet are the same as those in the GEPNet. The GNN module then passes the $q_{\rm GNN}^{(t)\backslash k}(x_k)$,  $k\in [K]$, to the estimation module. The estimation module of the GPICNet operates as described in Section \ref{sEst_BPIC} except for computing the soft symbol estimates $\hat{x}^{(t)}_k$ and their variances $v^{(t)}_k$ in \eqref{eA1_b0102} instead of \eqref{eA1_b0102_EP}, $k\in [K]$. Note that the use of the GNN module is to improve the accuracy of the posterior distribution approximation, as in the case of the GEPNet detector. The pseudo-code of the GPICNet detector is given in Algorithm \ref{A2}.

\begin{algorithm}
\caption{GPICNet detector }
\label{A2}
\begin{small} 
\begin{algorithmic}[1]
\State {\textbf{Input: } $\qH,\qy,\sigma^2 ,L,T$} 
\State {Initialization:  $\qg_k^{(0)}=\bold{0},\hat{\qx}^{(0)}=\bold{0}$}
	\For {$t=1,\dots,T$} 
	
	    \Statex \textbf{\quad\, The Observation Module:}
		\State{Compute  ${\qSigma}^{(t)}$ and ${\qmu}^{(t)}$  in \eqref{PIC_estim}}

		\Statex \textbf{\quad\, The GNN Module:}
		\State{Compute \eqref{concat2}}
		\If {$t=1$}
				\State{Compute $\qu_k^{(0)}$ in \eqref{GNN_init}, $k\in [K]$}
		\EndIf
		\For {$l=1,\dots,L$}
				\State{Compute $ \qm_{jk}^{(\ell)}$ in \eqref{factor_to_var}}, $ j,k\in [K],j\neq k$
				\State{Compute $\qm^{(\ell)}_k$ in \eqref{eq:m_k}, $k\in [K]$}
				\State{Compute $\qg_k^{(\ell)} $ and  $\qu_k^{(\ell)}$  in \eqref{GRU_MLP}}, $ k\in [K]$
		\EndFor
		
		\State{Compute $q_{\rm GNN}^{(t)\backslash k}(x_k)$  in \eqref{Readout}}, $ k\in [K]$
		\State{Compute \eqref{GNN_reset_index}}	
		
		\Statex \textbf{\quad\, The Estimation Module:}
		\State{Compute $\hat{x}_{k}^{(t)} $ and $v_{k}^{(t)} $   in \eqref{eA1_b0102}}, $ k\in [K]$
		\If {$t>1$}
    		\State{Update $\hat{x}_{k}^{(t)}$ and $v_{k}^{(t)} $   in \eqref{DSC}}, $ k\in [K]$
        \EndIf
	\EndFor
\State {\textbf{Return: }  Hard symbol estimates from $\left[\hat{x}_{1}^{(T)}, \dots,\hat{x}_{K}^{(T)} \right]$}
\end{algorithmic}
\end{small}
\end{algorithm}

\subsection{Permutation Equivariance}
\label{sectionPermutation}

In this section, we prove that the GEPNET and GPICNET detectors are permutation equivariant. If a permutation equivariant detector returns $\hat{\qx}^{(t)}$ for the user data $\qx$, channel matrix $\qH$ and received signal $\qy$, then it would return $\qPi^T\cdot\hat{\qx}'^{(t)}$ for $\qx'\triangleq\qPi^T\cdot\qx$, $\qH'\triangleq\qH\cdot\qPi$ and $\qy$, where $\qPi$ is a $K\times K$ permutation matrix. The GEPNet/GPICNet detector consists of the observation, GNN and estimation modules. Due to the transitivity of the permutation equivariance \cite{equivar2016}, it suffices to proof that each module is equivariant, therefore in the following  we prove the equivariance for each module separately.
\subsubsection{The observation module}
Operations in the observation module of the GEPNet are specified by \eqref{eA1_a0102} and \eqref{eA1_a0304}. By substituting $\qH'\triangleq\qH\cdot\qPi$ instead of $\qH$, $\qlambda'^{(t-1)}\triangleq\qPi^T\qlambda^{(t-1)}\cdot\qPi$ instead of $\qlambda^{(t-1)}$, and $\qgamma'^{(t-1)}\triangleq\qPi^T\qgamma^{(t-1)}$ instead of $\qgamma^{(t-1)}$ in \eqref{eA1_a0102}, we obtain $\qSigma'^{(t)}=\qPi^T\cdot\qSigma^{(t)}\cdot\qPi$ and $\qmu'^{(t)}=\qPi^T\cdot\qmu^{(t)}$. It can be seen that the main diagonal of the diagonal matrix $\qSigma'^{(t)}$ is equal to the permuted diagonal of $\qSigma^{(t)}$, where the permutation is defined by $\qPi^T$. The main diagonals of $\qlambda'^{(t)}$ and $\qlambda^{(t)}$ are connected in the same way. 
This and the independent computation of \eqref{eA1_a0304} for each user $k\in [K]$ lead to 
\begin{equation}
\label{eq:Pi_obs}
{\qx'}_{{\rm obs}}^{(t)}=\Pi^T\cdot \qx_{{\rm obs}}^{(t)}\quad {\rm and} \quad {\qv'}_{{\rm obs}}^{(t)}=\Pi^T\cdot \qv_{{\rm obs}}^{(t)}.
\end{equation}
Therefore, the observation module of the GEPNet ensures permutation equivariance. 
 The computations in the observation module of the GPICNet are also permutation equivariant, we skip the derivation for brevity as it can be proved by following the above mentioned steps.

\subsubsection{The GNN module}
According to Algorithms \ref{A1}-\ref{A2}, all computations for users $k\in [K]$ in the GNN module are performed independently using the same expressions, except for \eqref{factor_to_var}-\eqref{eq:m_k}. Therefore, we focus on \eqref{factor_to_var}-\eqref{eq:m_k}. Due to \eqref{concat}, \eqref{concat2}, and \eqref{eq:Pi_obs}, GEPNet's and GPICNet's node attributes satisfy $\qa'^{(t)}=\Pi^T\cdot\qa^{(t)}$. Let us define a function $\pi$ such that $\Pi_{k,\pi(k)}=1$ for any $k\in [K]$. 
Then $\qh'_{\pi(k)}=\qh_{k}$, and therefore the substitution of $\qh'_{\pi(k)}$ into  \eqref{GNN_init} results in the message $\qu'^{(0)}_{\pi(k)}=\qu^{(0)}_k$. Similarly,  $\qf'_{\pi(j)\pi(k)}=\qf_{jk}$. This leads to ${\qm'}_{\pi(j)\pi(k)}^{(\ell)}=\qm_{jk}^{(\ell)}$ in  \eqref{factor_to_var}. Note that \eqref{factor_to_var} uses the same MLP $\mathcal{D}$ for all $j$ and $k$. Since each column of a permutation matrix has a single unit, the function $\pi$ satisfies $\pi(j)=\pi(k)$ iff $j=k$, and therefore we have $\sum_{j\in [K]\setminus\{k\}} {\qm'}_{j\pi(k)}^{(\ell)}=\sum_{j\in [K]\setminus\{k\}} {\qm'}_{\pi(j)\pi(k)}^{(\ell)}=\sum_{j\in [K]\setminus\{k\}} \qm_{jk}^{(\ell)}$. As result, ${\qm'}_{\pi(k)}^{(\ell)}=\qm_{k}^{(\ell)}$ in \eqref{eq:m_k}, and consequently, ${\qm'}^{(\ell)}=\Pi^T\cdot \qm^{(\ell)}$. Therefore, the permutation equivariance is preserved for all the expressions up to \eqref{eq:m_k}. The remaining \eqref{GRU_MLP}-\eqref{GNN_reset_index} do not affect the permutation equivariance property, since the same operations are performed independently for each user.   

\subsubsection{The estimation module}
The estimation module also guarantees permutation equivariance, since it follows from Algorithms \ref{A1}-\ref{A2} that users $k\in [K]$ are independently processed using the same expressions.

Since all the modules in the proposed GEPNet/GPICNet detectors  are permutation equivariant, we conclude that the proposed detectors are also permutation equivariant that is robust to the user permutations.

\subsection{Computational Complexity Analysis}
\label{sectionComplAnalysis}

\newcolumntype{M}[1]{>{\centering\arraybackslash}m{#1}}

\begin{table*}\small
\centering
 \begin{threeparttable}
\begin{tabular}{|l|  M{36.5em} | M{4.5em}|  M{4.5em} |} \hline 
Detector & Number of multiplications &  $16$-QAM, $N=256,$ $K=128$  $(\times 10^{7})$  &  $16$-QAM,  $N=256,$ $K=256$ $(\times 10^{7})$ \\ \hline \hline 
AMP \cite{2009Donoho_ProcSci_AMP} &  $(4NK + 8N +6K + 4 M K) T$  & $0.1359$ & $0.2698$ \\ \hline
GNN \cite{AScotti_GNN_2020} &  $\left(1.5N+0.5NK+K+S_u\left( N_{h1} +N_{h2} +3 \right) + N_{h1}N_{h2} + M  \right)  K  + \left(   4 N_{h1}  S_u + 5 N_{h1} + N_{h2}\left( N_{h1} +S_u +2 \right) + 3 N_{h1}^2 \right) K T   $  & $2.4317$    & $5.2862$  \\ \hline
MMSE \cite{LMMSE} & $ K^3+K^2 (N+1) +NK$   & $0.6340$   & $3.3686$  \\ \hline
RE-MIMO \cite{2021_KPratik_TSP_REMimo} & $ 2(N +1) +  \left( 5 d_s (2 N +1)+ 4d_s^2  +2\right) K +  \left( 0.5 N (K+1) + M + 2 d_\phi d_k + d_\phi dv + d_\phi d_s + d_\phi d_s + \frac{5}{8} d_\psi^2 +1  \right)  KT $   & $159.3654$  & $322.9252$  \\  \hline
OAMPNet \cite{2018HHE_Globecom_OAMPNet}  & $ NK  (K-1)+ (K^3 + N^2K + NK^2 + 2NK + 12K + 4  M K + 2K +8 ) T$ & $15.1656$ & $ 52.1412$  \\ \hline
EP  \cite{Jespedes-TCOM14} & $ NK^2+ NK + \left( K^3 + K^2 + 13K + 2 M K \right)T$   & $2.5389$  & $18.5324$ \\ \hline
BPIC & $   NK^2- 6K + \Big( 17 +2 M + N \Big) K T $ & $0.5110$  & $1.8788$   \\    \hline 
GPICNet   & $ \left( 1.5N+ 1.5NK -5     +S_u\left( N_{h1} +N_{h2} +3 \right) + N_{h1}N_{h2} + M  \right) K + \left( 3NK+   2 M+10  +  \left(  N_{h1}  \left( 4  S_u + 5+ 3 N_{h1} \right)+ N_{h2} \left( N_{h1} +S_u +2 \right)  \right) L  \right) K  T $   & $5.1308$   & $11.525$  \\ \hline
GEPNet   &  $ \Big(2.5N+1.5NK+K+S_u \big( N_{h1} +N_{h2} +3 \big) + N_{h1}N_{h2} + M  \Big) K + 
\left(K^3 + K^2+13K +   2K M +  \Big (  N_{h1} ( 4  S_u + 5+ 3 N_{h1} )+ N_{h2} ( N_{h1} +S_u +2 )  \Big) K L  \right)  T $  & $7.1497$   & $28.1767$   \\ \hline
 \end{tabular}
\begin{tablenotes}
\item[*] The hyperparameters $d_s, d_\psi = d_s+M+N, d_\phi =d_\psi+1, d_v = d_k = d_\phi/n $ are  used in RE-MIMO detector, where $d_s$ and $n$ are set to $512$ and  $8$, respectively.
\end{tablenotes}
\caption{The computational complexity comparison}
\label{table_complex}
\end{threeparttable}
\end{table*}

In this section, we analyse the computational complexity of the GEPNet detector depicted in Alg. \ref{A1}.  We provide complexity in terms of the number of multiplications for the real-valued system \eqref{eII_1}. The corresponding complexity for the complex-valued system \eqref{eII_1a} can be easily obtained by substituting $K=2N_t$ and $N=2N_r$. We first compute $\qH^T \qH$ and $\qH^T \qy$ to be used in the GEPNet iterations, the complexity cost is $NK^2+NK$. The dominant complexity of the GEPNet detector per iteration is related to matrix inversion operation in \eqref{eA1_a01} that requires $K^3 + K^2+2K$ number of multiplications. Expressions \eqref{eA1_a0304}, \eqref{eA1_b0102}, \eqref{eA1_b0304}, and \eqref{eq:damping} are all related to matrix-vector multiplications and the cost is $2 \sqrt{M}K + 11K$.  The rest of the operations belong to the GNN computations, whose complexity is $ \Big(1.5N+0.5NK+K+S_u\left( N_{h1} +N_{h2} +3 \right) + N_{h1}N_{h2} + M  \Big) K  + \Big(   4 N_{h1}  S_u + 5 N_{h1} + N_{h2}\left( N_{h1} +S_u +2 \right) + 3 N_{h1}^2 \Big) K L   $.  As the GEPNet iterations are performed  $T$ times, the total computational complexity of the GEPNet detector is  as given in the Table \ref{table_complex}. To numerically illustrate the complexity, we set $N=256$, $K=128,256$, $16$-QAM, and  $T = 10$ and calculate the complexity of each detector in the last two columns of Table II. 
As expected, the proposed GEPNet has a higher complexity compared to the EP, but their complexity ratio decreases with increasing $K$. The proposed detectors provide a flexible complexity-performance trade-off: the complexity of the proposed detectors can be  reduced at the expense of performance by adjusting the layer sizes, i.e., $N_{h_1}$, $N_{h_2}$, $S_u$. 
This is in contrast to the near optimal sphere decoding  that can be realized only for MIMO systems with a small number of user antennas in which case the sphere decoding complexity grows as $O(K^3)$ \cite{Hassibi_2005}. For example, the sphere decoding complexity for $K = 32$, $N = 512$, and $16$-QAM is $2.1\times 10^7$ real-number multiplications \cite{2019SphereDecoder}, which is the sum of the initial radius computation complexity $1.9 \times 10^7$ real-number multiplications and the number of visited nodes $5.24\times 10^4$ multiplied by the average complexity of a single node processing $38$ real-number multiplications, assuming that each complex-number multiplication in \cite{2019SphereDecoder} requires four real-number multiplications. For the same MIMO configuration, the complexity of the GEPNet is $1.2\times 10^7$ real-number multiplications. However, the sphere decoding complexity is of the same order of magnitude as the GEPNet complexity only for a small $K$, since the complexity of the sphere decoding generally grows exponentially with $K$ as shown in \cite[Section III]{Hassibi_2005}. Even efficient node pruning techniques are unable to resolve the problem of exponentially growing size of the sphere decoding tree \cite{Hassibi_2005}, which makes the near optimal sphere decoder impractical when $K$ is large.

Unlike the GEPNet detector which needs to perform matrix inversion operations and therefore having cubical complexity with respect to the number of users, the low complexity GPICNet detector avoids such a high complexity by using MRC and PIC schemes, performed in \eqref{PIC_estim}, and the complexity cost is  $(2N+3)K$. Note that  $\qh_k^T \qh_k$, $(\qh_k^T \qh_k)^2$, and $\frac{\qh_k^T}{\qh_k^T \qh_k}$ for $k\in [K]$ are only computed once as their values do not change with the iterations.  The number of multiplications needed in \eqref{eA1_b0102_EP}, \eqref{DSC}, \eqref{DSC_error} is $ (2 M + 14) K $. 
Note that the PIC and DSC schemes are inactive at the first iteration. The rest of the operations in the GPICNet belong to the GNN computations and its total computational complexity is given in Table \ref{table_complex}.

\section{Training and Robustness of the Proposed Detectors}
\label{SectTrainingAndRobust}

In this Section, we describe a method that we use to train  the proposed GEPNet and GPICNet detectors. Once trained, the proposed detectors can be implemented in the systems with dynamic changes of the number of users. We further evaluate the robustness of the proposed  detectors with respect to the changes of the number of users. The obtained results allow us to reduce the training cost.

\subsection{Training Method}
\label{SectImplementation}

We implemented the proposed detectors in Python using PyTorch \cite{Pytorch}. The training process was divided into $600$ epochs. In each epoch, $1563$ batches of $64$ samples were generated, where a sample includes realizations of $\qx$, $\qH$ and $\qn$ satisfying \eqref{A2}. Thus, the total number of samples in each epoch is $100032$. We used QAM modulation with varying SNR values that are uniformly distributed in the range of $[{\sf SNR_{\rm min}},{\sf SNR_{\rm max}}]$. The proposed detectors were trained using Adam optimizer with the learning rate $0.0001$. We applied  the learning rate scheduler (PyTorch ReduceLROnPlateau) to adjust the learning rate based on the validation loss that was computed using additional 5000 samples in each epoch. The reducing factor of the learning rate is set to $0.91$. We used the total cross-entropy loss function expressed as 
\begin{equation}\label{Loss_func}
Loss= -\frac{1}{W}  \sum_{w=1}^W \sum_{k=1}^{K_{\rm train}}  \sum_{a \in \Omega}  \mathbb{I}_{x_k^{(w)}=a}
{\sf log} \left( q_{\rm GNN}^{(T) \backslash k} \left(a\right) \right),
\end{equation}
where $W$ is the number of training samples in each batch, $K_{\rm train}$ denotes the number of users in the training phase, 
$\qx^{(w)}\in \Omega^{K_{\rm train}}$ is the transmitted vector, the training label $\mathbb{I}_{x_k^{(w)}=a}$ is the indicator function that takes value one
if $x_k^{(w)}=a$ and zero otherwise,   
and $ q_{\rm GNN}^{(T) \backslash k} \left(a\right) $  is the corresponding probability estimate obtained by the GEPNet detector for the ${w}$-th training sample and $k$-th user. The loss in \eqref{Loss_func} is used to update the NN parameters using the backpropagation method \cite{Goodfellow-et-al-2016}. 
Note that the value of $K_{\rm train}$ may vary over batches. Specifically, all samples in each batch have the same $K_{\rm train}$, but samples from different batches may have different $K_{\rm train}$. We set the number of iterations $T=10$ for all the detectors during the training and testing except $T=15$ for the GPICNet during the testing.

In the testing phase, we first created a testing dataset by randomly generating $500000$ samples for each SNR point with the same $N$ and $M$ that were used in the training phase, and $K$ was set depending on the testing scenario. Finally, we computed the SER of all the trained detectors using the testing dataset.  A PyTorch implementation of the proposed GEPNet detector is available at https://github.com/GNN-based-MIMO-Detection/GNN-based-MIMO-Detection.
For comparison purpose, we used the implementation of  RE-MIMO and OAMPNet detectors provided in GitHub repository of \cite{2021_KPratik_TSP_REMimo}.
Note that the proposed detectors share the same NN parameters for all users, which implies that the number of the NN parameters to be optimized in the training phase does not grow with the number of users and thus ensures scalability.

\begin{figure*}
\centering
{\includegraphics[scale=0.48]{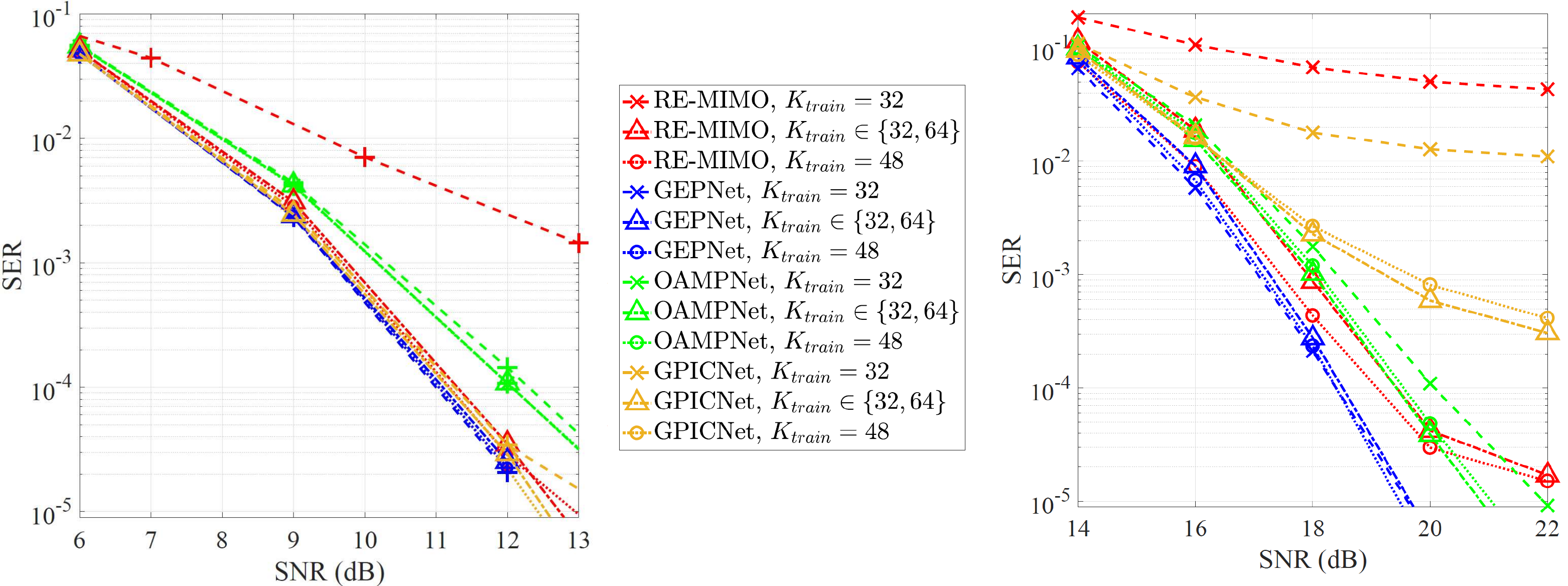}}
\caption{Robustness of the NN based detectors for  $N=64$ and  $K=48$ ; $4$-QAM, ${\sf SNR_{\rm min}}=3$, and ${\sf SNR_{\rm max}}=16$ (left hand side); $16$-QAM (right hand side), ${\sf SNR_{\rm min}}=14$, and ${\sf SNR_{\rm max}}=24$}

\label{Inter_Extra}
\end{figure*}

\subsection {Robustness to the Changes in the Number of Users}
\label{SectRobustness}

In this section, we investigate the robustness of the proposed detectors 
changes in the number of users. 
Typically, a detector is trained and tested for the same configuration $(N,K)$. This leads to the necessity to retrain the detector once the configuration has changed. The number of active users in uplink MU-MIMO systems may vary with the time.  Therefore, it is desirable to use a practical detector that is able to handle a varying number of users without retraining.  
The proposed detectors are expected to be robust due to
\begin{enumerate}
    \item Using the number of detector parameters independent from  the number of users.
    \item Representing each user by the corresponding node in the GNN so that all nodes are specified by the same neural networks (NN) defined in equations \eqref{GNN_init}-\eqref{factor_to_var}, \eqref{GRU_MLP}-\eqref{Readout}. Therefore, all users share the same NN parameters, and scaling of the proposed detectors in terms of the number of users is straightforward.
\end{enumerate}

We evaluate the robustness of the GEPNet, GPICNet, OAMPNet and RE-MIMO detectors to the changes in $K$ in two scenarios: 
\begin{enumerate}
\item Interpolation scenario, where we train the 
detectors for two different numbers of users $K_{\rm train} \in \{a,b\}$, and then test the detectors in a system with a number of users $K$ in the range of $a$ and $b$. \item Extrapolation scenario, where we train the 
detectors for a certain number of users $K_{\rm train}$, and then test the detectors in a system with a larger number of users $K>K_{\rm train}$. 
\end{enumerate}
As a baseline for comparison, we provide 
results for the detectors that are trained and tested for the same number of users,
i.e., $K_{\rm train}=K$.
The results for $a=32$, $b=64$, $N=64$ and $K=48$ are given in Fig. \ref{Inter_Extra}. 
It can be seen that the curves $K_{\rm train}\in\{32, 64\}$ are close to the curves $K_{\rm train}=48$
for all the detectors. 
However, the curve $K_{\rm train}=32$ is able to approach the curve $K_{\rm train}=48$ only for the GEPNet detector, while there is a significant gap between the curves for all the other detectors, i.e., these detectors show a significant performance degradation. Therefore, we conclude that all the considered detectors are  robust to the changes in the number of users in the interpolation scenario, but only the GEPNet detector demonstrates the robustness in the extrapolation scenario. This indicates that the GEPNet detector is able to cope with the varying number of users without retraining. 

Since the proposed GEPNet and GPICNet detectors demonstrate a good robustness in the interpolation scenario, we propose to train them only for $K_{\rm train}\in\{\min (\qK),\max (\qK)\}$, where $\qK$ is a set of all $K$ of interest. This would lead to a substantial complexity reduction compared to the training for all $K_{\rm train}\in\qK$.


\section{Simulation Results}\label{Simulation}

In this section, we first evaluate the posterior distribution approximation of the proposed detectors.
We then analyse the performance of our proposed detectors by comparing their SER performance with that of the state-of-the-art MP and NN based detectors. More specifically, the SER is evaluated with respect to the channel condition number, SNR, and transmit-to-receive antennas ratio.  The MP detectors considered in this paper are the AMP \cite{2009Donoho_ProcSci_AMP}, MMSE \cite{LMMSE}, BPIC \cite{AKosasih}, OAMP \cite{Ma-17ACCESS}, and EP \cite{Jespedes-TCOM14} detectors. 
The weighting coefficients in the EP and GEPNet detectors are set to $0.9$ and $0.7$, respectively. Besides the proposed NN based GEPNet and GPICNet detectors, we also consider the state-of-the-art NN based GNN \cite{AScotti_GNN_2020},  OAMPNet \cite{2018HHE_Globecom_OAMPNet} and RE-MIMO \cite{2021_KPratik_TSP_REMimo} detectors.
Since we focus on MU-MIMO systems where the transmit-to-receive antennas ratio is high, we consider $\alpha \triangleq \frac{K}{N} \in \left[ \frac{1}{2}, 1 \right]$.
The proposed GEPNet and GPICNet detectors only need to be trained with $K_{\rm train} \in \{\frac{1}{2} N, N \}$, as discussed in the previous section. The RE-MIMO \cite{2021_KPratik_TSP_REMimo} was trained with $K_{\rm train} \in \{\frac{1}{2}N,\frac{1}{2}N+1, \dots, N \}$, as suggested in \cite{2021_KPratik_TSP_REMimo}. The OAMPNet and GNN detectors were trained  with the same number of users as in the testing phase, i.e., $K_{\rm train}=K$, as suggested by their authors. 
We also provide the ML detection results. 
The ML detector was implemented by using Gurobi optimizer \cite{Gurobi} to solve the quadratic minimization problem \eqref{eq:ML}. Specifically, the Gurobi optimizer employs an advanced version of the widely known branch-and-cut method \cite{enwiki:1037121170}. Unless otherwise mentioned, $500000$ channel realizations were used to compute the SER. We employ $4$-QAM, $16$-QAM, and $64$-QAM modulation.

\subsection{Improvement of the Posterior Distribution Approximation by using the GNN Framework}

\begin{figure*}
\centering
\subfloat[SNR = $9$ dB and $K=8$]
{\includegraphics[scale=0.35]{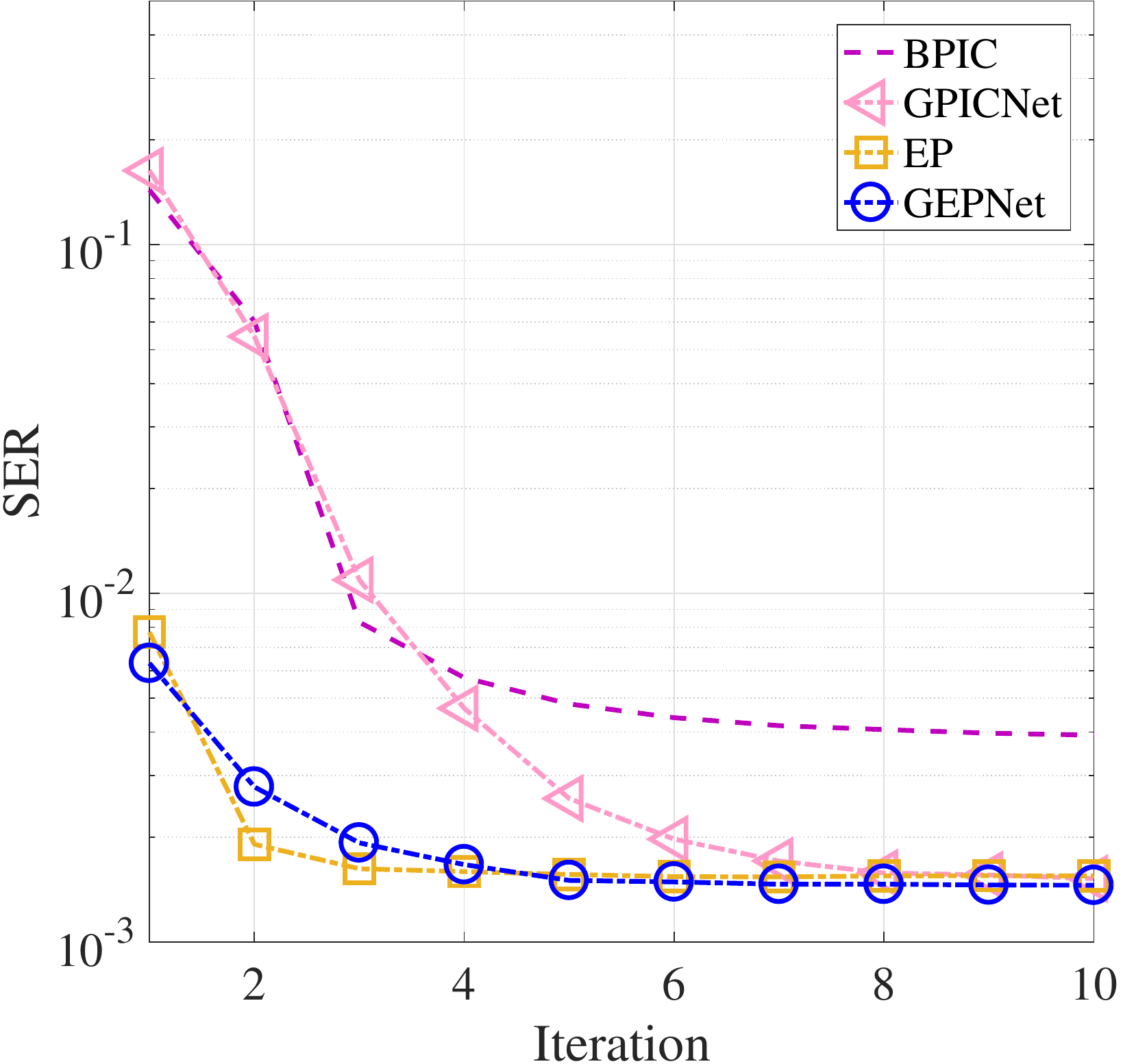}}\hfill
\centering
\subfloat[SNR = $11$ dB and $K=12$]
{\includegraphics[scale=0.35]{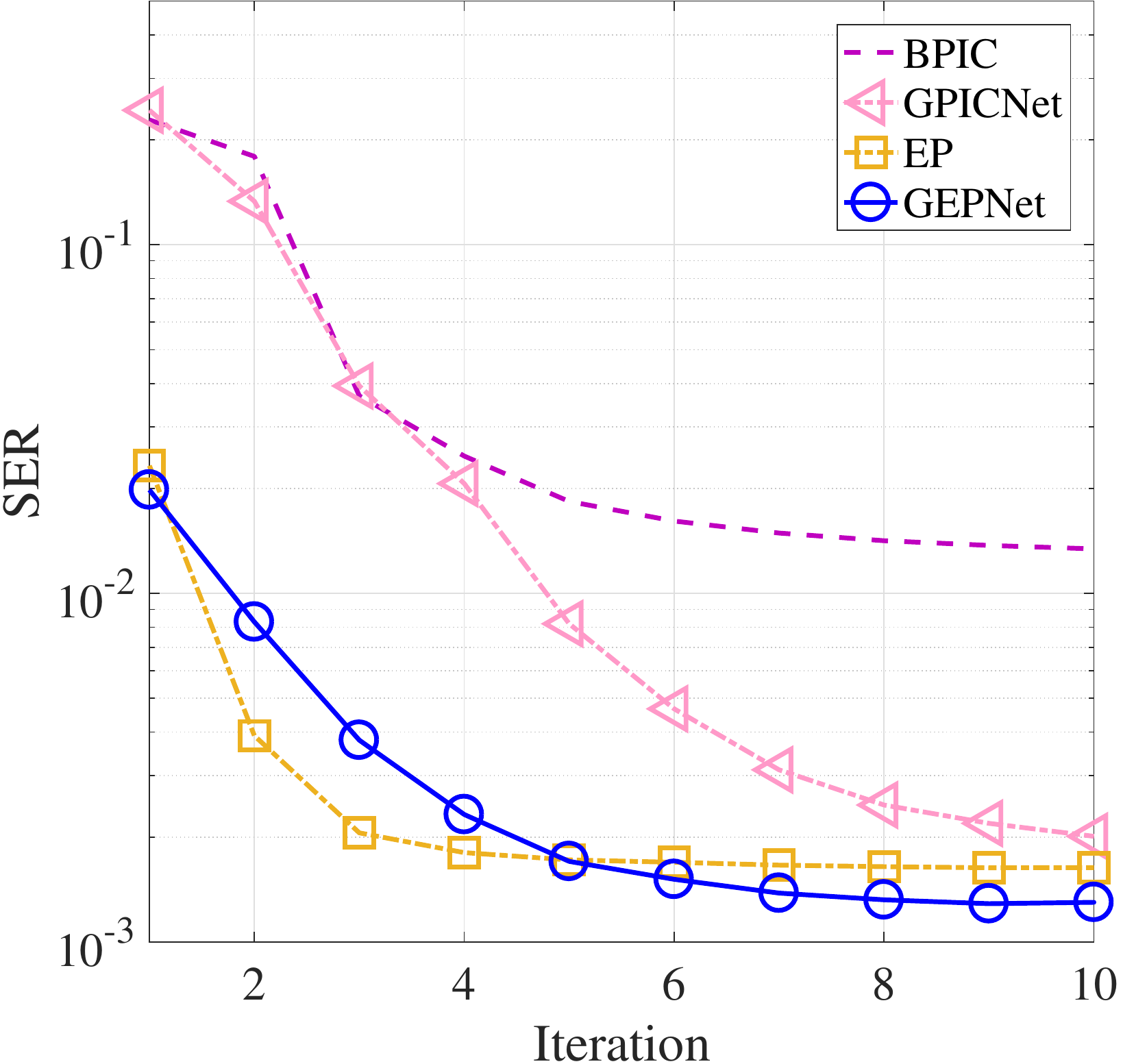}}\hfill
\centering
\subfloat[SNR = $12.5$ dB and $K=16$]
{\includegraphics[scale=0.35]{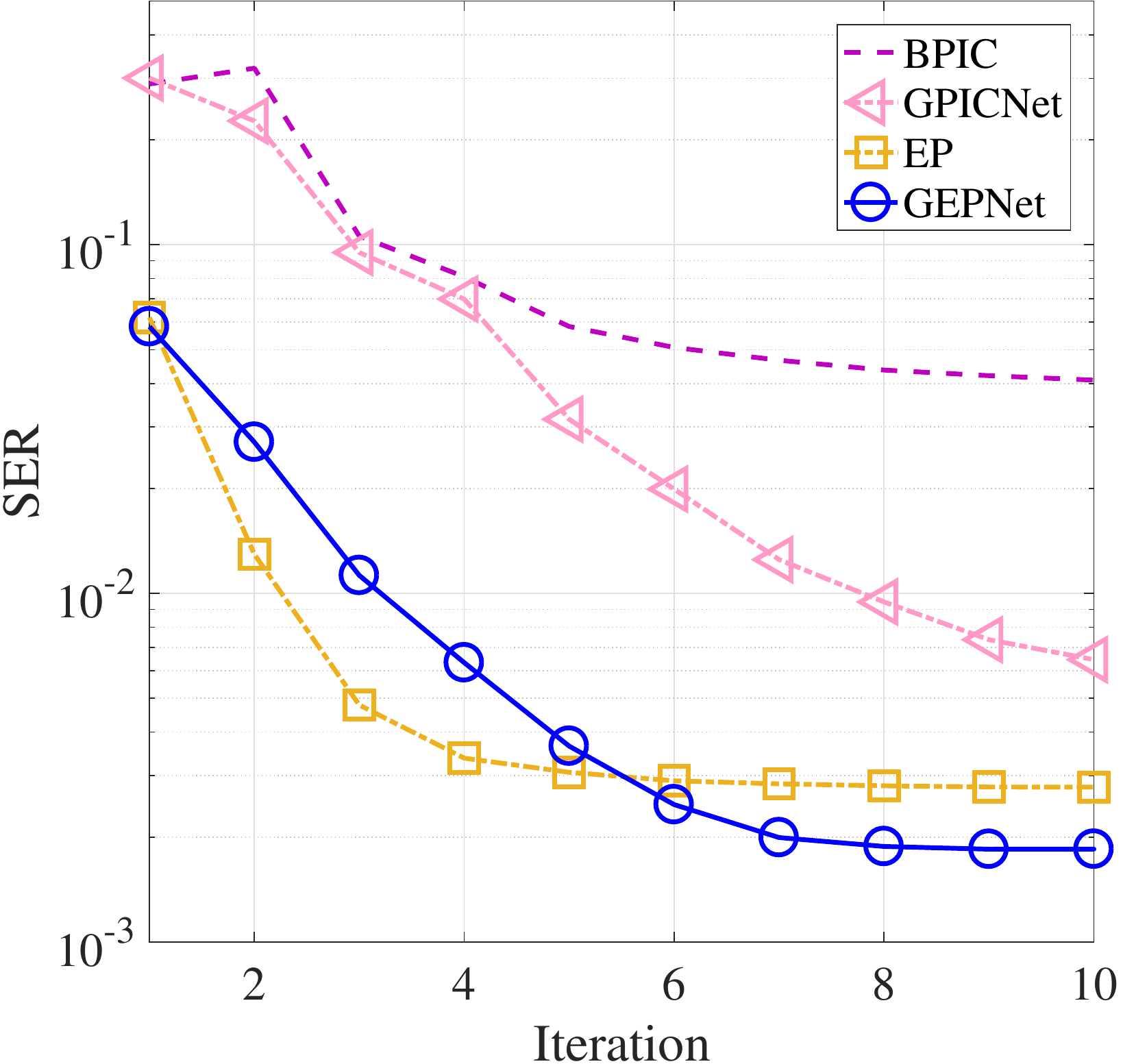}}
\caption{The convergence of the detectors for $4$-QAM and $N=16$}
\label{error_analysis_prop_dets}
\end{figure*}

\begin{figure*}
\centering
\subfloat[SNR = $9$ dB and $K=8$]
{\includegraphics[scale=0.35]{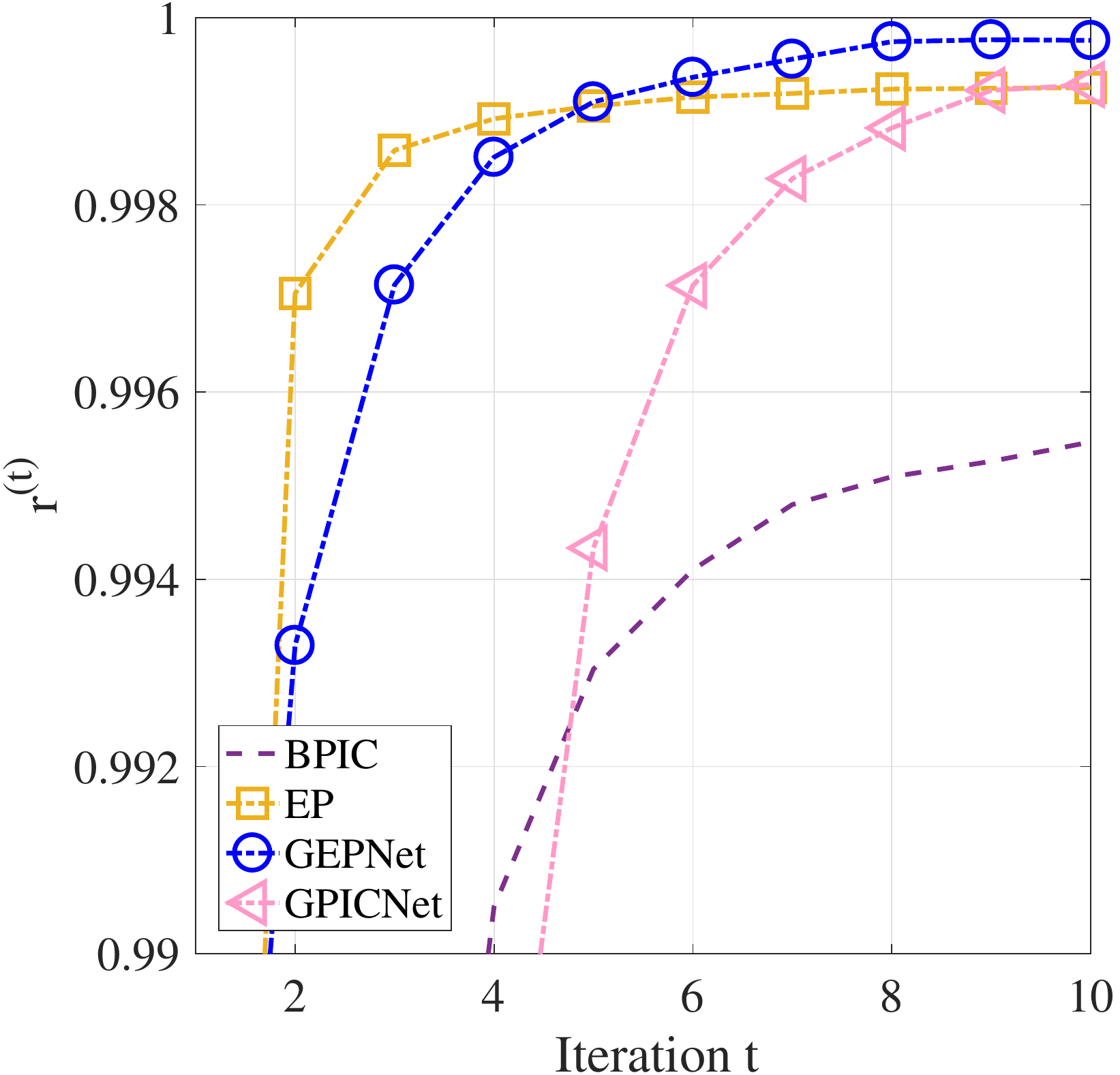}}\hfill
\centering
\subfloat[SNR = $11$ dB and $K=12$]
{\includegraphics[scale=0.35]{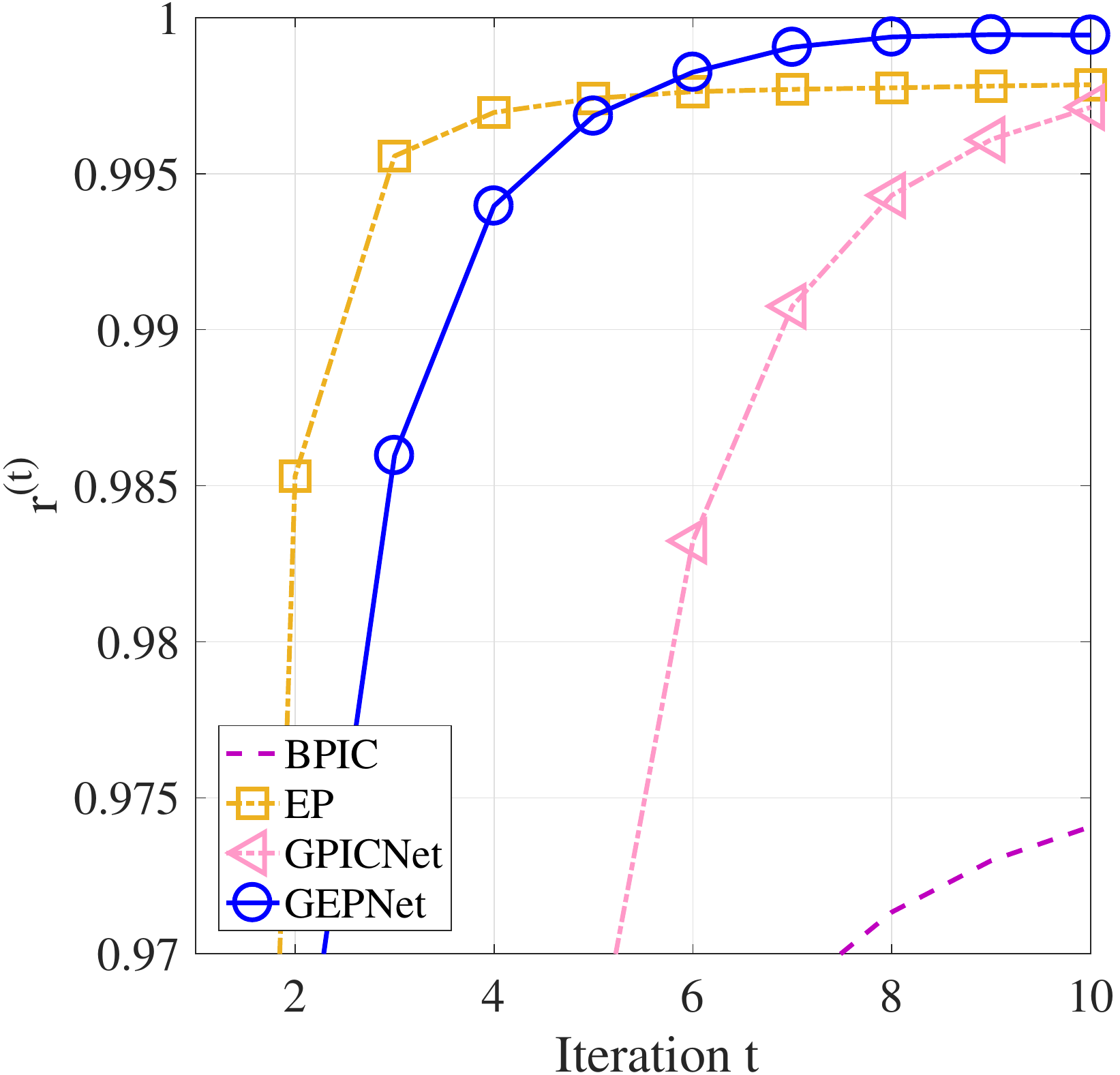}}\hfill
\centering
\subfloat[SNR = $12.5$ dB and $K=16$]
{\includegraphics[scale=0.35]{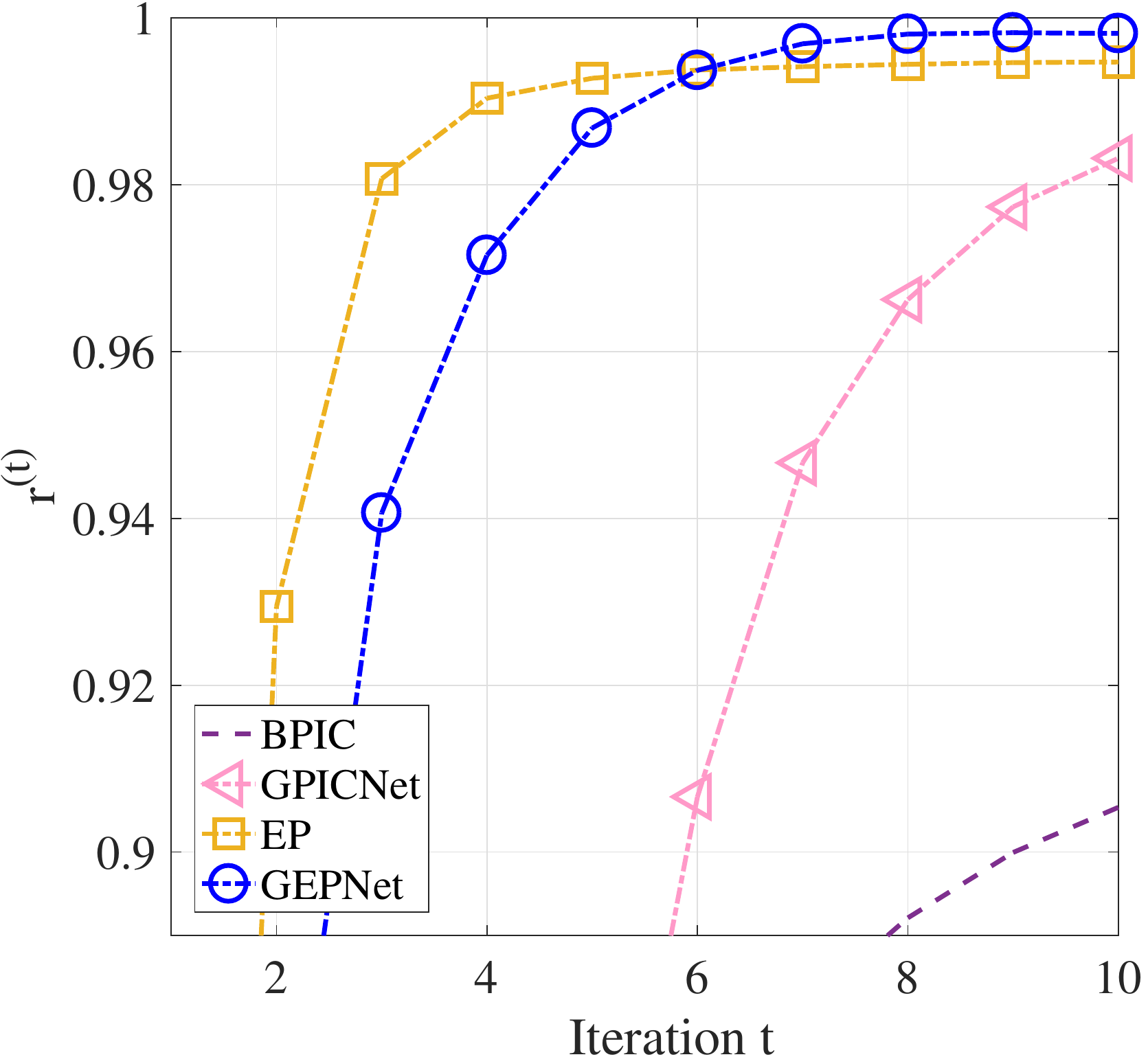}}
\caption{The average probability ratio for $4$-QAM and $N=16$}
\label{ratio_analysis_prop_dets}
\end{figure*}

\begin{figure*}
\centering
\subfloat[SNR = $13$ dB, $4$-QAM, $N=16, \text{ and } K=16$]
{\includegraphics[scale=0.35]{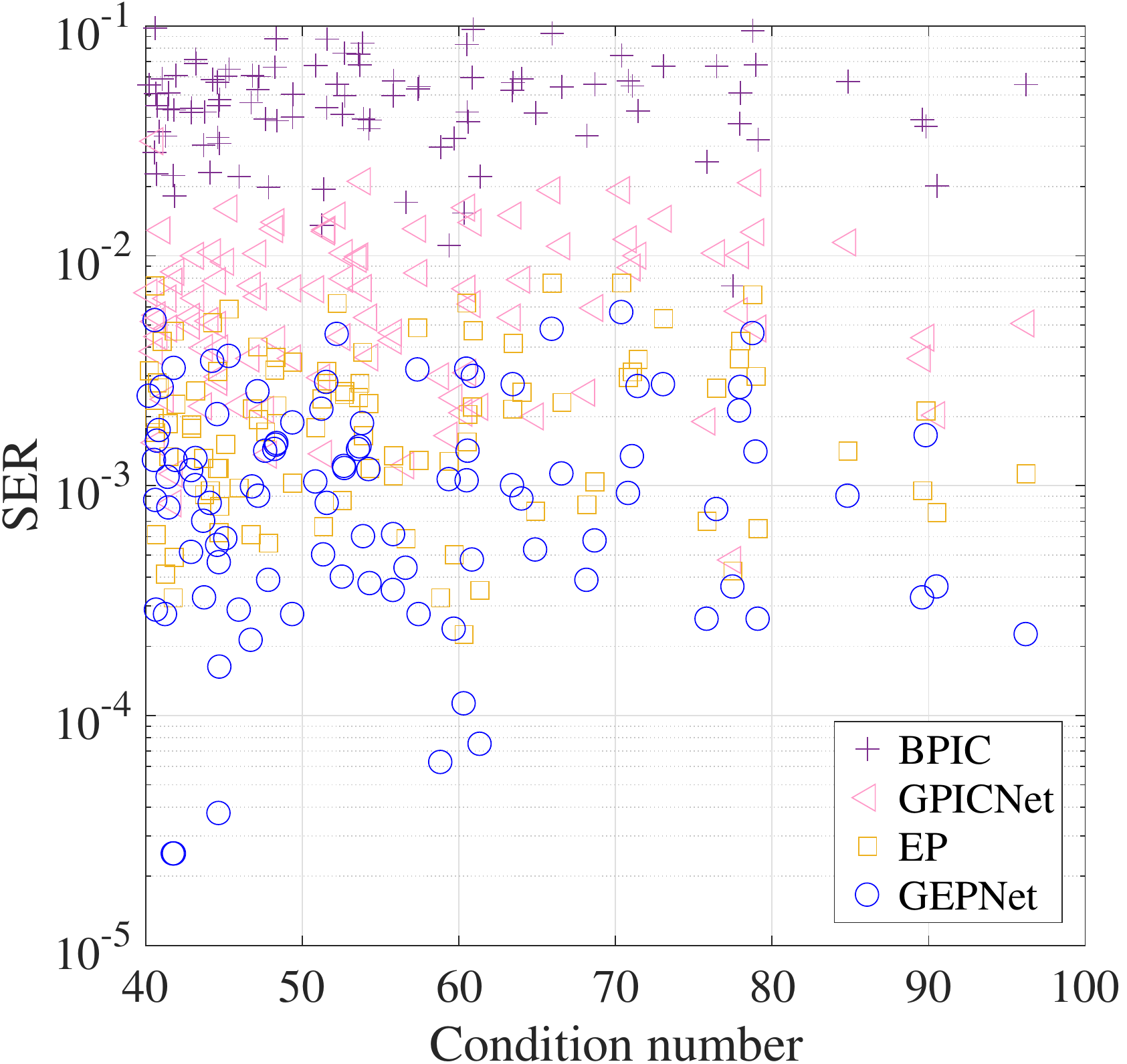}}\hfill
\centering
\subfloat[SNR = $11$ dB, $4$-QAM, $N=64, \text{ and } K=64$]
{\includegraphics[scale=0.35]{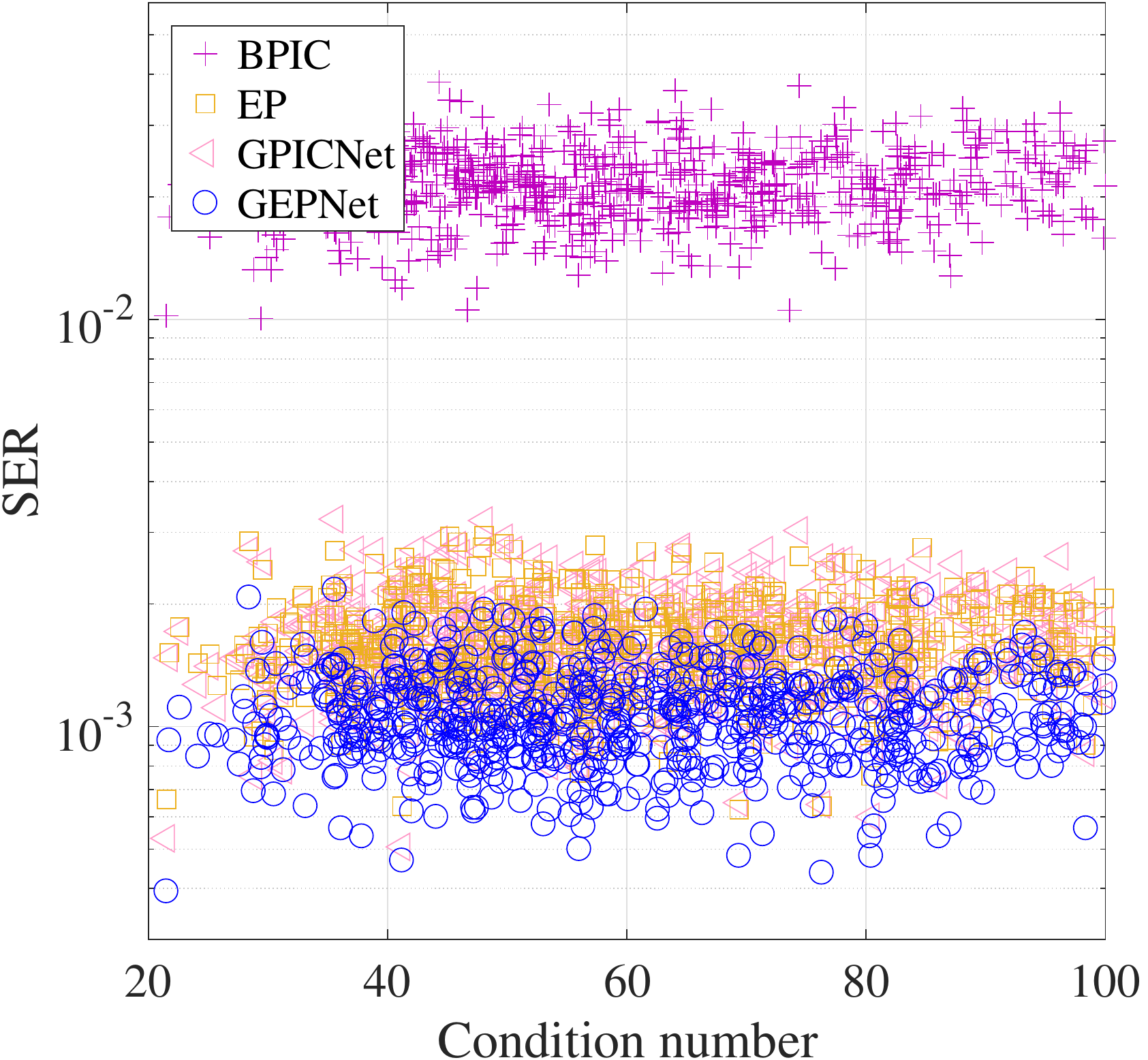}}\hfill
\centering
\subfloat[SNR = $20$ dB, $16$-QAM, $N=64, \text{ and } K=64$]
{\includegraphics[scale=0.35]{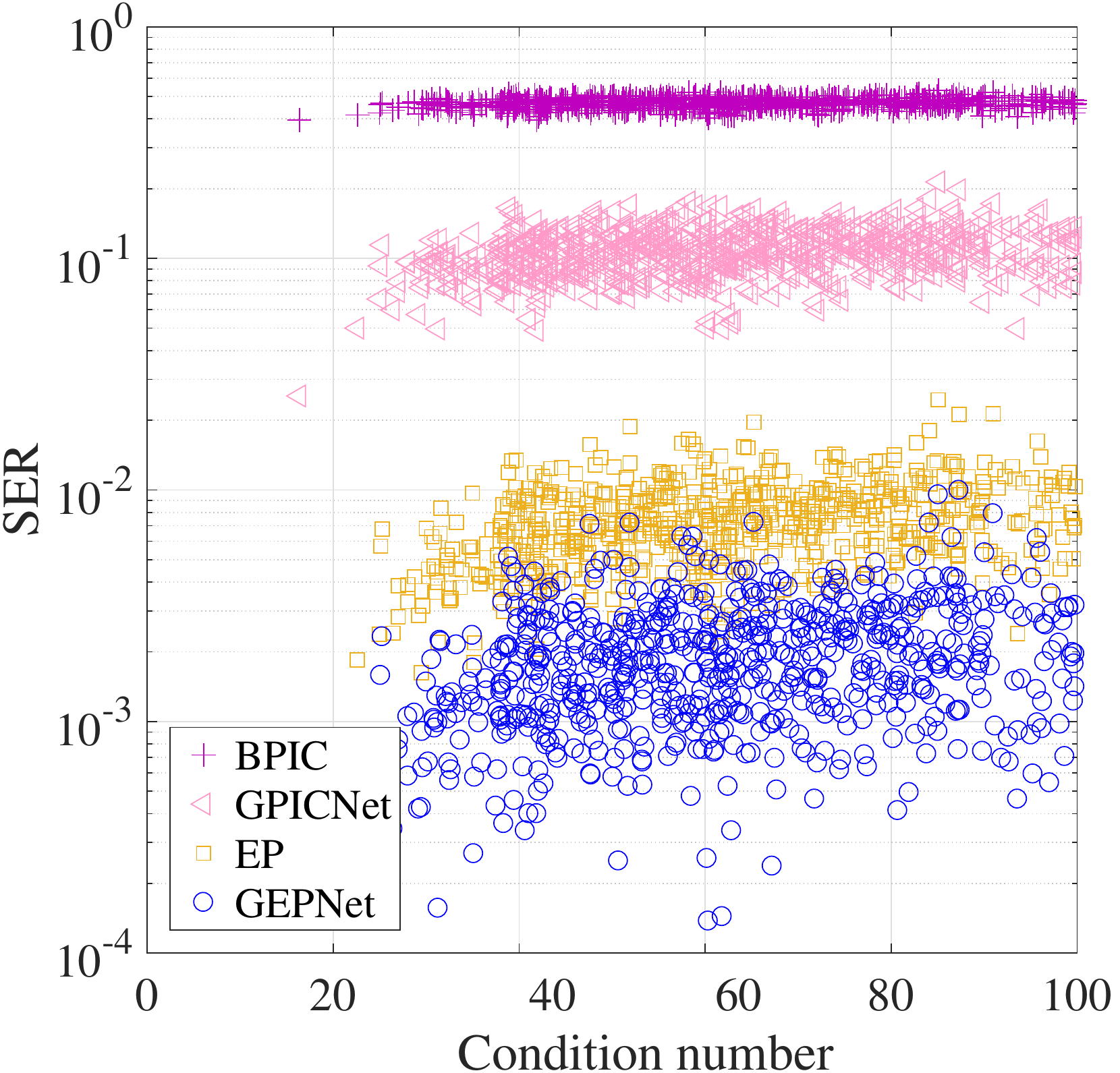}}
\caption{The SER performance comparison with respect to the channel condition number}
\label{Channel_condition}
\end{figure*}

In this section, we show that the proposed GEPNet and GPICNet detectors improve the accuracy of the posterior distribution approximation compared to the EP and BPIC detectors, respectively, by computing the SER and $r^{(t)}$, defined in Section \ref{Sect_Analysis_Gaussian}. Note that the SNR values are set so that the ML detector can achieve the SER of $10^{-3}$. 
Figs. \ref{error_analysis_prop_dets}-\ref{ratio_analysis_prop_dets} show the SER and $r^{(t)}$ for the EP, GEPNet, BPIC and GPICNet detectors.  
It is evident that the GEPNet and GPICNet detectors exhibit a lower SER and a higher $r^{(t)}$ compared to the EP and BPIC detectors, respectively. Therefore, we conclude that the GEPNet and GPICNet detectors provide more accurate posterior distribution approximation.

\subsection{SER vs Channel Condition Number}

The symbol detection is especially challenging in the case of ill-conditioned channel matrices. A matrix is ill-conditioned if it has a high condition number, which is defined as the ratio between the maximum and minimum singular values. Note that the channel matrix coefficients in \eqref{A2} 
follow a Gaussian distribution with zero mean and variance $1/N$.  Since the channel matrix coefficients are random variables, the corresponding condition number is also a random variable. Thus, there is a certain  probability that a particular random realization of channel matrix $\qH$ would be ill-conditioned.
We perform simulations for SER vs channel condition number by generating $5000$ random realizations of symbols $\qx$ and noise $\qn$ for each of $1000$ channel matrix realizations $\qH$. The SER is calculated for each condition of channel realization.  
Fig. \ref{Channel_condition} shows the SER performance of the BPIC, EP, GPICNet, and GEPNet detectors with respect to the channel matrix condition number. The SNR is set so that the GEPNet detector achieves the SER of $10^{-3}$. It can be seen that the integration of the GNN with the BPIC and EP detectors allows the detectors to deal better with ill-conditioned channel matrices. 

\subsection{SER vs SNR}

\begin{figure*}
\centering
\subfloat[$K=8$]
{\includegraphics[scale=0.35]{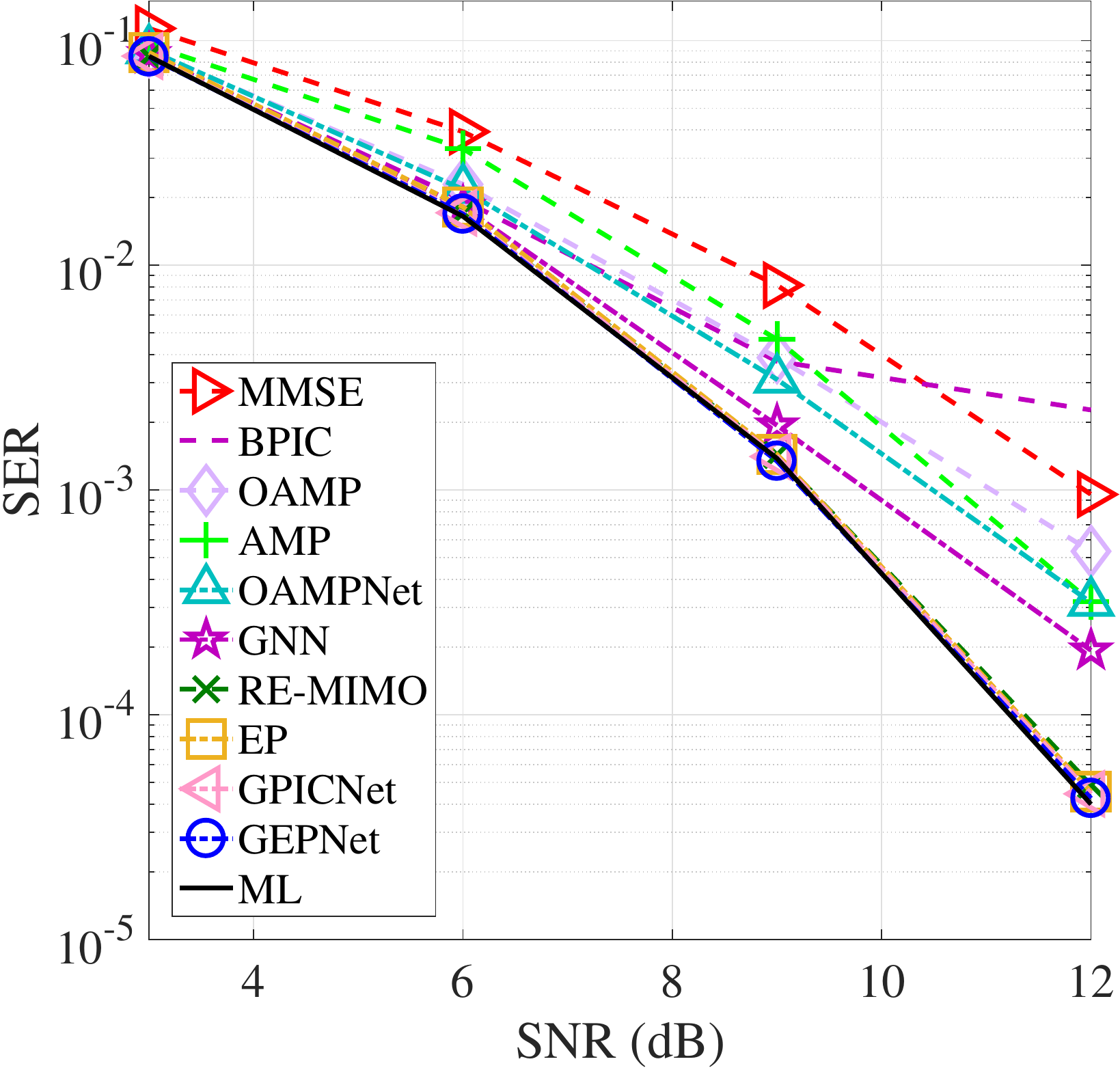}}\hfill
\centering
\subfloat[$K=12$]
{\includegraphics[scale=0.35]{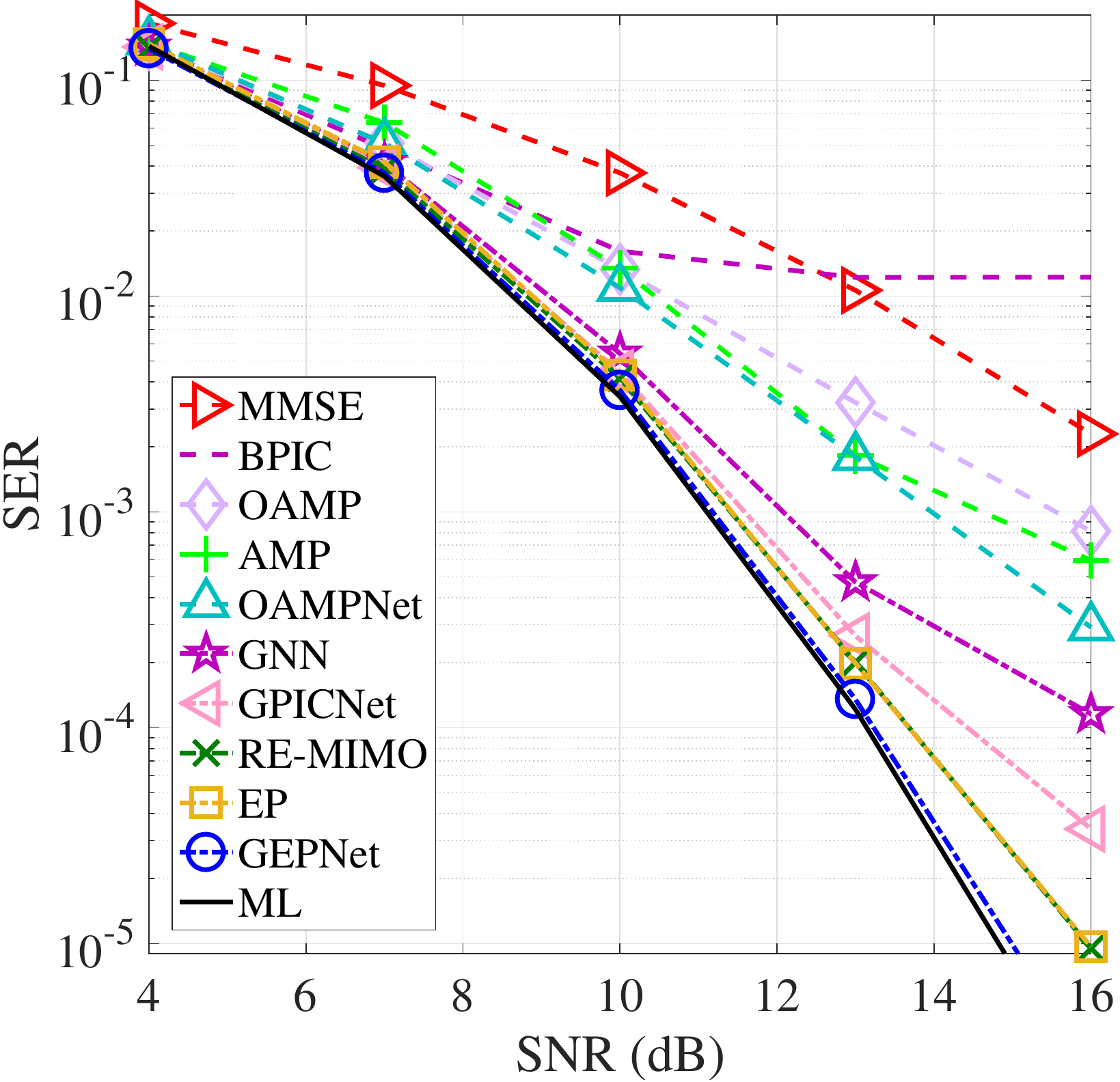}}\hfill
\centering
\subfloat[$K=16$]
{\includegraphics[scale=0.35]{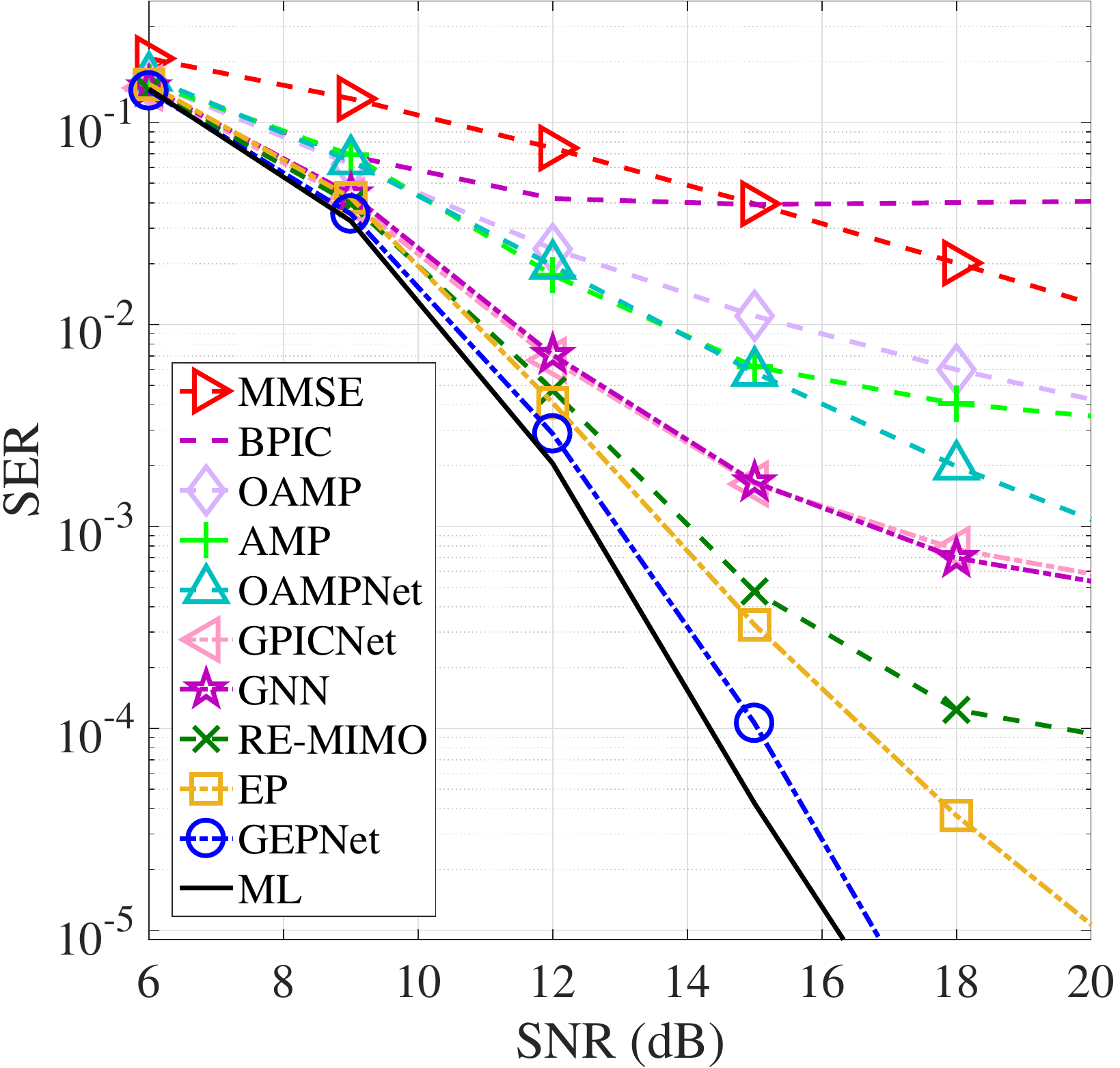}}\hfill
\caption{The SER performance comparison for $4$-QAM, $N=16$,  ${\sf SNR_{\rm min}}=3$, and ${\sf SNR_{\rm max}}=21$}
\label{SER1}
\end{figure*}

\begin{figure*}
\centering
\subfloat[$K=32$]
{\includegraphics[scale=0.35]{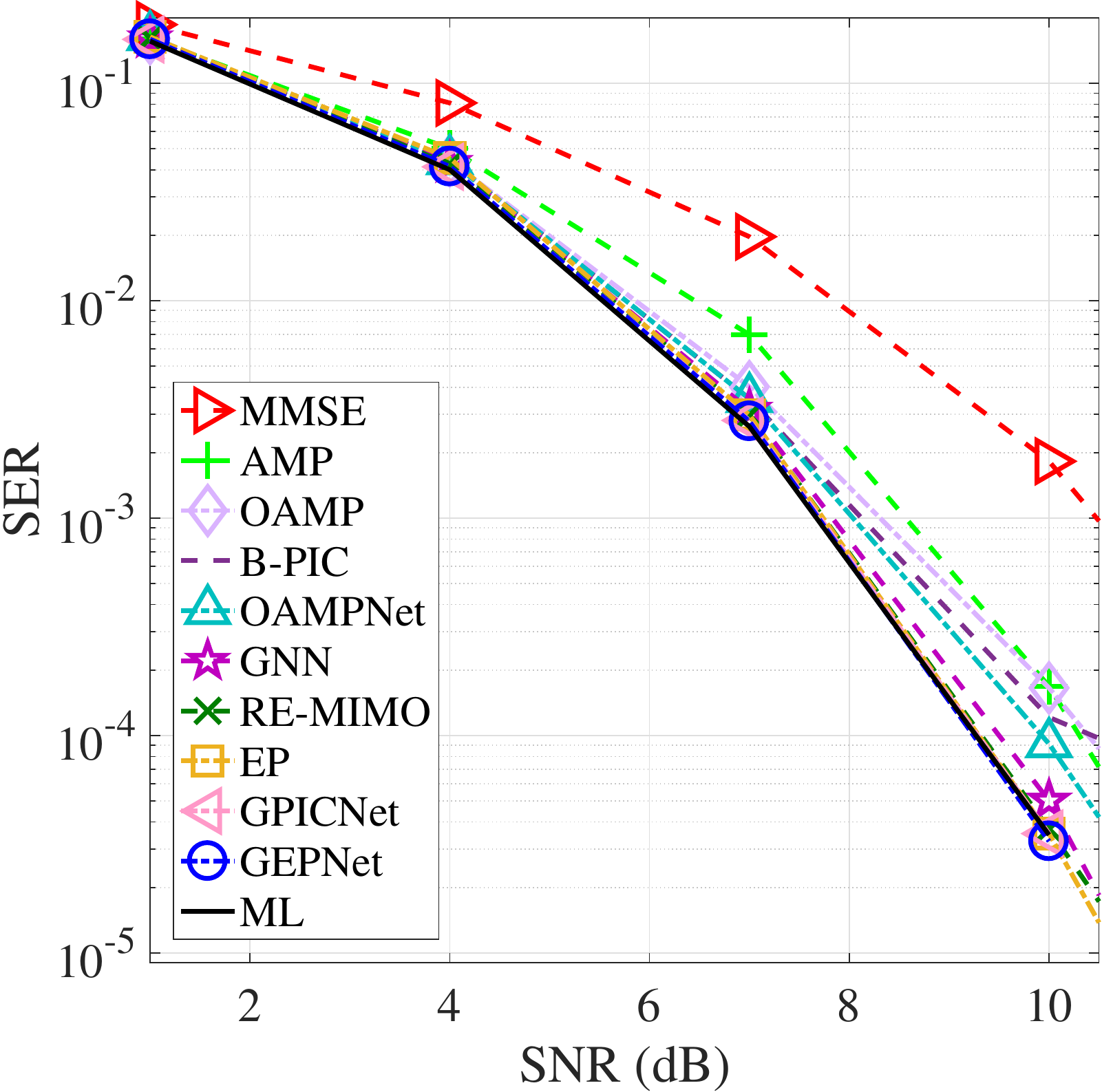}}\hfill
\centering
\subfloat[$K=48$]
{\includegraphics[scale=0.35]{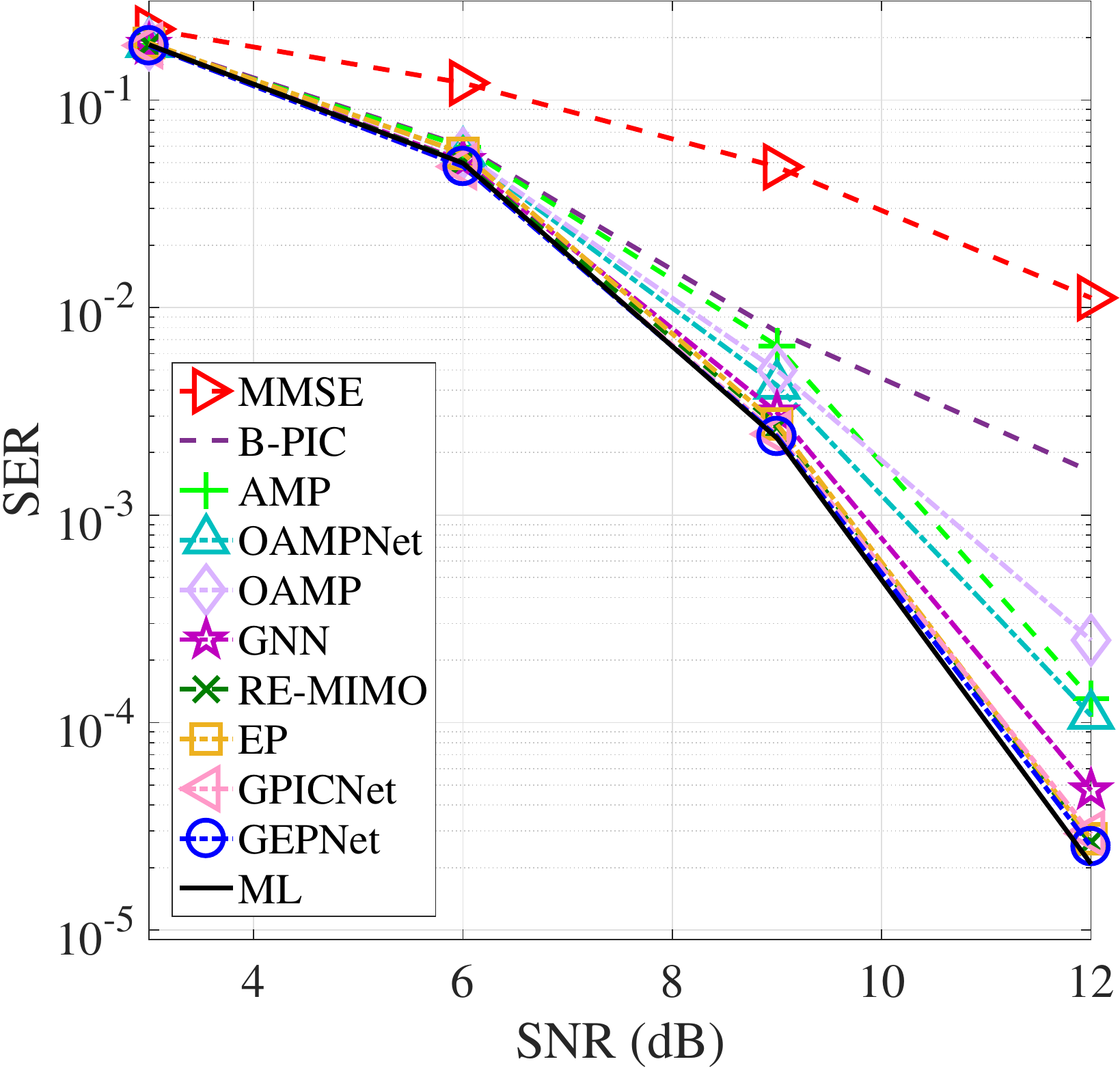}}\hfill
\centering
\subfloat[$K=64$]
{\includegraphics[scale=0.35]{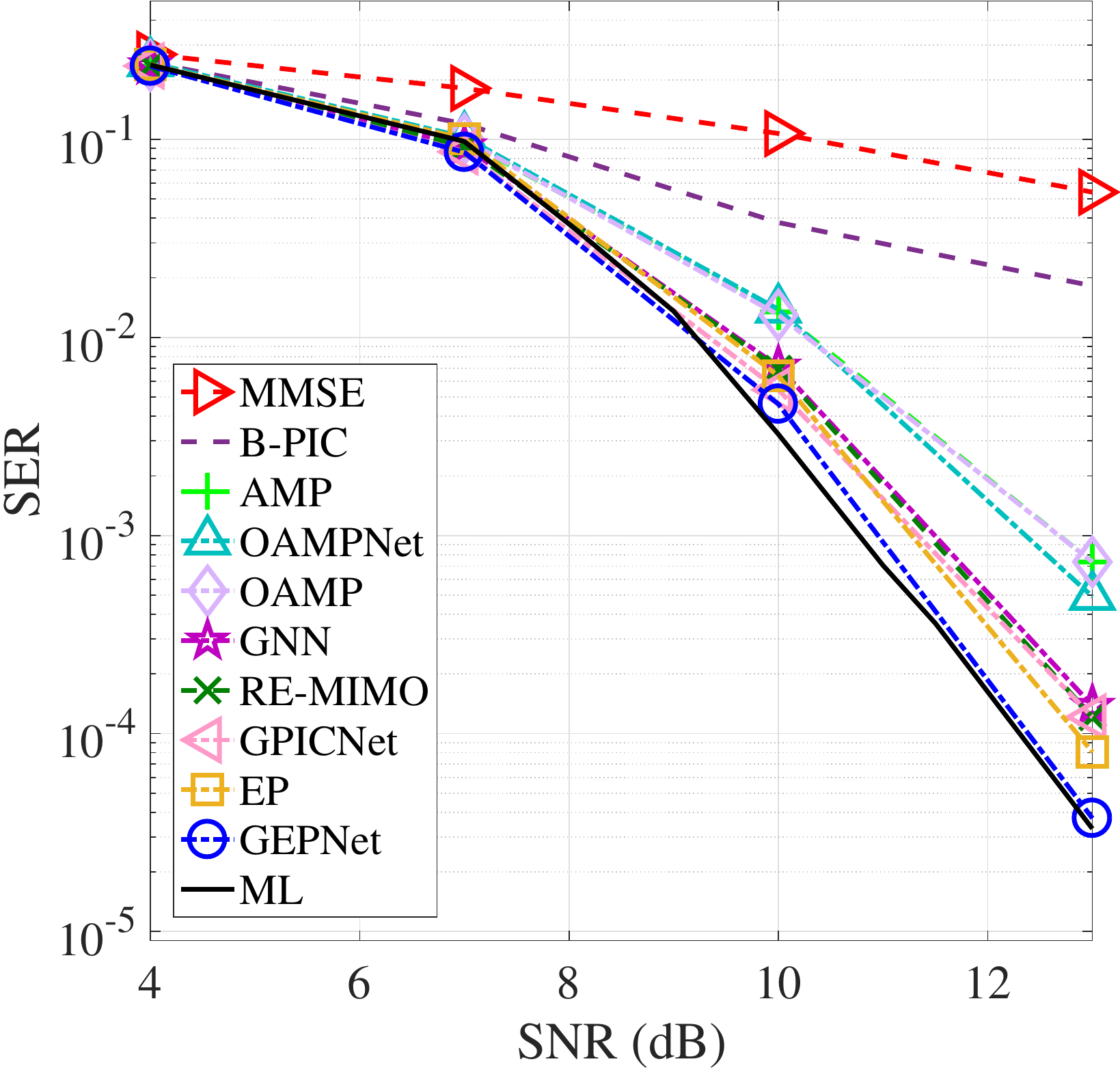}}\hfill
\caption{The SER performance comparison for $4$-QAM, $N=64$,  ${\sf SNR_{\rm min}}=1$, and ${\sf SNR_{\rm max}}=13$}
\label{SER2}
\end{figure*}

\begin{figure*}
\centering
\subfloat[$K=32$]
{\includegraphics[scale=0.35]{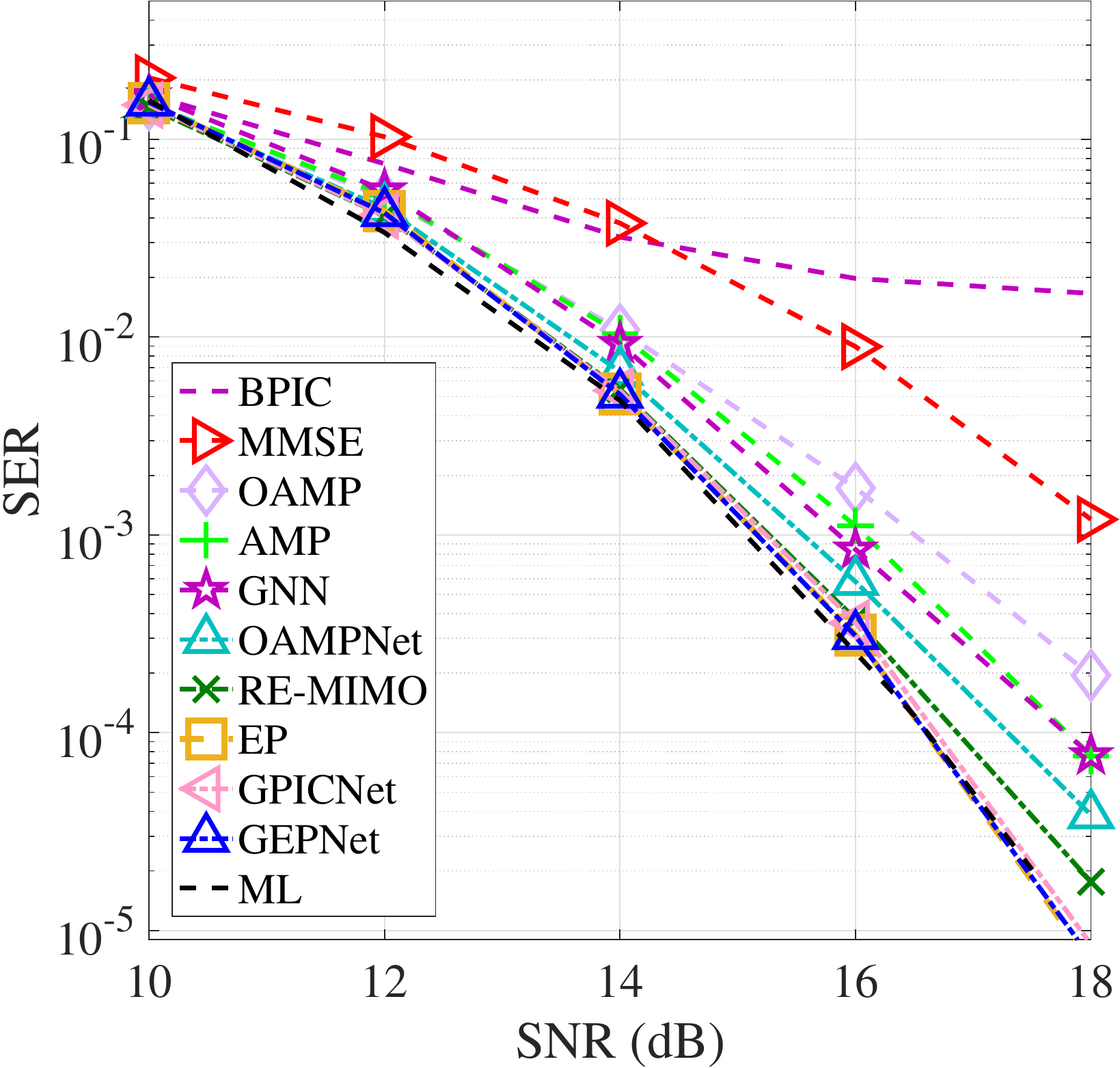}}\hfill
\centering
\subfloat[$K=48$]
{\includegraphics[scale=0.35]{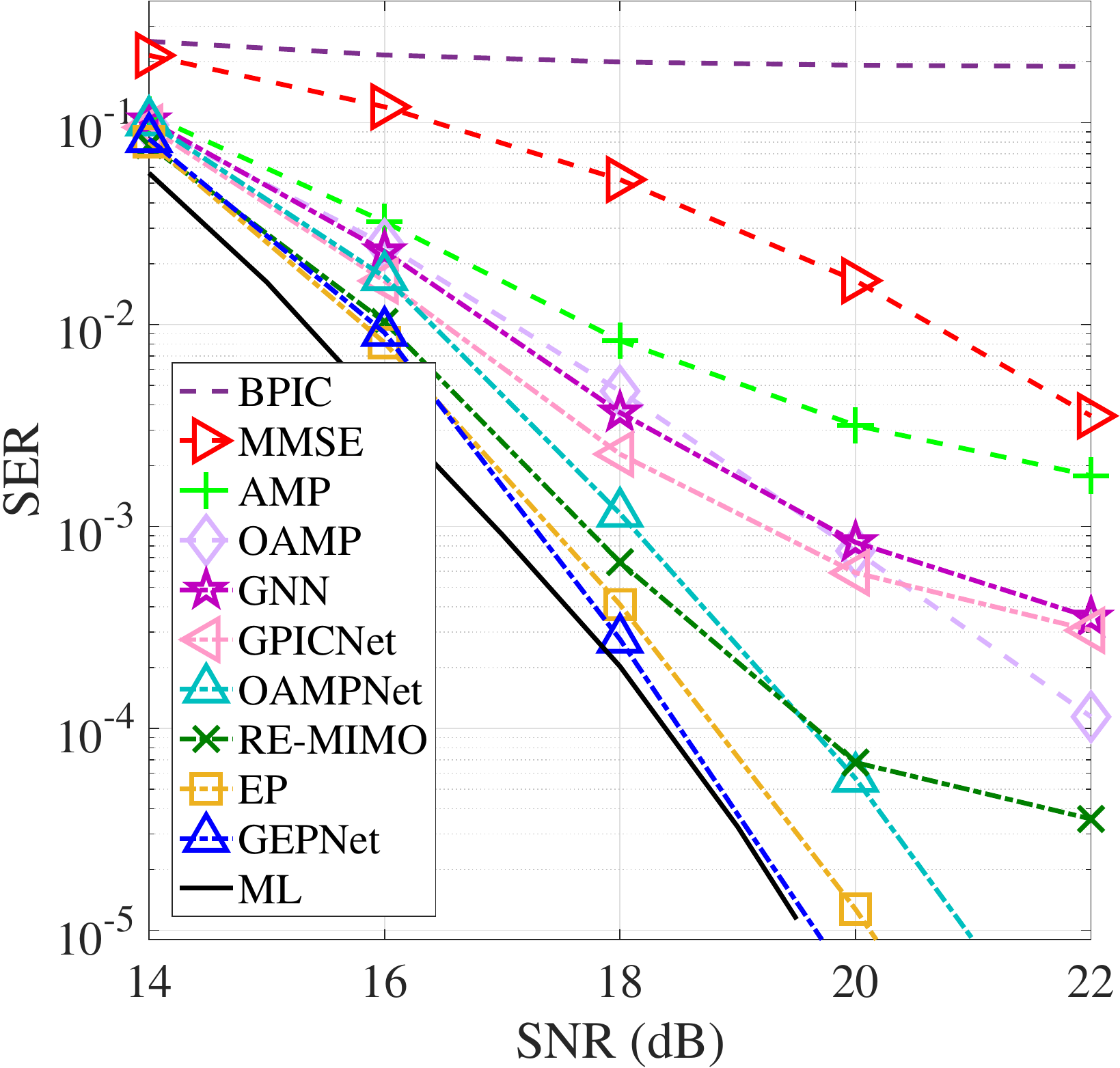}}\hfill
\centering
\subfloat[$K=64$]
{\includegraphics[scale=0.35]{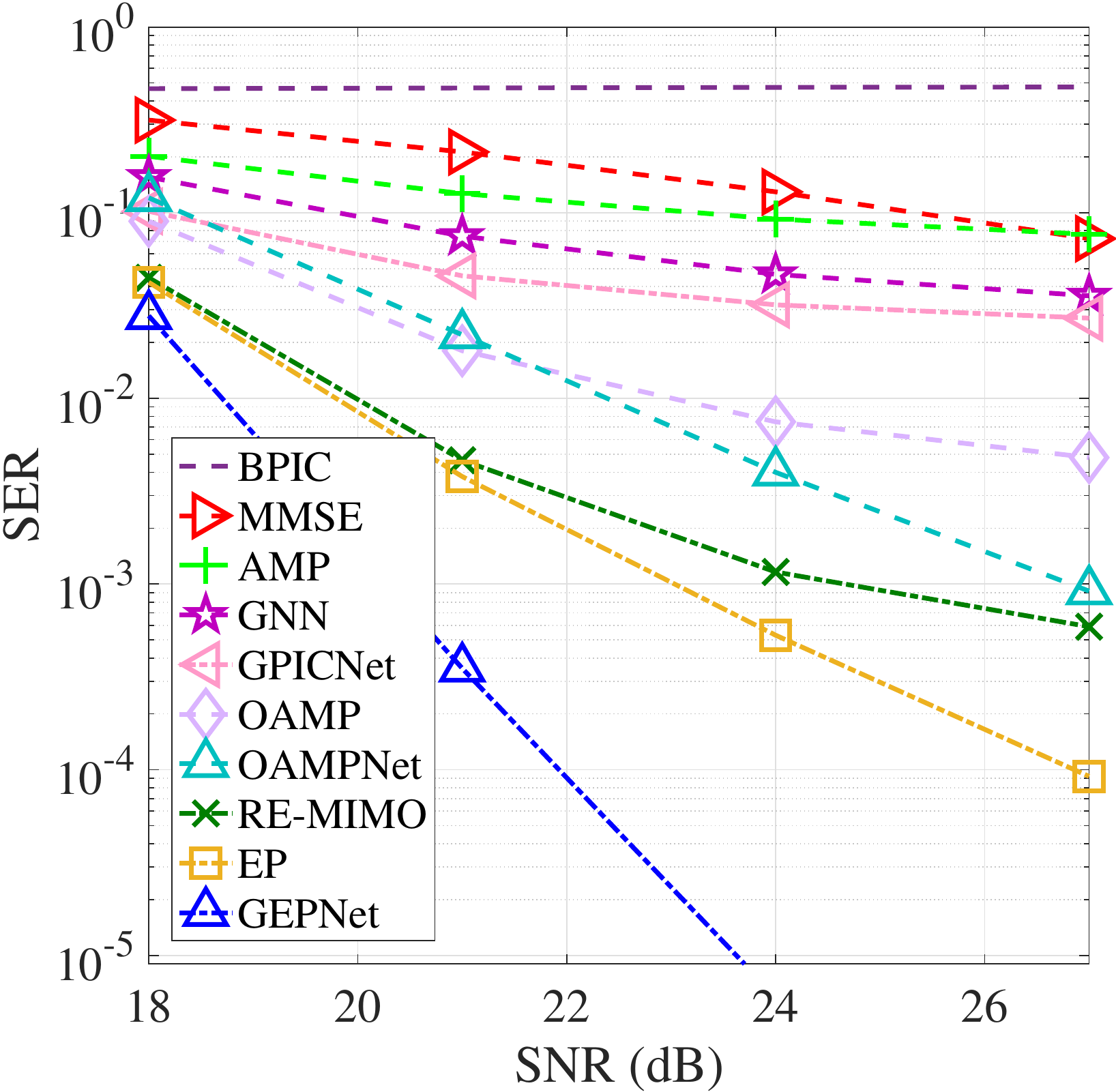}}\hfill
\caption{The SER performance comparison for $16$-QAM and $N=64$,  ${\sf SNR_{\rm min}}=10$, and ${\sf SNR_{\rm max}}=27$}
\label{SER3}
\end{figure*}

\begin{figure*}
\centering
\subfloat[$N=8,K=8$]
{\includegraphics[scale=0.35]{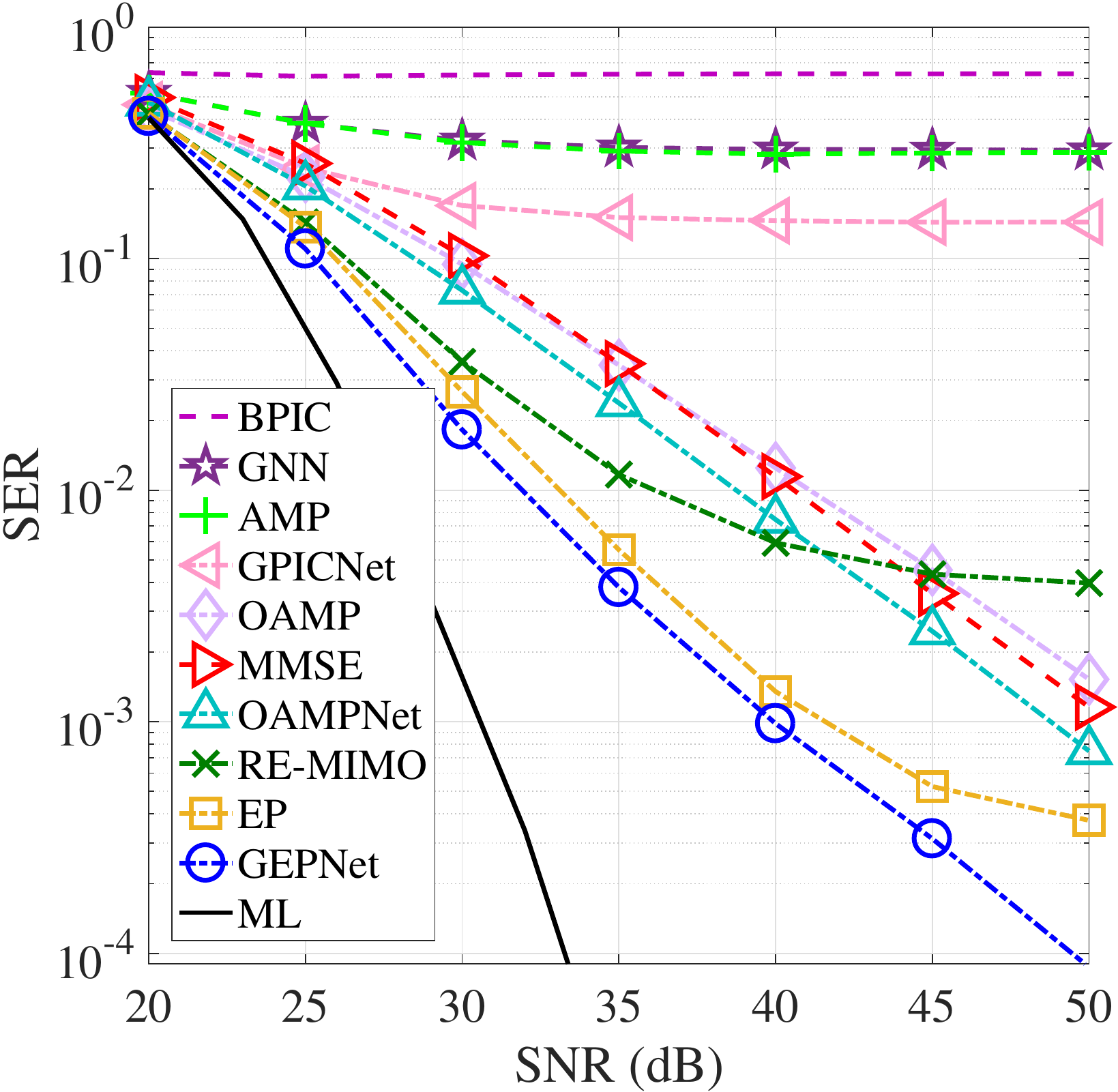}}\hfill
\centering
\subfloat[$N=16,K=8$]
{\includegraphics[scale=0.35]{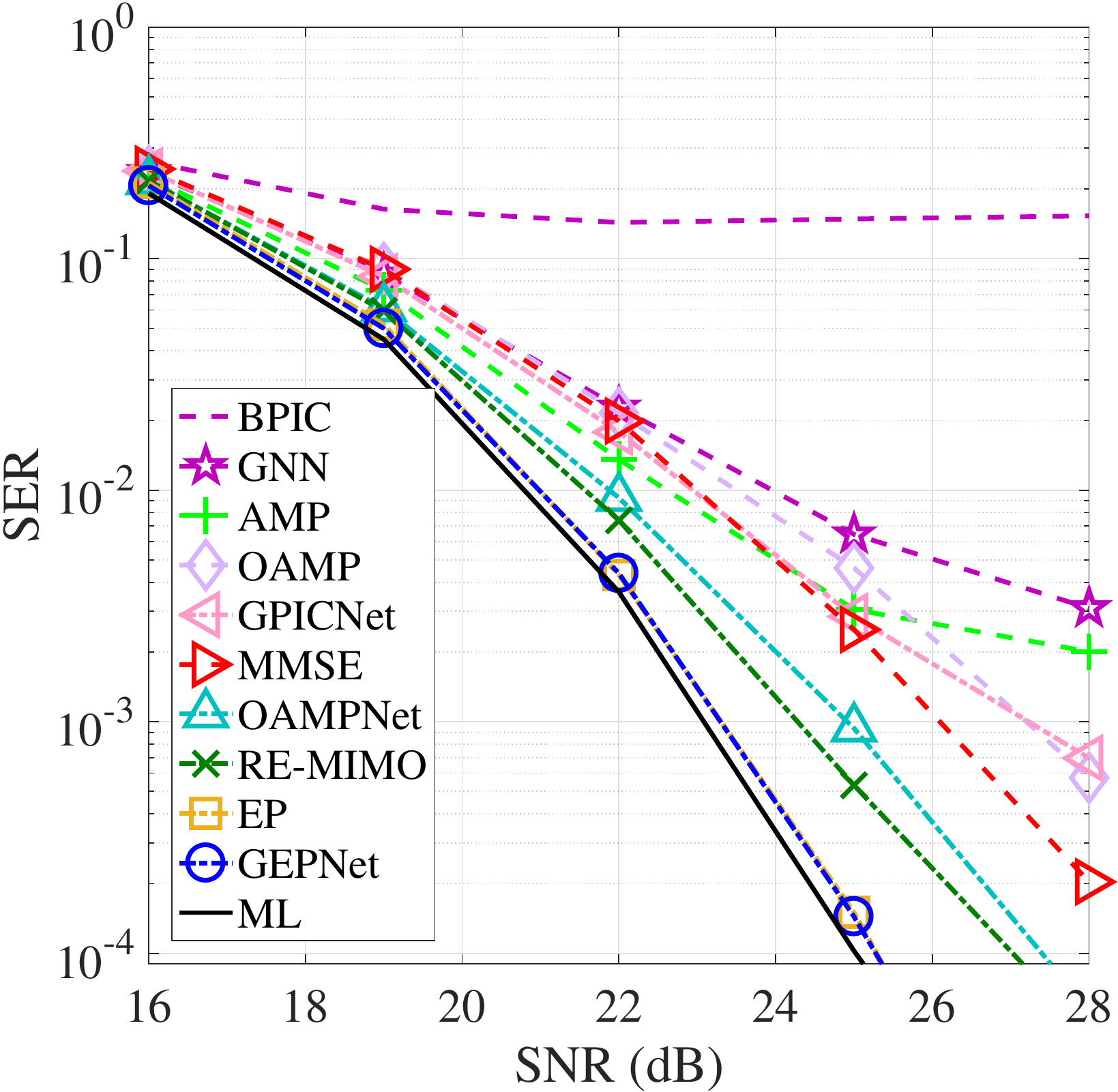}}\hfill
\centering
\subfloat[$N=16,K=16$]
{\includegraphics[scale=0.35]{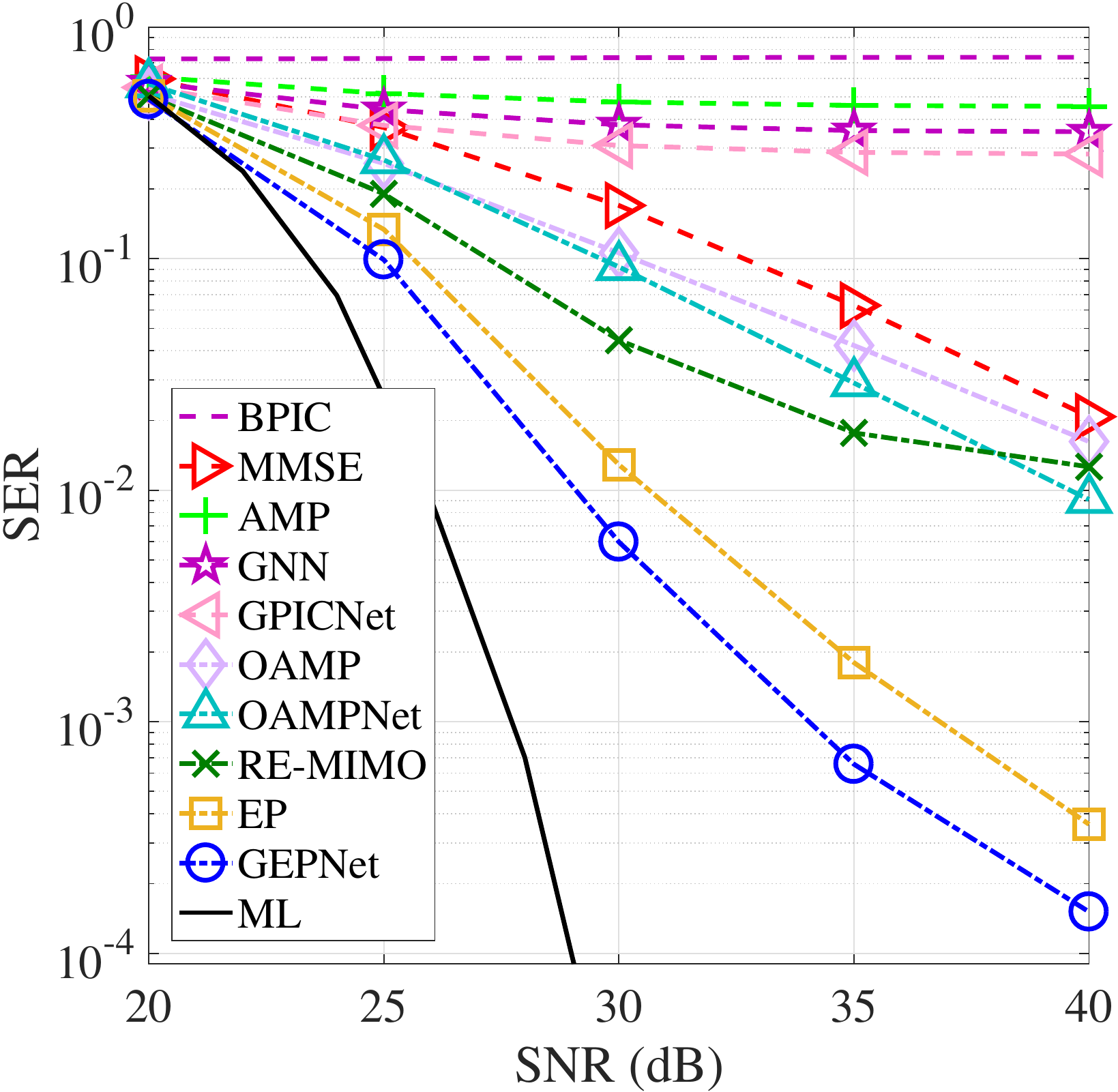}}\hfill
\caption{The SER performance comparison for $64$-QAM in small MIMO configurations, ${\sf SNR_{\rm min}}=20$, and ${\sf SNR_{\rm max}}=51$}
\label{SER_revision}
\end{figure*}

\begin{figure*}
\centering
\subfloat[SNR = $13$ dB, $4$-QAM,  and $N=16$]
{\includegraphics[scale=0.35]{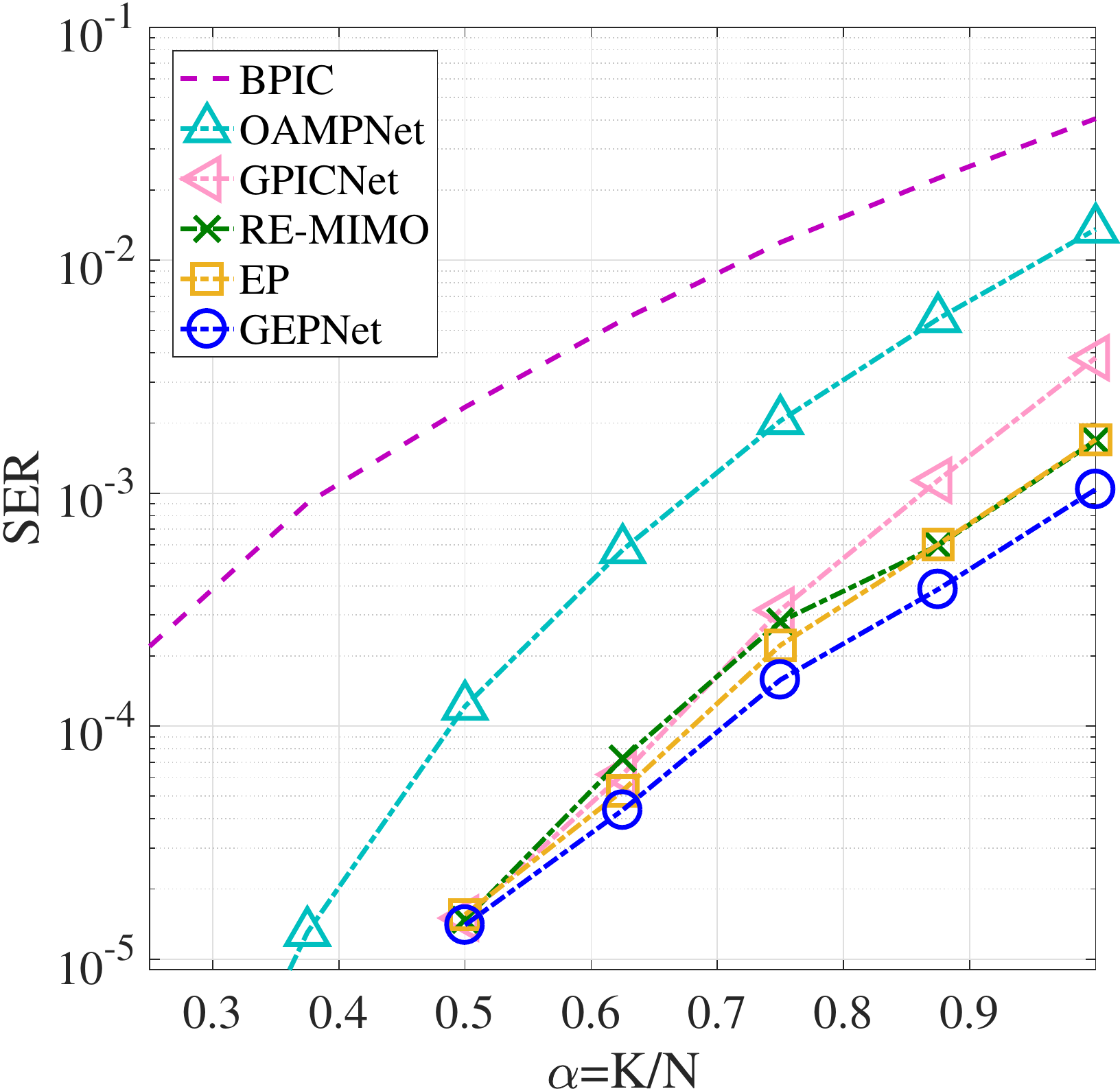}}\hfill
\centering
\subfloat[SNR = $11$ dB, $4$-QAM,  and $N=64$]
{\includegraphics[scale=0.35]{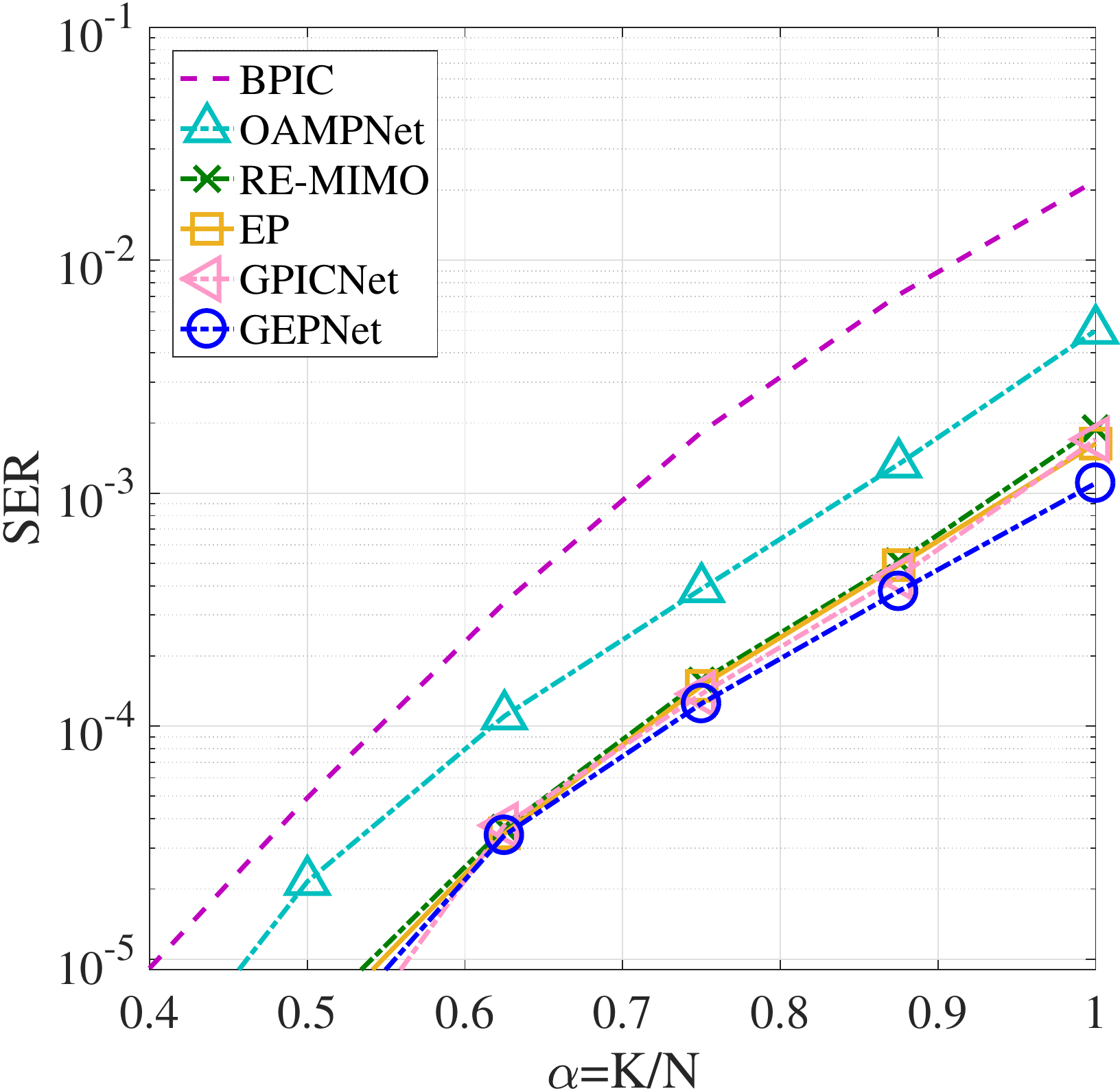}}\hfill
\centering
\subfloat[SNR = $20$ dB, $16$-QAM,  and $N=64$]
{\includegraphics[scale=0.35]{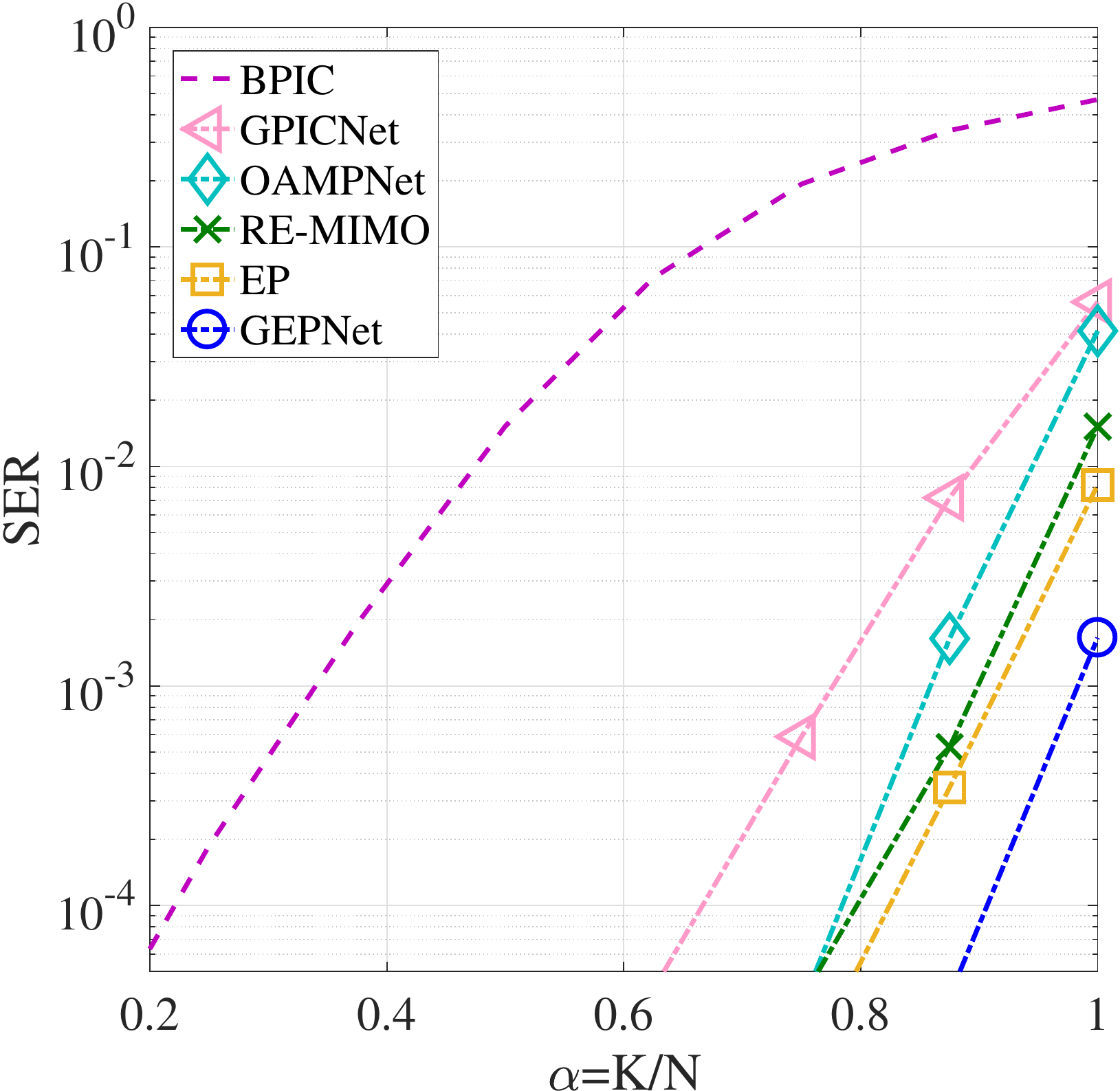}}
\caption{The SER performance comparison with respect to the transmit-to-receive antennas ratio}
\label{SER_ratio}
\end{figure*}

We first evaluate the SER performance of the proposed detectors with $4$-QAM modulation scheme for $N=16$ and $K\in \{8,12,16\}$, as illustrated in Fig. \ref{SER1}. The GEPNet closely approaches the ML performance for the transmit-to-receive antennas ratio $\frac{K}{N} =  \frac{8}{16},\frac{12}{16},\frac{16}{16}$. This is in contrast to the other MP \cite{2009Donoho_ProcSci_AMP,AKosasih,Ma-17ACCESS,Jespedes-TCOM14} and NN based \cite{AScotti_GNN_2020,2018HHE_Globecom_OAMPNet,2021_KPratik_TSP_REMimo} detectors that show a significant performance gap when  $\frac{K}{N}$ is $1$. The GPICNet detector achieves a similar performance to the GEPNet detector  when $\frac{K}{N}$ is $0.5$. However, its performance degrades when the number of users increases, as expected since the GPICNet detector balances the complexity and performance. Nevertheless, the GPICNet provides a great advancement in terms of the  performance compared to its predecessor BPIC detector. We then investigate the performance of the detectors in systems with a higher number of receive antennas, i.e., $N=64$, as illustrated in Fig. \ref{SER2}. We observe that the proposed detectors for $N=64$ behave in the same way as for $N=16$. 
Finally, to better understand the behaviour of the detectors with a higher order QAM modulation, we perform simulations with $N=64$, $K\in \{32,48,64\}$ and  $16$-QAM modulation scheme. The results in Figs. \ref{SER3}a and b indicate that the GEPNet can achieve a near ML performance, while in Fig. \ref{SER3}c the GEPNet is shown to outperform the state-of-the-art MU-MIMO detectors by more than $4$dB at SER of $10^{-4}$. Note that we are not able to generate  the ML detection result for $K=64$ with $16$-QAM due to prohibitive computational complexity. The GPICNet achieves an excellent detection performance  when $\frac{K}{N}$ is  $0.5$.  
We further test the proposed detectors in small MIMO systems with a high order QAM modulation scheme, i.e., $64$-QAM. Low dimensional MIMO systems (e.g., $N=K=16$) are currently of high interest, especially for $5$G smartphones \cite{Serghiou_2020}. The results are depicted in Fig. \ref{SER_revision} where the GEPNet  outperforms the EP by around $3$ dB for $N=K=16$ MIMO configuration at the SER of $10^{-3}$.

\subsection{SER vs Transmit-to-Receive Antennas Ratio}

We performed simulations to analyse the performance of the detectors in terms of the SER with respect to the transmit-to-receive antennas ratio $\alpha$. The MMSE, OAMP, AMP, and GNN detectors are excluded from the comparison, since their performance is worse than that of the EP, OAMPNet, and RE-MIMO detectors, as shown in the previous section. The BPIC detector is included, since the gap between the GPICNet and BPIC gives an idea how significantly the GNN can improve the performance. The SNR is set so that the GEPNet detector achieves the SER of $10^{-3}$ when $\alpha=1$.  The results are depicted in Fig. \ref{SER_ratio}. It can be seen that the GEPNet detector outperforms the other MU-MIMO detectors. Compared to the BPIC detector, the GPICNet detector has two times higher number of users that can be served simultaneously by $N$ receive antennas. In practice, this means that the GPICNet doubles the multiplexing gain of the BPIC detector.

\section{Conclusion}
\label{sConclusion}

In this paper, we analysed the accuracy of the IGA-based posterior distribution approximation  in the state-of-the-art MP detectors, specifically, the EP, OAMP and BPIC detectors. Motivated by this analysis, we developed a GNN-based framework to improve the accuracy of the posterior distribution approximation by fine-tuning the cavity distribution in the MP detectors that rely on the IGA. We focused on the improvement of the high-performance EP and low-complexity BPIC detectors. We introduced an additional GNN module in these detectors to fine-tune the cavity distribution. The resulting detectors are referred to as the GEPNet and GPICNet detectors. We proved that the proposed detectors are robust to user permutations and changes in the number of users. Simulation results show that the GEPNet detector can achieve a near-ML performance in various system configurations and significantly outperforms the EP detector, while the GPICNet detector doubles the multiplexing gain of the BPIC detector.

\section*{Acknowledgment}

This research was supported by the University of Sydney Research Training Program scholarship, Australian Research Council Laureate Fellowship grant number FL160100032, and ARC Discovery Project grant number DP210103410.

{\renewcommand{\baselinestretch}{1.1}
\begin{footnotesize}
\bibliographystyle{IEEEtran}
\bibliography{IEEEabrv,myBib}
\end{footnotesize}}

\begin{IEEEbiography}
 [{\includegraphics[width=1in,height=1.25in,clip,keepaspectratio]{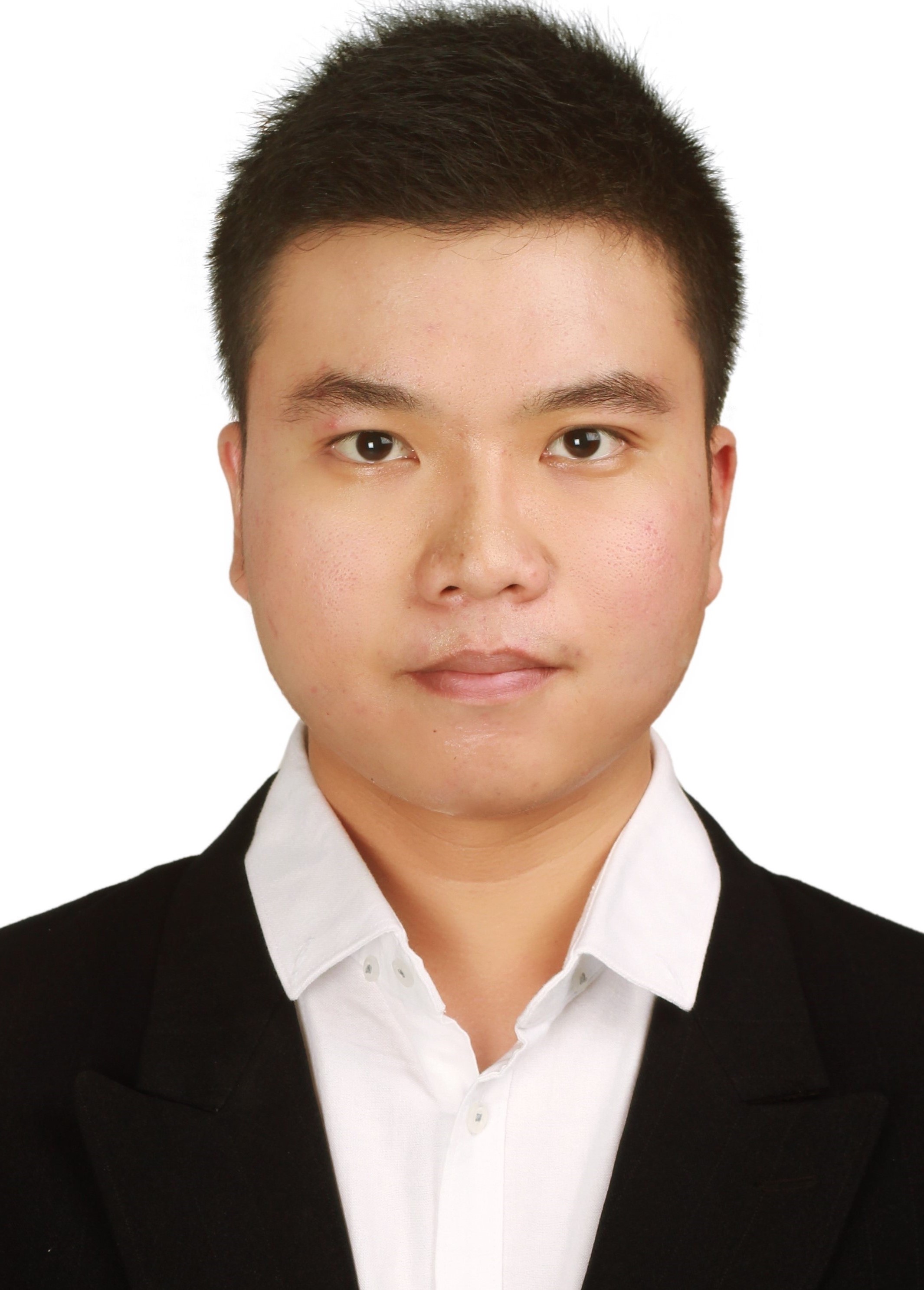}}]{Alva Kosasih}(S’19)
 received the B.Eng. and M.Eng. degrees both in electrical engineering from Brawijaya University, Indonesia, in 2013 and 2017,
respectively; and the M.S. degree in communication engineering from National Sun Yat-sen University, Taiwan, in 2017. He is currently pursuing the Ph.D degree in the School of Electrical and Information Engineering, the University of Sydney, Australia. His research interests
include massive MIMO systems and the application of  machine learning in wireless communications.
\end{IEEEbiography}

\begin{IEEEbiography}
 [{\includegraphics[width=1in,height=1.25in,clip,keepaspectratio]{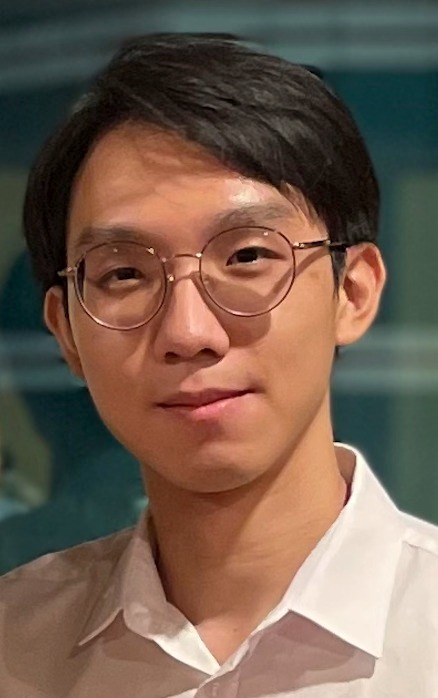}}]{Vincent Onasis}
Vincent Onasis is a final year undergraduate student at the University of Sydney, Australia. He is a Dalyell Scholar pursuing the B. Eng. Degree in software engineering and will graduate at the end of 2022. His interest lies in practical application of software engineering in solving real world challenges, such as the implementation of machine learning in telecommunication networks.
\end{IEEEbiography}

\begin{IEEEbiography}
 [{\includegraphics[width=1in,height=1.25in,clip,keepaspectratio]{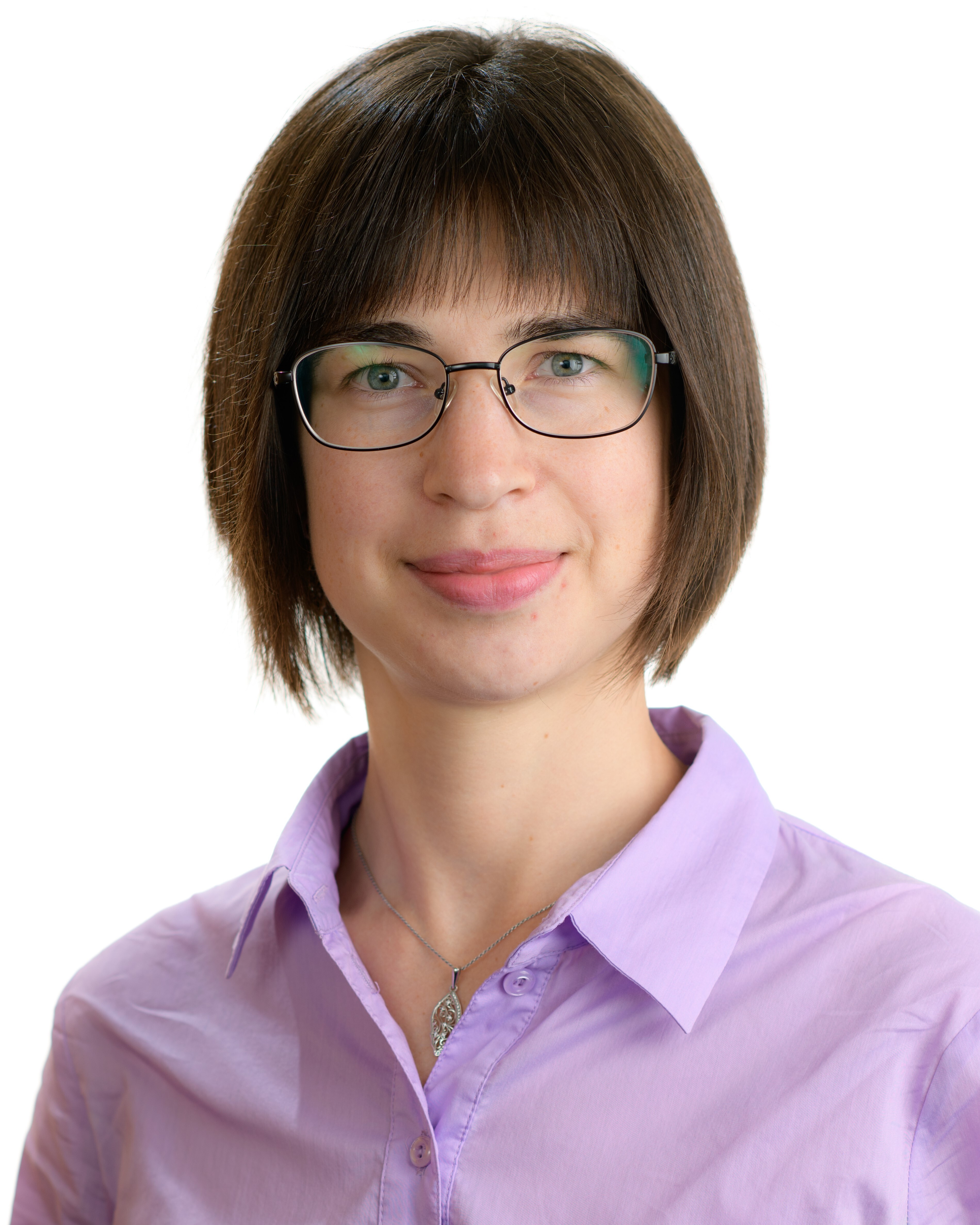}}]{Vera Miloslavskaya}
received B.Sc., M.Sc. and PhD degrees from Peter the Great St. Petersburg Polytechnic University (SPbPU) in 2010, 2012 and 2015, respectively. Her
research interests include coding theory and its applications in telecommunications and storage systems. She is currently a Postdoctoral Research Associate in Telecommunications in the School of Electrical and Information Engineering at the University of Sydney.
\end{IEEEbiography}

\begin{IEEEbiography}
[{\includegraphics[width=1in,height=1.25in,clip,keepaspectratio]{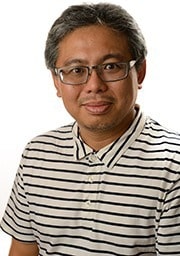}}]{Wibowo Hardjawana }(M'09) 
received the Ph.D. degree in electrical engineering from The University of Sydney, Australia, in 2009. He was an Australian Research Council Discovery Early Career Research Award Fellow and is now Senior Lecturer with the School of Electrical and Information Engineering, The University of Sydney. Prior to that he was Assistant Manager at Singapore Telecom Ltd, managing core and radio access networks. His current research interests are in 5/6G cellular radio access and wireless local area networks, with focuses in system architectures, resource scheduling, interference, signal processing and the development of corresponding standard-compliant prototypes.

\end{IEEEbiography}

\begin{IEEEbiography}
[{\includegraphics[width=1in,height=1.25in,clip,keepaspectratio]{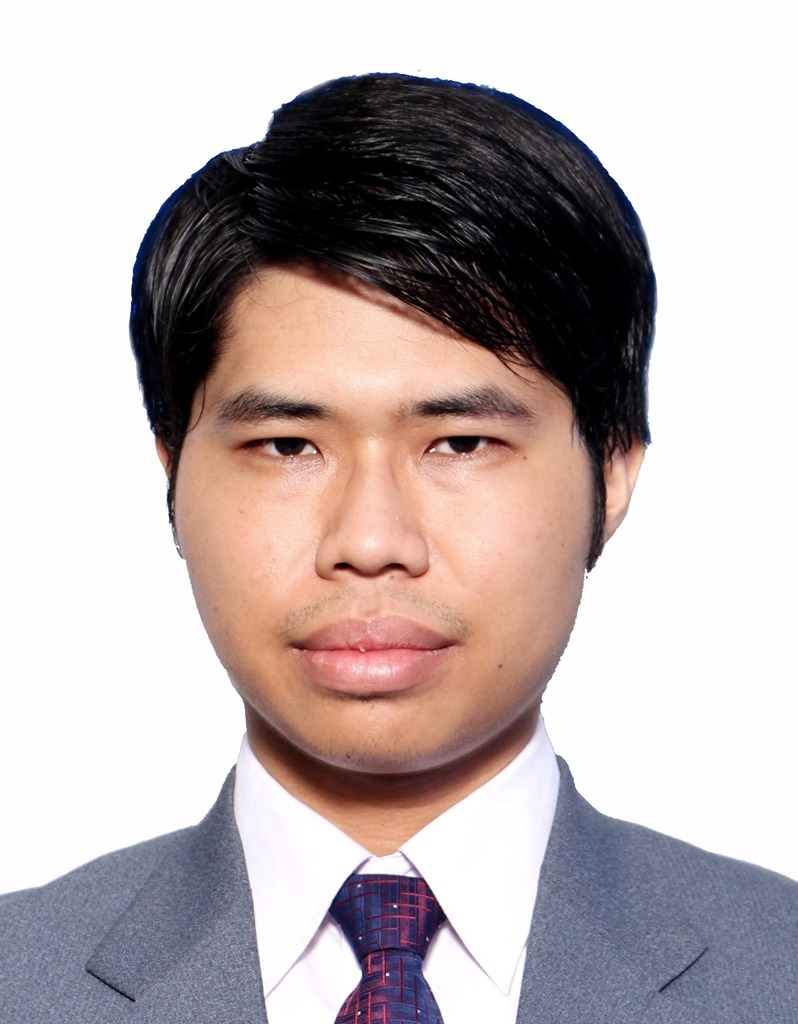}}]{Victor Andrean} 
received B.Eng. in electrical engineering from the University of Brawijaya, Indonesia, in 2016 and M.Sc. degree from Department of Electrical Engineering at National Taiwan University of Science and Technology, Taipei city, Taiwan, in 2019. He is currently pursuing a Ph.D. degree in the same department. His current research interests include deep learning and domain adaptation. He is currently also working at Pangea as a data scientist.

\end{IEEEbiography}

\begin{IEEEbiography}
 [{\includegraphics[width=1in,height=1.25in,clip,keepaspectratio]{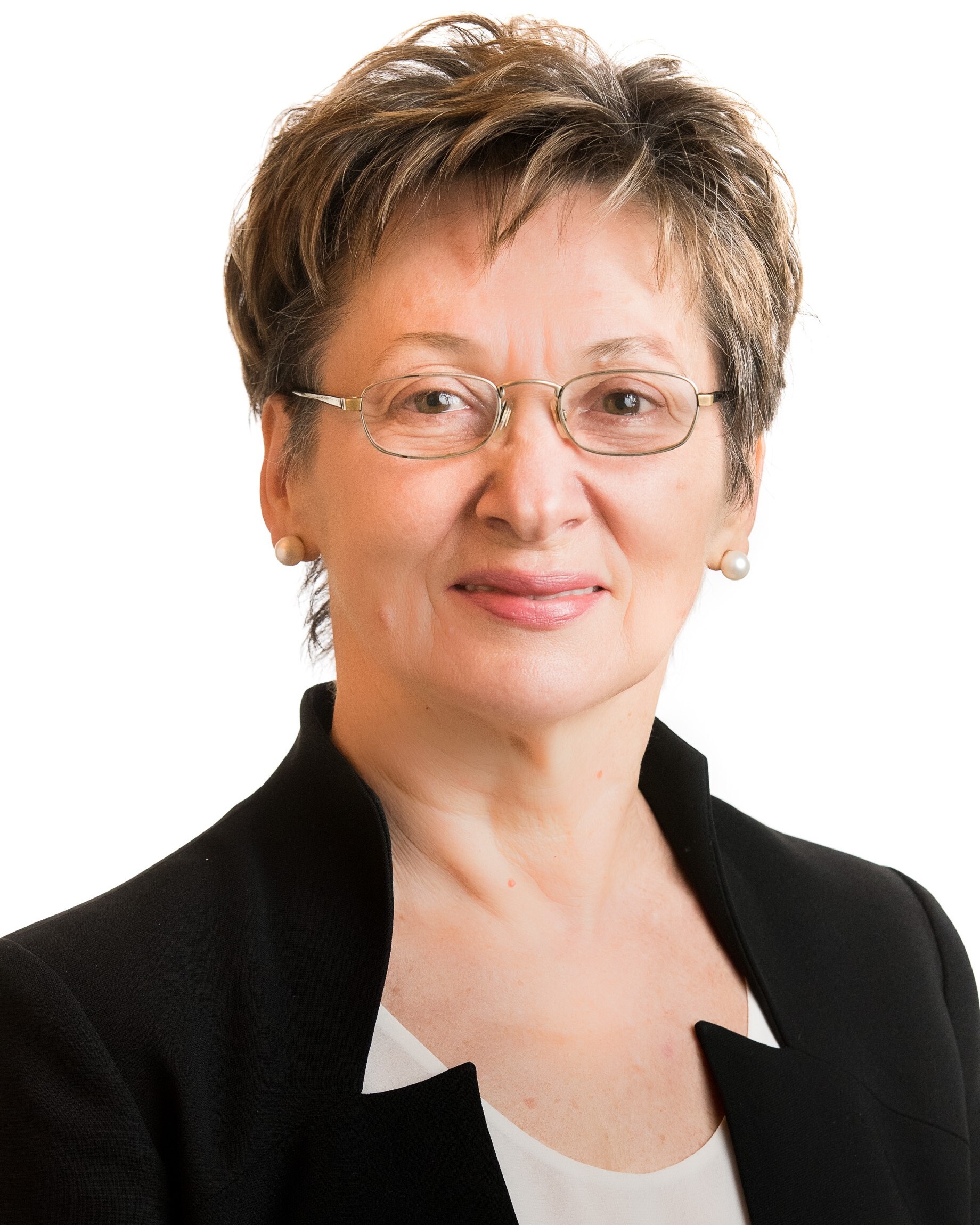}}]{Branka Vucetic}(Life Fellow, IEEE)
is an ARC Laureate Fellow and Director of the Centre of Excellence for IoT and Telecommunications at the University of Sydney. 
Her current research work is in wireless networks and the Internet of Things. In the area of wireless networks, she works on ultra-reliable low-latency communications (URLLC) and system design for millimetre wave frequency bands. In the area of the Internet of Things, Vucetic works on providing wireless connectivity for mission critical applications. Branka Vucetic is a Fellow of IEEE, the Australian Academy of Technological Sciences and Engineering and the Australian Academy of Science. The work of Branka Vucetic was supported in part by the Australian Research Council Laureate Fellowship grant number FL160100032.
\end{IEEEbiography}

\end{document}